\documentclass[letterpaper,english,prd,preprint,nofootinbib,showpacs,preprintnumbers,fleqn,floatfix]{revtex4}
\usepackage[T1]{fontenc}
\usepackage[latin1]{inputenc}
\usepackage{color}
\usepackage{graphicx}
\usepackage{amssymb}

\makeatletter

\usepackage{graphicx}
\usepackage{here}

\renewcommand{\vec}[1]{{\bf #1}}
\def\D0{D\O~}

\usepackage{babel}
\makeatother
\begin{document}

\newcommand{\SLASH}[2]{\makebox[#2ex][l]{$#1$}/}
\newcommand{\Dslash}{\SLASH{D}{.5}\,}
\newcommand{\kslash}{\SLASH{k}{.15}}
\newcommand{\pslash}{\SLASH{p}{.2}}
\newcommand{\qslash}{\SLASH{q}{.08}}

\newcommand{\prescr}[1]{{}^{#1}}
\newcommand{\ra}{\rightarrow}
\newcommand{\alphas}{\alpha_{s}}
\newcommand{\ttbar}{t\overline{t}}
\newcommand{\qqbar}{q\overline{q}}
\newcommand{\bcc}{\begin{center}}
\newcommand{\ecc}{\end{center}}
\newcommand{\beqn}{\begin{eqnarray}}
\newcommand{\eeqn}{\end{eqnarray}}
\newcommand{\beqy}{\begin{equation}}
\newcommand{\eeqy}{\end{equation}}
\newcommand{\mt}{m_t}
\newcommand{\mste}{m_{\tilde t_1}}
\newcommand{\mstz}{m_{\tilde t_2}}
\newcommand{\ste}{\tilde t_1}
\newcommand{\stz}{\tilde t_2}
\newcommand{\mstl}{m_{\tilde t_L}}
\newcommand{\mstr}{m_{\tilde t_R}}
\newcommand{\stl}{\tilde t_L}
\newcommand{\str}{\tilde t_R}
\newcommand{\msbe}{m_{\tilde b_1}}
\newcommand{\msbz}{m_{\tilde b_2}}
\newcommand{\msbl}{m_{\tilde b_L}}
\newcommand{\msbr}{m_{\tilde b_R}}
\newcommand{\astop}{A_{\tilde t}}
\newcommand{\phist}{\Phi_{\tilde t}}
\newcommand{\thetast}{\theta_{\tilde t}}
\newcommand{\tanb}{\tan{\beta}}
\newcommand{\qqa}{\mbox{$q\overline{q}$ annihilation}}
\newcommand{\ggf}{\mbox{gluon fusion}}
\newcommand{\sd}{\hat s}
\newcommand{\td}{\hat t}
\newcommand{\lt}{\lambda_t}
\newcommand{\ltb}{\lambda_{\bar t}}
\newcommand{\tr}{{\cal T}\!r}


\preprint{hep-ph/0703016}

\title{SUSY QCD one-loop effects in (un)polarized top-pair production at
hadron colliders}

\author{Stefan Berge,$^{1}$%
\footnote{E-mail: berge@physik.rwth-aachen.de%
} Wolfgang Hollik,$^{2}$%
\footnote{E-mail: hollik@mppmu.mpg.de%
} Wolf M. Mosle,%
\footnote{now at MicroStrategy, Inc.%
} and Doreen Wackeroth,$^{3}$%
\footnote{E-mail: dow@ubpheno.physics.buffalo.edu%
}}

\affiliation{$^{1}$Institut f\"ur Theoretische Physik, RWTH Aachen,
D-52056 Aachen, Germany\\
$^{2}$Max-Planck-Institut für Physik
(Werner-Heisenberg-Institut),
Föhringer Ring 6, D-80805 Munich, Germany\\
$^{3}$Department of Physics, University at Buffalo, The State University of New York,
Buffalo, NY 14260-1500, U.S.A.
}

\begin{abstract}
We study the effects of ${\cal O}(\alpha_s)$ supersymmetric QCD
(SQCD) corrections on the total production rate and kinematic
distributions of polarized and unpolarized top-pair production in $pp$ and $p \bar p$
collisions.  At the Fermilab Tevatron $p\bar p$ collider, top-quark pairs are
mainly produced via quark-antiquark annihilation, $q\bar q\to t\bar
t$, while at the CERN LHC $pp$ collider gluon-gluon scattering, $gg\to
t\bar t$, dominates. We compute the complete set of ${\cal
O}(\alpha_s)$ SQCD corrections to both production channels
and study their dependence on the parameters of the Minimal
Supersymmetric Standard Model. In particular, we discuss
the prospects for observing strong, loop-induced SUSY
effects in top-pair production at the Tevatron Run~II and the LHC. 
\end{abstract}

\date{\today{}}

\maketitle

%
%
%
%
\section{Introduction}
\label{sec:Introduction}
Since the discovery of the top quark at the Fermilab Tevatron $p \bar
p$ collider in 1995~\cite{Abe:1995hr,Abachi:1995iq}, the top-physics
program has shifted to precisely studying its properties.  The high
top-quark yields at the Tevatron Run~II and the CERN LHC $pp$ collider
open a new, rich field of top-quark phenomenology, which may enable a
precision physics program with top-quark observables, in the spirit of
the successful LEP/SLC studies of precision $Z$ boson
observables~\cite{lepewwg:2005di}.  

The Standard Model (SM) of electroweak and strong interactions has
seen impressive experimental confirmation.  However, the Higgs boson,
which is predicted by the SM as a direct consequence of mass
generation via spontaneous electroweak symmetry breaking (ESB), has so
far eluded direct observation.  Moreover, the many shortcomings of the
SM suggest that it may indeed be a low-energy limit of a more
fundamental theory.  One of the most promising candidates for a theory
beyond the SM is Supersymmetry (SUSY) (for a review see, e.g.,
Refs.~\cite{Fayet:1976cr,Nilles:1983ge,Haber:1984rc,Wess:1992cp,Yao:2006px}).
Supersymmetry solves the finetuning problem, allows for gauge coupling
unification, provides a dark matter candidate, predicts a light Higgs
boson and agrees with precision electroweak measurements.  However, no
direct experimental evidence of SUSY has been found yet.

The large mass of the top quark suggests that it plays a special role
in the mechanism of ESB, and that new physics connected to ESB may be
found first through precision studies of top-quark observables.
Deviations of experimental measurements from the SM predictions,
including electroweak and QCD corrections, could indicate new
non-standard top production or decay mechanisms.  If supersymmetric
particles are detected at the LHC, the comparison of precisely
measured top-quark observables with their predictions including SUSY
loop effects may yield additional information about the underlying
model, that may not be accessible otherwise.  Measuring precisely the
properties of the top quark therefore is an important goal at the
Tevatron Run~II and the LHC. To fully exploit the potential of these
colliders for precision top-quark physics, it is crucial that
predictions for top-quark observables include higher-order corrections
within the SM and beyond.  In this paper we study the impact of SUSY
QCD (SQCD) one-loop corrections within the Minimal Supersymmetric SM
(MSSM) on strong top-pair production at both the Tevatron Run~II and the LHC.
  
Presently, the total top-pair production cross section has been
measured at the Tevatron~\footnote{See www-cdf.fnal.gov and
  www-d0.fnal.gov for most recent results from the CDF and \D0
  collaborations, respectively.}  with a relative uncertainty of
$\Delta\sigma_{t\bar t}/\sigma_{t\bar t}=12 \%$ (with ${\cal L}=760 \;
{\rm pb}^{-1}$)~\cite{Boisvert:2006kw} and is in good agreement with
the theoretical QCD
prediction~\cite{Kidonakis:2003qe,Cacciari:2003fi}.  It is anticipated
that at the Tevatron Run~II the top-pair production cross section will
be measured with a relative uncertainty of $\Delta \sigma_{t \bar
  t}/\sigma_{t\bar t} \approx 10\%$ (with ${\cal L}=2 \;
\mbox{fb}^{-1}$). At the LHC, the goal is to measure $\sigma_{t\bar
  t}$ ultimately with a relative uncertainty of $< 5\%$.  The current
experimental uncertainty still leaves room for SUSY loop effects.
Through the virtual presence of supersymmetric particles in quantum
corrections, the measurement of the cross section, kinematic
distributions and the extraction of masses and couplings may be
affected. Among the top-quark observables under study,
parity-violating asymmetries in the production of left and
right-handed top quarks are of special interest. They have the
potential to provide a clean signal of non-SM physics: QCD preserves
parity and the SM induced asymmetries are too small to be observable,
at least at the Tevatron $p \overline{p}$
collider~\cite{Kao:1999kj,Bernreuther:2006vg}.  The produced top
quarks decay almost entirely into a bottom quark and a $W$ boson
before they can hadronize~\cite{Bigi:1986jk} or flip their spins.  The
spin correlation of the top-pair system will therefore be preserved
and can be measured by studying angular distributions of the decay
products~\cite{Kuhn:1983ix}.  In order to determine the sensitivity of
top-quark observables to SUSY loop effects, it is necessary to
calculate the radiative corrections to the top production and decay
processes, both within the SM and its supersymmetric extension, and to
implement these calculations in a Monte Carlo program. The latter will
allow an efficient determination of those observables that are most
sensitive to supersymmetric particles, after taking into account the
detector response.

In this paper, we present the calculation of the complete SQCD ${\cal
  O}(\alpha_s)$ corrections to the main production channels of strong
top-pair production at the Tevatron and LHC, $q\bar q \to t\bar t$
($q\bar q$ annihilation) and $gg \to t \bar t$ (gluon fusion), and
present a detailed study of their numerical impact on top-quark
observables in both unpolarized and polarized $t \bar t$ production
(the initial-state quarks and gluons are unpolarized). We consider the
MSSM with CP-conserving couplings.  The inclusion of the top-quark
decays and implementation in a Monte Carlo program for off-shell $t
\bar t$ production is work in progress~\cite{Beneke:2000hk}, but
beyond the scope of this paper.

The SM predictions to both polarized and unpolarized $t\bar t$
production at hadron colliders are under good theoretical control: The
Born-level amplitudes to $q\bar q$ annihilation and gluon fusion were
first considered in
Refs.~\cite{Georgi:1978kx,Combridge:1978kx,Gluck:1977zm,Babcock:1977fi,Hagiwara:1978hw,Jones:1977di}.
At next-to-leading order (NLO), the SM QCD corrections for the total
cross section and spin-independent kinematic distributions have been
presented in
Refs.~\cite{Nason:1987xz,Beenakker:1988bq,Altarelli:1988qr,Beenakker:1990ma,Mangano:1991jk,Frixione:1995fj},
electroweak corrections have been studied in
Refs.~\cite{Beenakker:1993yr,Kao:1997bs,Kuhn:2005it,Bernreuther:2005is,Moretti:2006nf,Moretti:2006ea,Kuhn:2006vh}
and soft gluon resummation and threshold effects in
Refs.~\cite{Bonciani:1998vc,Kidonakis:2001nj,Kidonakis:2003qe,Kidonakis:2004hr,Banfi:2004xa}.
To measure spin correlations and asymmetries at the Tevatron and LHC,
higher-order corrections to polarized $t\bar t$ production need to be
known as well: The NLO QCD and electroweak corrections for the
polarized production have been calculated in
Refs.~\cite{Bernreuther:2000yn,Bernreuther:2001rq,Bernreuther:2001bx,Bernreuther:2004jv}
and
Refs.~\cite{Kao:1994rn,Kao:1997bs,Kao:1999kj,Bernreuther:2005is,Bernreuther:2006vg},
respectively.

A number of studies of higher-order corrections within the MSSM are
also available: The SUSY electroweak one-loop corrections to
unpolarized on-shell $t \bar t$ production at hadron colliders have
been calculated in Refs.~\cite{Yang:1996dm, Kim:1996nz, Hollik:1997hm}
($q\bar q$ annihilation) and Ref.~\cite{Hollik:1997hm} (gluon fusion),
and the same set of corrections for polarized top-pair production has
been considered in Ref.~\cite{Kao:1999kj}.  A study of leading
logarithmic SUSY electroweak corrections can be found in
Ref~\cite{Beccaria:2004sx} and of Yukawa corrections in a
two-Higgs-doublet model is given in Ref.~\cite{Stange:1993td}. The NLO
SQCD corrections to the $gg \to t \bar t$ subprocess without
polarization have been presented in Refs.~\cite{Zhou:1997fw,Yu:1998xv}
and in Ref.~\cite{Yu:1998xv} for the scattering of polarized protons.
The unpolarized $q\bar q$ annihilation subprocess has been
investigated at NLO SQCD by a number of groups: In
Ref.~\cite{Kim:1996nz} the corrections with the assumption of
negligible box contributions have been calculated, whereas in
Refs.~\cite{Li:1996jf, Li:1995fj} the gluon self-energy and crossed
box diagram contributions have been neglected.  The full NLO SQCD
corrections to $q\bar q\to t\bar t$ have been examined in
Refs.~\cite{Alam:1996mh, Sullivan:1996ry, Wackeroth:1998wm}.  However,
the results of these different calculations do not agree.  The origin
of this disagreement has been found~(see footnote in
Ref.~\cite{Wackeroth:1998wm}), and is due to a difference in the
signs of the direct and crossed box diagrams (see
Fig.~\ref{fig:qqttboxes}), leading to large numerical differences that
are especially pronounced in case of comparable small top squark and
gluino masses. However, a reliable, complete and detailed study of the
impact of NLO SQCD corrections on $t\bar t$ production based on the
correct result is not available in the literature.

This paper represents the first detailed study of the numerical impact
of NLO SQCD corrections on both polarized and unpolarized strong
$t\bar t$ production at hadron colliders, that considers both main
production channels, $q \bar q$ annihilation and gluon fusion.  We
present numerical results for the total production cross section, the
invariant $t\bar t$ mass and top transverse momentum distributions,
and polarization asymmetries, at both the Tevatron Run~II and the LHC,
respecting current experimental bounds on the masses of the
supersymmetric particles. We provide analytic expressions for the
complete NLO SQCD corrections to polarized $t\bar t$ production and
compare our results with the literature where available. Our results
have been thoroughly cross-checked by performing at least two
independent calculations, based on analytic methods and the
FeynArts~\cite{Hahn:2000kx, Kublbeck:1990xc} and
FormCalc~\cite{Hahn:2004rf,Hahn:1998yk} packages.

The paper is structured as follows: In Section~\ref{sec:nlosqcd} we
present the analytic results for the partonic $q\bar q,gg \to t \bar
t$ scattering processes, specify our renormalization procedure and our
convention of the squark mixing parameters and choice of MSSM input
parameters.  Section~\ref{sec:results} contains a detailed study of
the numerical impact of the NLO SQCD corrections on polarized and
unpolarized top-pair production at the Tevatron RUN~II and LHC.  We
first discuss their impact in detail at parton level and then present
their most pronounced effects to the observable, hadronic cross
sections, followed by a discussion of their prospects to be observable
at these colliders. We also include a comparison with available
results in the literature.  Finally, a summary of our results and
conclusion can be found in Section~\ref{sec:conclusion}. Details of
the calculation and explicit analytic expressions are provided in the
appendices. We also made available a Fortran code~\footnote{The
  Fortran code is provided at
  http://ubpheno.physics.buffalo.edu/$\sim$dow/ppttsqcd.} that
calculates the cross sections presented in this paper to polarized and
unpolarized top-pair production at hadron colliders at leading-order
(LO) QCD and NLO SQCD.


\section{Strong top-pair production at NLO SQCD}\label{sec:nlosqcd}

\begin{figure}[h]
\begin{center}
\hspace*{0.5cm}
\includegraphics[width=4.6cm,
  keepaspectratio]{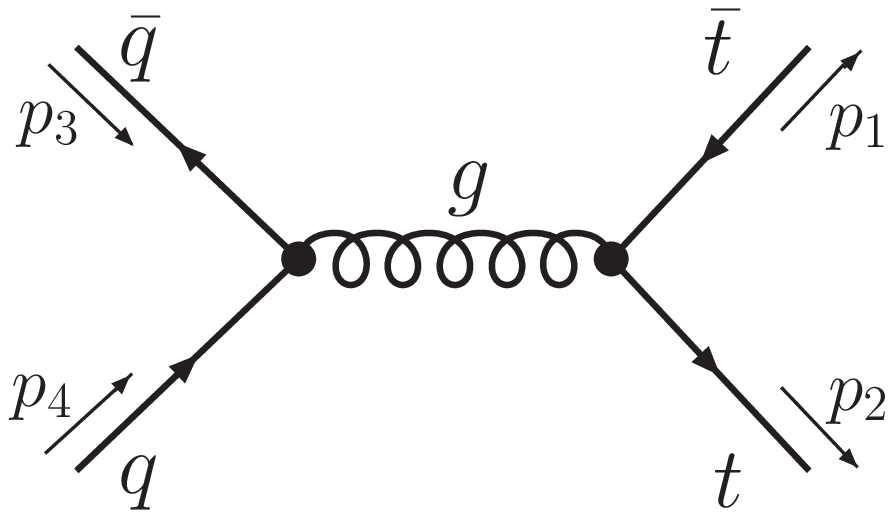}
\\[-6pt]
\end{center}
\mbox{}\\[-44pt]
\caption{\emph{Feynman diagram to the 
$q\overline{q} \to t\overline{t}$ subprocess at LO QCD.}}%
\label{fig:qqbarlo}
\end{figure}

\begin{figure}[h]
\begin{center}
\hspace*{0.5cm}
\setlength{\unitlength}{1cm}
\begin{picture}(15,2.5)
\put(0.6,0){\includegraphics[width=4.0cm,
  keepaspectratio]{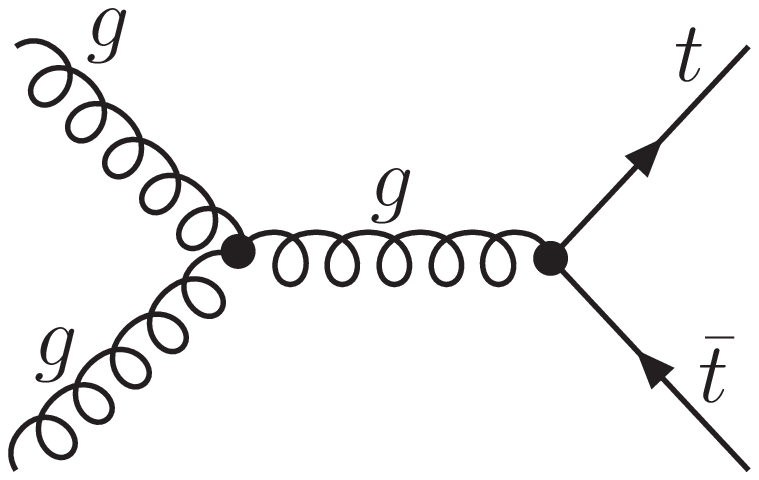}}
\put(5.3,0){\includegraphics[width=3.6cm,
  keepaspectratio]{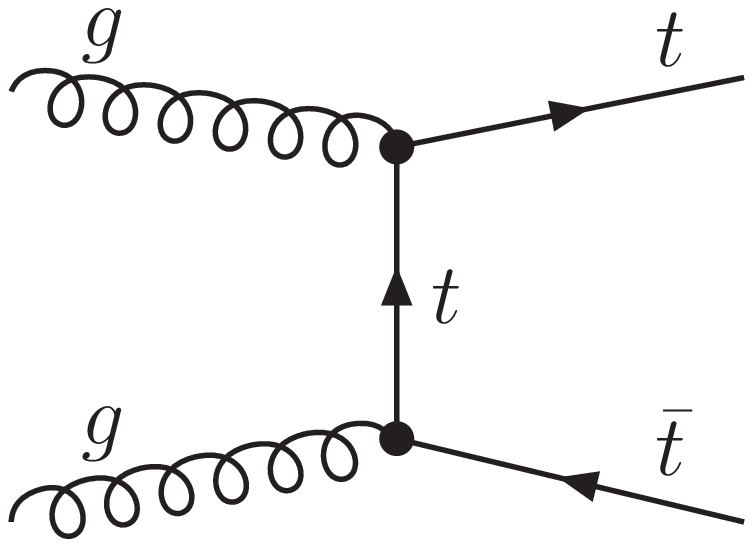}}
\put(9.9,0){\includegraphics[width=3.7cm,
  keepaspectratio]{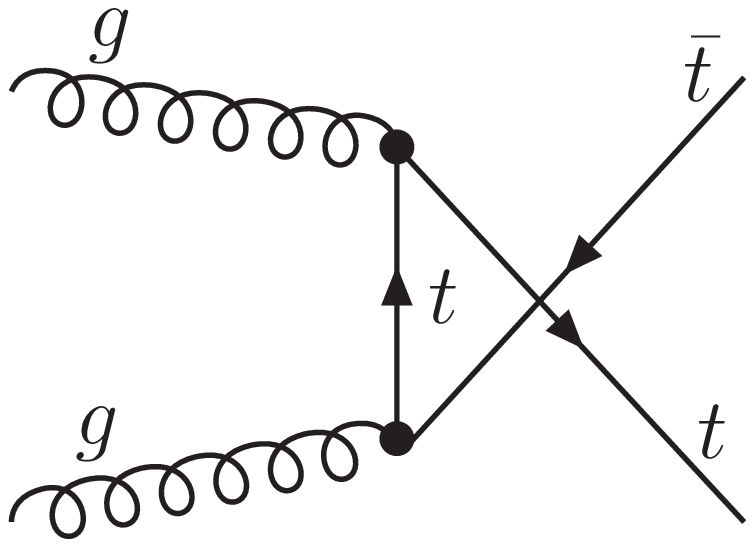}}
\end{picture}
\hspace*{-.05cm}(a)\hspace*{4.15cm}(b)\hspace*{4.1cm}(c)
\end{center}
\mbox{}\\[-44pt]
\caption{\emph{Feynman diagrams to the $s$ (diagram (a)), $t$ (diagram (b)) and $u$ (diagram(c)) 
channels of the
$gg \to t\overline{t}$ subprocess at LO QCD.}}%
\label{fig:ggfusionlo}
\end{figure}

At the Tevatron the main production mechanism for the strong
production of top-quark pairs is the annihilation of a quark-antiquark
pair
\[ q(p_4) + \overline{q}(p_3) \rightarrow t(p_2,\lambda_t) + 
\overline{t}(p_1,\lambda_{\bar t}) \; ,\] 
whereas at the LHC the
top-quark pairs are mainly produced via the fusion of two gluons
\[ g(p_4) + g(p_3) \rightarrow t(p_2,\lambda_t) + \overline{t}(p_1,\lambda_{\bar t}) \; ,\]
where $p_i, i=1,\cdots,4$ are the four-momenta of the in- and
outcoming particles and $\lambda_t(\lambda_{\overline t})=\pm 1/2$
denotes the top(antitop) helicity state. The corresponding Feynman
diagrams at LO QCD are shown in
Figs.~\ref{fig:qqbarlo},~\ref{fig:ggfusionlo}. The partonic
differential cross sections to the $\qqa$ and $\ggf$ processes for
polarized top-quark pairs at NLO SQCD can be written as
\begin{eqnarray}
\label{eq:parton}
d \hat \sigma_{q\bar q,gg}^{NLO}(\hat t,\hat s,\lambda_t,\lambda_{\bar t})
&=& d \hat \sigma_{q\bar q,gg}^{LO}(\hat t,\hat s,\lambda_t,\lambda_{\bar t})+ \delta d \hat \sigma_{q\bar q,gg}(\hat t,\hat s,\lambda_t,\lambda_{\bar t}) \nonumber\\
&=& \frac{d\Phi_{2\to 2}}{8 \pi^2\hat s} \left[\overline{\sum} |{\cal M}_{B}^{q\bar q,gg}|^2 + 
2 {\cal R}e \; \overline{\sum}
(\delta {\cal M}_{q\bar q,gg}^{SQCD} \times {\cal M}_{B}^{q\bar q,gg *}) \right]\; ,
\end{eqnarray}
where $\hat s=(p_1+p_2)^2=(p_3+p_4)^2, \hat
t=(p_3-p_1)^2=(p_4-p_2)^2=m_t^2-\hat s (1-\beta_t\cos\hat \theta)/2$
are Mandelstam variables with $\hat \theta$ denoting the scattering
angle in the parton center-of-mass system (CMS) and
$\beta_t=\sqrt{1-4\mt^2/\sd}\,$ is the top-quark velocity. The matrix
elements squared are averaged over initial-state spin and color
degrees of freedom and summed over final-state color degrees of
freedom.  The phase space of the $2\to 2$ scattering process,
$d\Phi_{2\to 2}$, as usual reads
\begin{equation}
\label{eq:phasespace}
\int d\Phi_{2\to 2}=\int \frac{d^3 \vec{p}_1}{2p_1^0 } \frac{d^3 \vec{p}_2}{2 p_2^0} \delta^{(4)}(p_3+p_4-p_1-p_2)=\frac{\beta_t}{8} \int_0^{2\pi} d\phi^* \int_{-1}^{1} d\cos\hat \theta\;  ,
\end{equation} 
where the phase space has been evaluated in the parton CMS with
$\phi^*$ denoting the azimuthal angle.  The matrix elements, ${\cal
  M}_{B}^{q\bar q,gg}$ and $\delta {\cal M}_{q\bar q,gg}^{SQCD}$,
describe respectively the LO QCD and one-loop SQCD contributions to
$\qqa$ and $\ggf$. The explicit expressions for ${\cal M}_{B}^{q\bar
  q,gg}$ and the spin and color averaged transition amplitude squared,
$\overline{\sum} |{\cal M}_{B}^{q\bar q,gg}|^2$, can be found in
Ref.~\cite{Beenakker:1993yr} (Sec.~2.1 and 2.2) for unpolarized and in
~\cite{Kao:1999kj} (Appendix A) for polarized top-pair production.

The observable hadronic differential cross sections are obtained by
convoluting the partonic cross sections of Eq.~(\ref{eq:parton}) with
parton distribution functions (PDFs)
\begin{eqnarray}
\label{eq:hadwq}
d\sigma_{LO,NLO}(S,\lambda_t,\lambda_{\bar t}) &=& \sum_{ij=q\bar
q,gg}\frac{1}{1+\delta_{ij}} \int_0^1 dx_1 dx_2\nonumber\\
&\times&  \left[f_i(x_1,\mu_F)
f_j(x_2,\mu_F) d\hat \sigma_{ij}^{LO,NLO}(\alpha_s(\mu_R),\hat s,\hat
t,\lambda_t,\lambda_{\bar t}) + i \leftrightarrow j
\right] 
\end{eqnarray}
with $S=\hat s /(x_1 x_2)$.  In the numerical evaluation, we use the
CTEQ6L1 (LO) set of parton distribution functions~\cite{Lai:1999wy}
with the QCD factorization ($\mu_F$) and renormalization scale
($\mu_R$) chosen to be $\mu_F=\mu_R=m_t$. Accordingly, the cross
section is evaluated using the one-loop evolution of the strong coupling
constant with $n_f=5$ light flavors, $\Lambda^5_{QCD} = 0.165$ GeV and
$\alpha_s^{LO}(M_Z)=0.130$, which yields $\alpha_s^{LO}(\mt) = 0.1176$.
 
In the remaining parts of this section we will present the explicit
expressions for the partonic NLO SQCD differential cross sections,
$d\hat \sigma_{q\bar q}^{NLO}$ and $d\hat \sigma_{gg}^{NLO}$, to
polarized top-quark pair production.  The corresponding cross sections
for unpolarized $t\bar t$ production can be obtained by summing over
the top and antitop-quark helicity states, $\sum_{\lt,\ltb=\pm1/2}
d\hat \sigma^{LO,NLO}_{q\bar q,gg}(\lt,\ltb)$.

\mbox{}\\[-50pt]

\subsection{NLO SQCD corrections to $\qqa$ and $\ggf$}
\label{sec:parton}

\begin{figure}[h]
\begin{center}
\hspace*{0.5cm}
\includegraphics[width=4.0cm,
  keepaspectratio]{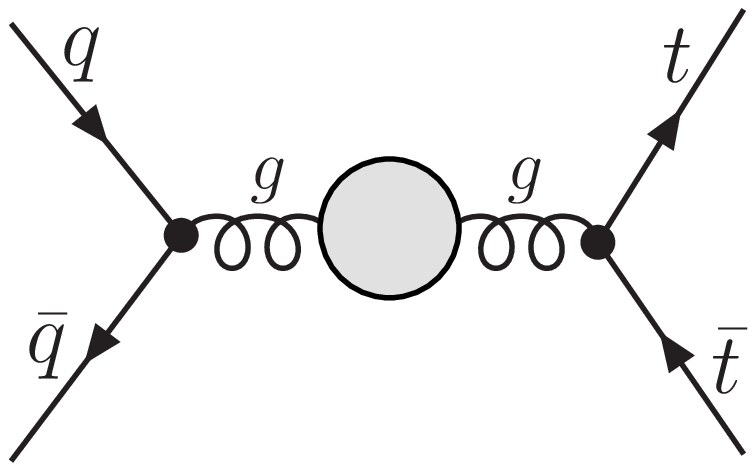}
\hspace*{.15cm}
\includegraphics[width=4.0cm,
  keepaspectratio]{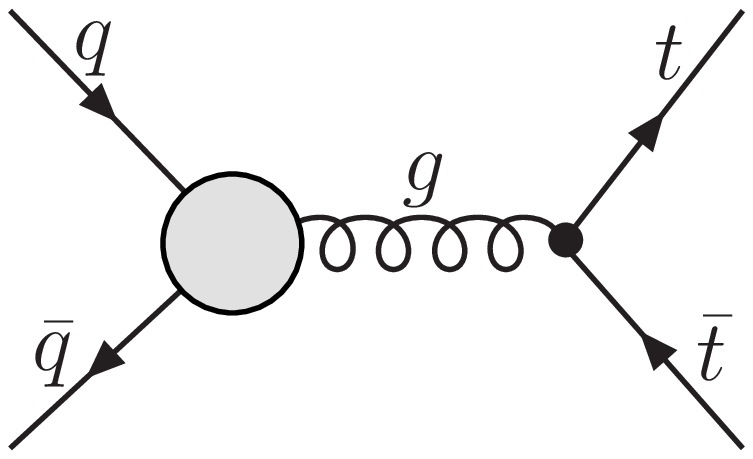}
\hspace*{.15cm}
\includegraphics[width=4.0cm,
  keepaspectratio]{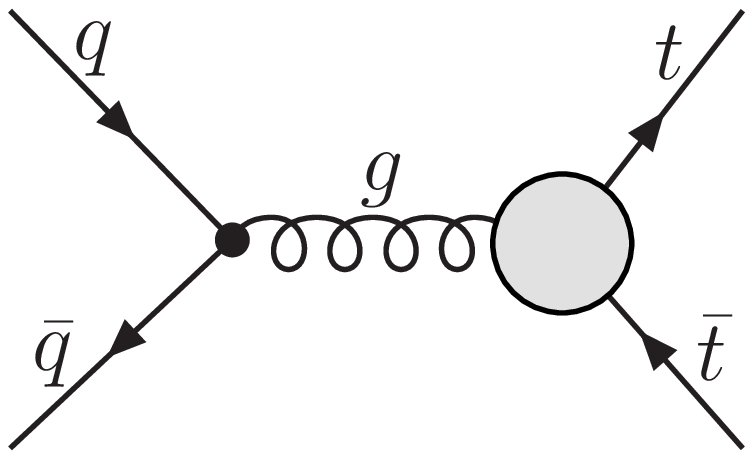}\\[-6pt]
\hspace*{.55cm}(a)\hspace*{3.95cm}(b)\hspace*{4.0cm}(c)
\end{center}
\mbox{}\\[-44pt]
\caption{\emph{Generic self-energy (diagram (a)) and vertex (diagrams (b) and (c)) corrections to $q\overline{q} \to t\overline{t}$ at NLO SQCD. The corrections to the gluon propagator and the $gq\bar q$ vertex are explicitly shown in Figs.~\ref{fig:gself_renorm} and~\ref{fig:gtt_renorm}, respectively.}}%
\label{fig:qqtt_gen_vertself}
\end{figure}

The SQCD ${\cal O}(\alpha_s)$ corrections modify the tree-level $g
t\bar t, ggg$ and $gq \bar q$ vertices and the gluon propagator through
the virtual presence of gluinos ($\tilde g$), squarks ($\tilde
q_{L,R}$), stops ($\tilde t_{L,R}$) and sbottoms ($\tilde b_{L,R}$),
i.e.~the superpartners of the gluon, the left and right-handed light
quarks, top and bottom quarks, respectively, 
as shown in Figs.~\ref{fig:qqtt_gen_vertself}-\ref{fig:ggtt_boxes}.  
For completeness, we also provide the corresponding Feynman rules in
Appendix~\ref{sec:feynman_rules}.
%
%

\begin{figure}[h]
\begin{center}
  \includegraphics[width=4.9cm,
  keepaspectratio]{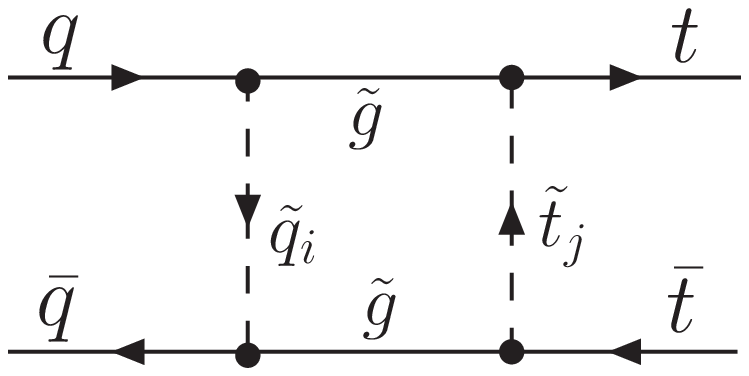}
\hspace*{1.5cm}
  \includegraphics[width=4.9cm,
  keepaspectratio]{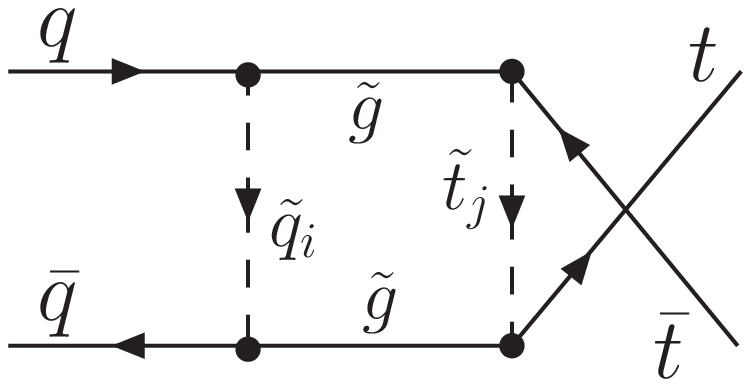}\\
\vspace*{-2pt}
\hspace*{-.2cm}(a)\hspace*{6.2cm}(b)\\
\end{center}
\vspace*{-20pt}
\caption{\emph{(a) Direct box diagram  
and (b) crossed box diagram contributing to
$q\overline{q}\to t\overline{t}$ at NLO SQCD. Graphs
containing squarks/stops are summed over the squark/stop mass eigenstates i,j=L,R
(no mixing), i,j=1,2 (with mixing).}}\label{fig:qqttboxes}
\end{figure}
%
%
\begin{figure}[h]
\begin{center}
\hspace*{0.5cm}
\includegraphics[width=3.4cm,
  keepaspectratio]{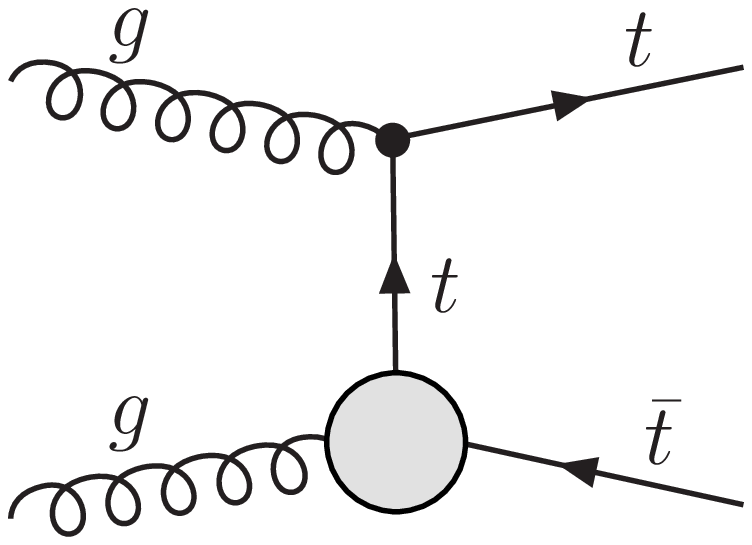}
\hspace*{.15cm}
\includegraphics[width=3.4cm,
  keepaspectratio]{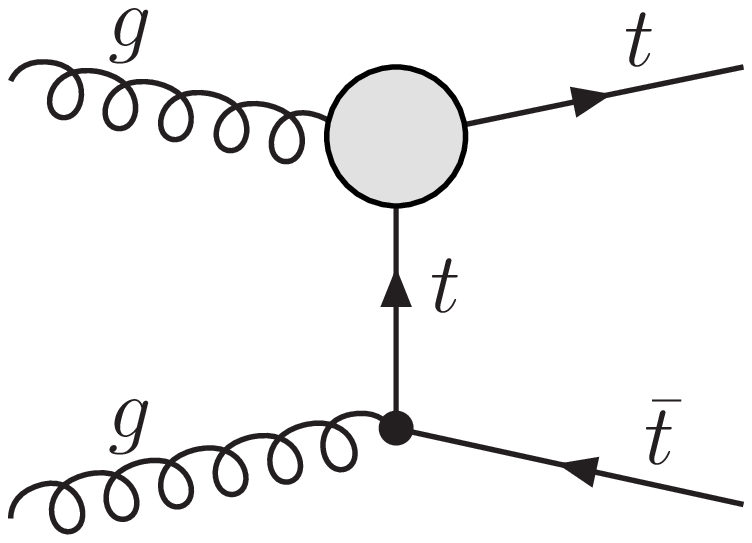}
\hspace*{.15cm}
\includegraphics[width=3.4cm,
  keepaspectratio]{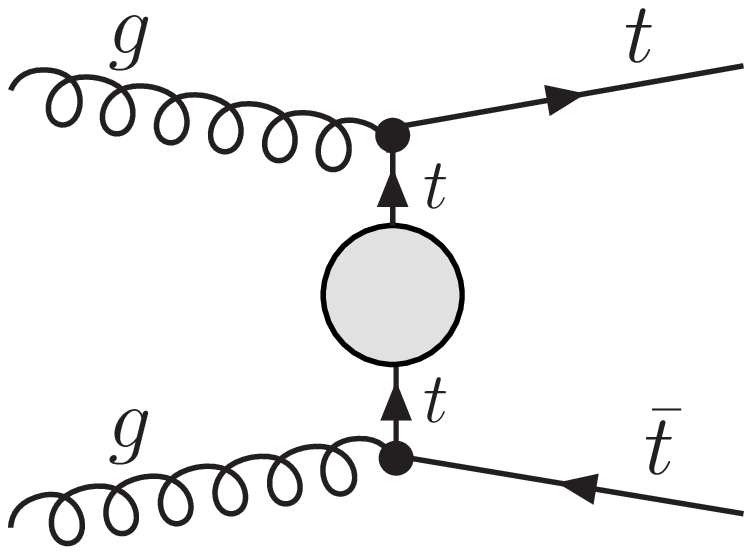}
\hspace*{.0cm} $\begin{array}{l} \mbox{\hspace*{.8cm}crossed}\\[-14pt] \mbox{{\bf \large +}} \\[-16pt]  \mbox{\hspace*{.8cm}diagrams}\\[63pt]\end{array}$
\\[-43pt]
\hspace*{.5cm}(a)\hspace*{3.35cm}(b)\hspace*{3.35cm}(c)\hspace*{2.5cm}
\\[25pt]
\hspace*{.5cm}
\includegraphics[width=3.4cm,
  keepaspectratio]{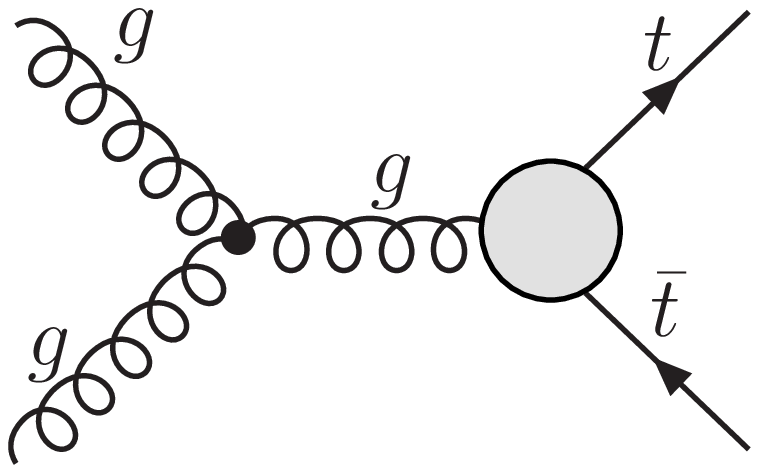}
\hspace*{.15cm}
\includegraphics[width=3.4cm,
  keepaspectratio]{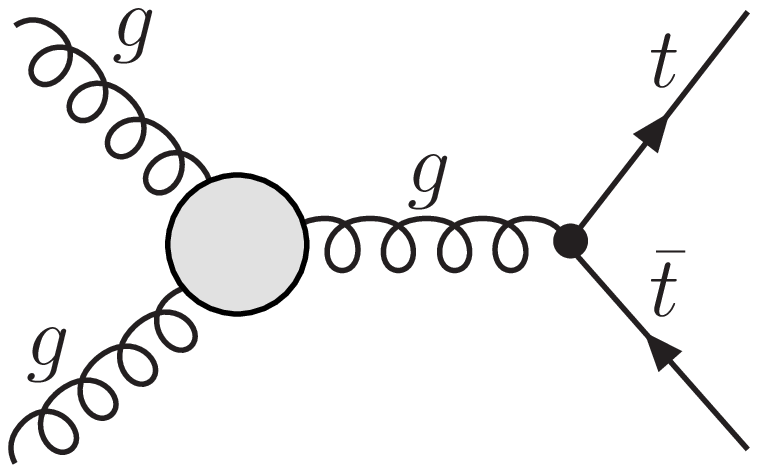}
\hspace*{.15cm}
\includegraphics[width=3.4cm,
  keepaspectratio]{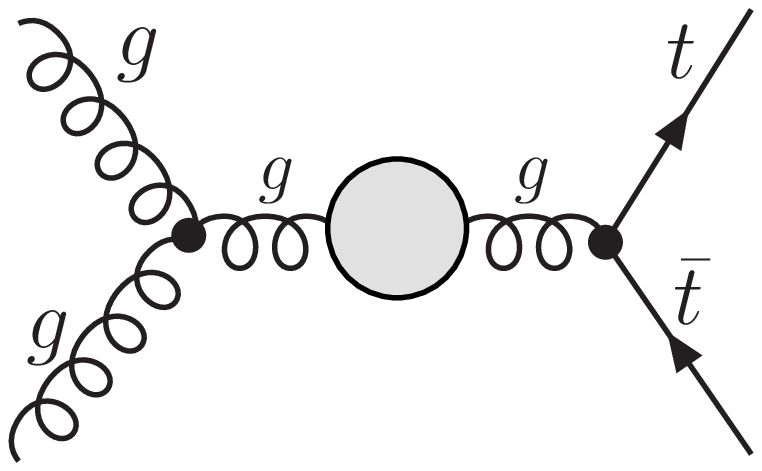}
\hspace*{2.7cm}
\end{center}
\mbox{}\\[-39pt]
\caption{\emph{Generic vertex and self-energy corrections to the
$t(u)$ channel (diagrams (a)-(c)) and the $s$ channel (lower row) of the $gg \to t\overline{t}$ subprocess 
at NLO SQCD. The corrections to the $gt\bar t$ vertex, 
the quark and gluon propagators, and the $ggg$ vertex are explicitly shown in Figs.~\ref{fig:gtt_renorm},~\ref{fig:qself_renorm},~\ref{fig:gself_renorm}, and~\ref{fig:ggg_renorm}, respectively.}}%
\label{fig:ggtt_gen_vertself}
\end{figure}
%
%
\begin{figure}[h]
\begin{center}
\includegraphics[width=3.7cm,
  keepaspectratio]{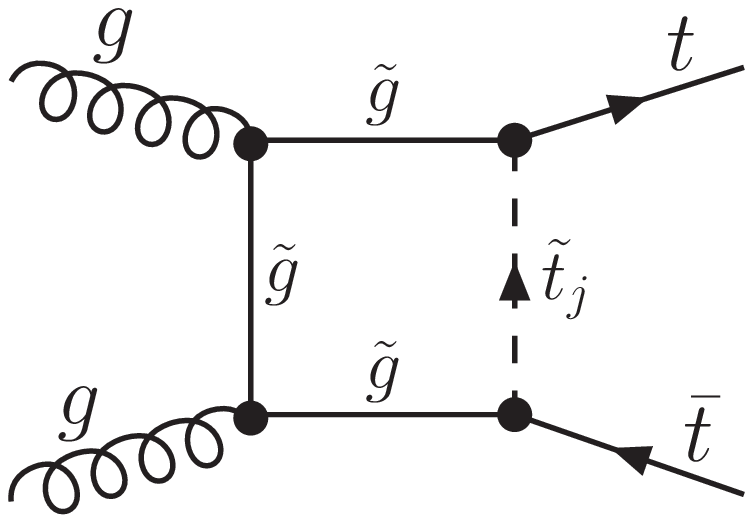}
\hspace*{.2cm}
\includegraphics[width=3.7cm,
  keepaspectratio]{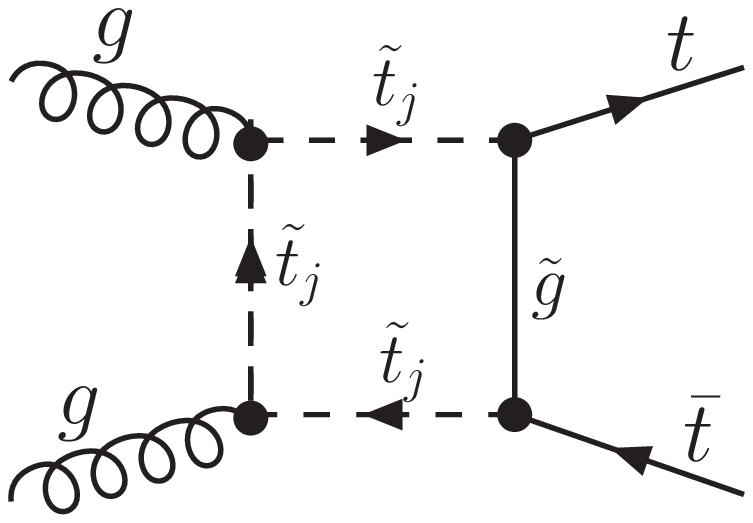}
\hspace*{.1cm} $\begin{array}{c} \mbox{{\bf \large + }\ crossed diagrams} \\[54pt]\end{array}$
\hspace*{.7cm}
\\[-30pt]
\hspace*{.5cm}(a)\hspace*{3.35cm}(b)\hspace*{5.5cm}
\\[25pt]
\hspace*{1.3cm}
\includegraphics[width=3.7cm,
  keepaspectratio]{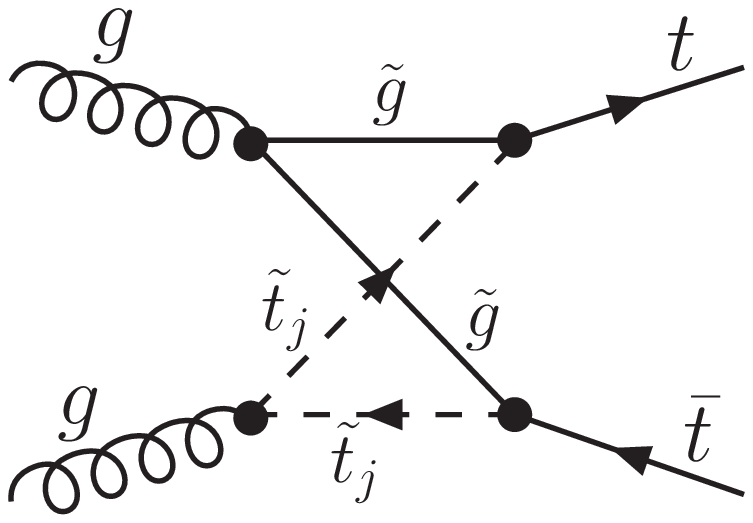}
\hspace*{.2cm}
\includegraphics[width=3.7cm,
  keepaspectratio]{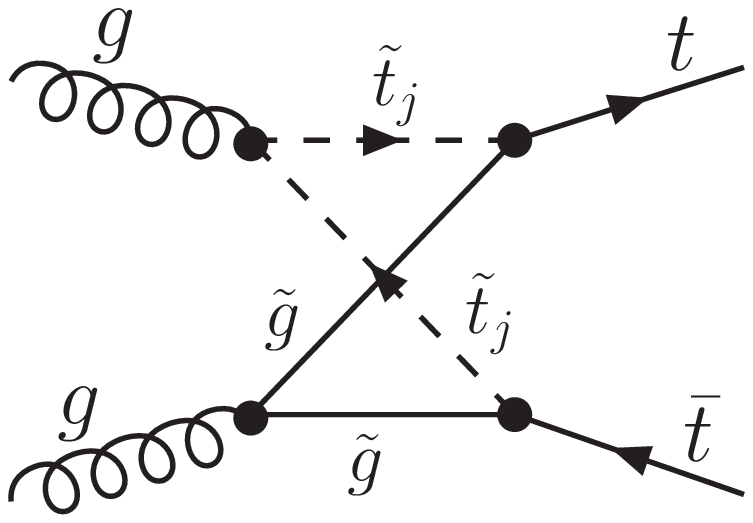}
\hspace*{.2cm}
\includegraphics[width=3.7cm,
  keepaspectratio]{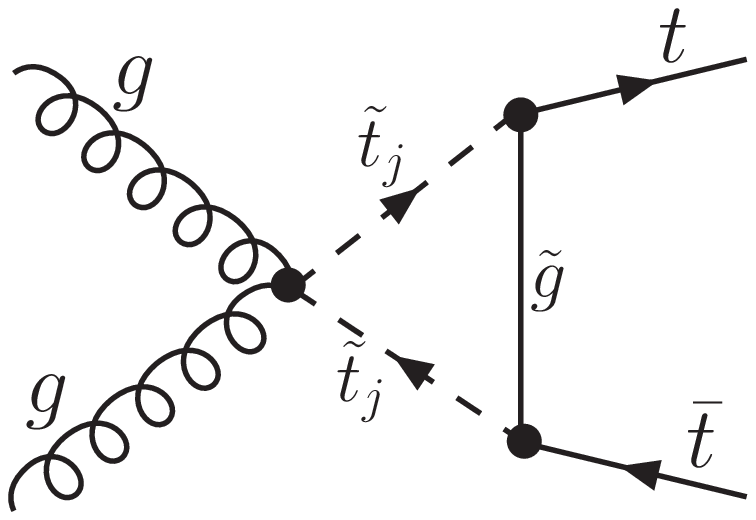}
\hspace*{2.3cm}
\end{center}
\mbox{}\\[-35pt]
\hspace*{.7cm}(c)\hspace*{3.42cm}(c)\hspace*{3.15cm}(d)\hspace*{2.cm}
\\[4pt]
\caption{\emph{Box corrections to $gg \to t\overline{t}$ 
at NLO SQCD. The box diagrams $d=a,b$ contribute to both
the $t$ and $u$(crossed) channels, whereas
the box diagrams $d=c$ and the box-triangle correction (diagram (d))
involving a quartic $gg\tilde t \tilde t$ interaction only contribute
to the $t$ channel. Graphs
containing stops are summed over the stop mass eigenstates j=L,R
(no mixing), j=1,2 (with mixing)}}%
\label{fig:ggtt_boxes}
\end{figure}

We closely follow Refs.~\cite{Beenakker:1993yr,Kao:1999kj} and write
the NLO SQCD matrix elements, $\delta M^{SQCD}_{q\bar q,gg}$, of
Eq.~(\ref{eq:parton}) in terms of form factors that describe the SQCD
one-loop corrections. In case of gluon fusion,
these form factors multiply so-called standard matrix elements (SMEs)
that contain the information about the Dirac matrix structure.  After
the interference with the Born matrix elements, the SMEs are written in
terms of scalar products involving the external four-momenta and the
top/antitop spin four-vectors, $s_{t,\bar t}^\mu$.  The latter are defined
after choosing the axes along which the $t$ and $\bar t$ spins are
decomposed, as described in Appendix~A of Ref.~\cite{Kao:1999kj}.
As studied in Ref.~\cite{Bernreuther:2004jv}, for instance, the
freedom in the choice of the spin axes can be used to increase spin
correlations at hadron colliders.  Here we choose the helicity basis
where the spin is quantized along the particle's direction of motion.
Using the helicity basis, the NLO SQCD contribution to the polarized
partonic cross sections of Eq.~(\ref{eq:parton}) read as follows (with
$z=\cos\hat\theta$):\\
\underline{$\qqa$}:
\begin{eqnarray}
\label{eq:qqannihilation}
\lefteqn{ 2 {\cal R}e \; \overline{\sum} (\delta {\cal M}_{q\bar
q}^{SQCD} \times {\cal M}_{B}^{q\bar q*}) = } \nonumber\\ & &
\overline{\sum} \mid {\cal M}^{q\bar q}_{B} \mid^2 \frac{\alpha_s}{2
\pi} {\cal R}e \left(F_V(\sd,m_q=0)+F_V(\sd,m_t)-\hat \Pi(\sd)\right)+
\nonumber\\ & & \frac{4 \pi \alpha_s^3}{9} {\cal R}e \left(\beta_t^2
(1-z^2) (1+4 \lt \ltb) F_M(\sd,m_t) - 2 (\lt-\ltb) \left[2 z
G_A(\sd,m_q=0)+ \right. \right. \nonumber\\ & & \left. \left. \beta_t
(1+z^2) G_A(\sd,m_t) \right]\right) + \frac{32 \pi \alpha_s^3}{9 \sd}
{\cal R}e \left( (-1) \frac{7}{3} B_t-\frac{2}{3} B_u
\right)(\sd,\hat t,\lt,\ltb) \; , 
\end{eqnarray}
where the $gt\bar t$ and $gq\bar q$ vertex and the gluon self-energy
corrections shown in Fig.~\ref{fig:qqtt_gen_vertself} are parametrized
respectively in terms of UV finite (after renormalization) form
factors $F_V$, $F_M$, $G_A$, and the subtracted gluon vacuum
polarization $\hat \Pi(\sd) \equiv\Pi(\sd)-\Pi(0)$. $F_V$, $F_M$, and
$G_A$ denote the vector, magnetic and axial vector parts of the
vertex correction. The direct and crossed box contributions to $\qqa$
shown in Fig.~\ref{fig:qqttboxes} are described by $B_t$ and $B_u$,
respectively.  Explicit expressions for these form factors and the box
contributions are
provided in Appendix~\ref{sec:qqtt_analytic} and $\hat\Pi(\sd)$ is given in Eq.~(\ref{eq:subgluon}).\\
\underline{$\ggf$}:
\begin{eqnarray}
\label{eq:ggfusion}
\lefteqn{2 {\cal R}e \; \overline{\sum} (\delta {\cal M}_{gg}^{SQCD} \times {\cal M}_{B}^{gg*})=} \nonumber \\ &&
\frac{4\pi\alpha_s^3}{64} 
\; 2\, {\cal R}e \left\{\sum_{j=1,2,3}
\left( c^s(j) \frac{1}{\hat s} 
\left[ M^{V,t}_{\lambda_t\lambda_{\bar t}}(2,j) \, 
\left(F_V(\hat s,m_t)-\hat \Pi(\sd)+\rho_{2}^{V,s}(\sd)\right)
\right. \right.\right.
\nonumber\\
&+& \left. \left. \left. 
M^{V,t}_{\lambda_t\lambda_{\bar t}}(12,j) \, 
\frac{(\hat t-\hat u)}{2m_t^2} F_M(\hat s,m_t)
+M^{A,t}_{\lambda_t\lambda_{\bar t}}(2,j) \, G_A(\hat s,m_t)
\right]\right.\right.
\nonumber\\
&+& \left. \left. c^t(j) \sum_{i=1,\ldots,7 \atop 11,\ldots,17} 
\left[ M^{V,t}_{\lambda_t\lambda_{\bar t}}(i,j) \, 
\left(\frac{\rho_i^{V,t}(\hat t,\hat s)}{\hat t-m_t^2}
+\frac{\rho_i^{\Sigma,t}(\hat t,\hat s)}{(\hat t-m_t^2)^2}
\right) 
+ M^{A,t}_{\lambda_t\lambda_{\bar t}}(i,j) \, 
\left(\frac{\sigma_i^{V,t}(\hat t,\hat s)}{\hat t-m_t^2}
+\frac{\sigma_i^{\Sigma,t}(\hat t,\hat s)}{(\hat t-m_t^2)^2}
\right) \right] \right.\right.
\nonumber\\
&+ & \left. \left. c^u(j) \sum_i 
\left[ M^{V,u}_{\lambda_t\lambda_{\bar t}}(i,j)
\,\left(\frac{\rho_i^{V,u}(\hat u,\hat s)}{\hat u-m_t^2}
+\frac{\rho_i^{\Sigma,u}(\hat u,\hat s)}{(\hat u-m_t^2)^2}
\right) 
+M^{A,u}_{\lambda_t\lambda_{\bar t}}(i,j)
\,\left(\frac{\sigma_i^{V,u}(\hat u,\hat s)}{\hat u-m_t^2}
+\frac{\sigma_i^{\Sigma,u}(\hat u,\hat s)}{(\hat u-m_t^2)^2}
\right) \right] \right. \right. 
\nonumber \\
&+& 
\left. \left. \frac{1}{2}\sum_{d=a,b,c} c^t_d(j) \sum_i
\left[  M^{V,t}_{\lambda_t\lambda_{\bar t}}(i,j) \, 
\rho_{i,d}^{\Box,t}(\hat t,\hat s)+
M^{A,t}_{\lambda_t\lambda_{\bar t}}(i,j) \, 
\sigma_{i,d}^{\Box,t}(\hat t,\hat s)\right] \right. \right.
\nonumber \\
&+& 
\left. \left. \frac{1}{2}\sum_{d=a,b} c^u_d(j) \sum_i
\left[  M^{V,u}_{\lambda_t\lambda_{\bar t}}(i,j) \, 
\rho_{i,d}^{\Box,u}(\hat u,\hat s)+
M^{A,u}_{\lambda_t\lambda_{\bar t}}(i,j) \, 
\sigma_{i,d}^{\Box,u}(\hat u,\hat s)\right] \right) \right. 
\nonumber \\
&+& \left. \frac{11}{18} \left[M^{V,t}_{\lambda_t\lambda_{\bar t}}(12,2)+
M^{V,t}_{\lambda_t\lambda_{\bar t}}(12,3) \right]\rho_{12}^{\Box,gg\tilde t \tilde t}(\sd)
\right\} \; ,
\end{eqnarray}
where $i$ numerates the 14 SMEs, $M_i^{(V,A),(t,u)}$ of Eqs.~(B1), (B2) in
Ref.~\cite{Beenakker:1993yr}, and $j=1,2,3$ the $s,t,u$ channel of the
Born matrix element to the gluon-fusion subprocess.  For convenience,
the color factors $c^{s,t,u}(j)$ of Ref.~\cite{Beenakker:1993yr} and
$c^{t,u}_d(j)$ are both provided in Appendix~\ref{sec:color}.  The
coefficients to the SMEs describe the parity conserving
($\rho_i^{(V,\Sigma),(t,u)},\rho_{i,d}^{\Box,(t,u)},\rho_{12}^{\Box,gg\tilde
  t \tilde t}$) and parity violating
($\sigma_i^{(V,\Sigma),(t,u)},\sigma_{i,d}^{\Box,(t,u)}$) parts of the
SQCD one-loop corrections to the
$t$ and $u$ channels of the gluon-fusion subprocess, which consist of  
\begin{itemize}
\item
vertex corrections that modify the $gt\bar t$ vertex ($\rho_i^{V,(t,u)},\sigma_i^{V,(t,u)}$),
shown in Fig.~\ref{fig:ggtt_gen_vertself} (diagrams (a) and (b)),
\item
top-quark self-energy corrections that modify the top-quark propagator ($\rho_i^{\Sigma,(t,u)},\sigma_i^{\Sigma,(t,u)}$), shown in Fig.~\ref{fig:ggtt_gen_vertself} (diagram (c)), 
\item
box contributions ($\rho_{i,d}^{\Box,(t,u)},\sigma_{i,d}^{\Box,(t,u)}$), 
shown in Fig.~\ref{fig:ggtt_boxes} (diagrams (a), (b) and (c)), and
\item
the box-triangle correction ($\rho_{12}^{\Box,gg\tilde t \tilde t}$), shown in 
Fig.~\ref{fig:ggtt_boxes} (diagram (d)).
\end{itemize}
The SQCD one-loop corrections to the $s$ channel of the $gg \to
t\bar t$ subprocess, shown in Fig.~\ref{fig:ggtt_gen_vertself} (lower row), are parametrized
in terms of the $gt\bar t$ vertex form factors ($F_V,F_M,G_A$), the
coefficient describing the $ggg$ vertex corrections ($\rho_2^{V,s}$),
and the subtracted gluon self-energy ($\hat \Pi(\hat s)$).  All coefficients
to the SMEs are UV finite after applying the renormalization procedure.
Their explicit expressions are provided in
Appendix~\ref{sec:ggtt_analytic}.  The expressions for the
SMEs after the interference with the Born matrix element,
$M^{(V,A),(t,u)}_{\lambda_t\lambda_{\bar t}}(i,j)$, can be found in
Appendix~C of Ref.~\cite{Kao:1999kj}.

Before we turn to the discussion of the numerical impact of these
higher-order SQCD corrections on $t\bar t$ cross sections, we describe
in detail in the following sections our choice of the renormalization
scheme and the MSSM input parameters.

\subsection{Renormalization scheme}
\label{sec:renormalization}

\begin{figure}[h]
\vspace*{.9cm}
  \begin{tabular}{@{\hspace{-.06cm}}p{3.cm}p{.35cm}p{3.cm}p{.35cm}p{3.cm}p{.35cm}p{2.65cm}p{.32cm}p{2.7cm}}
\vspace*{-1.5cm}
\includegraphics[width=3.2cm,
  keepaspectratio]{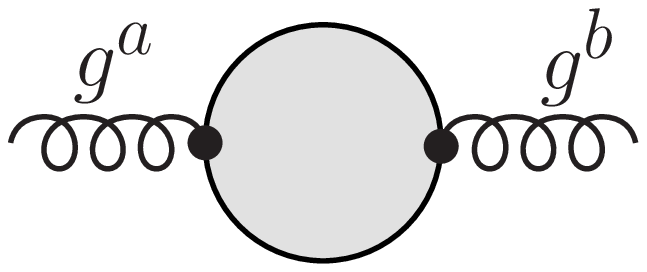}
&
$\begin{array}{c} \mbox{\bf  =} \\[50pt]\end{array}$
&
\vspace*{-1.68cm}
\includegraphics[width=3.2cm,
  keepaspectratio]{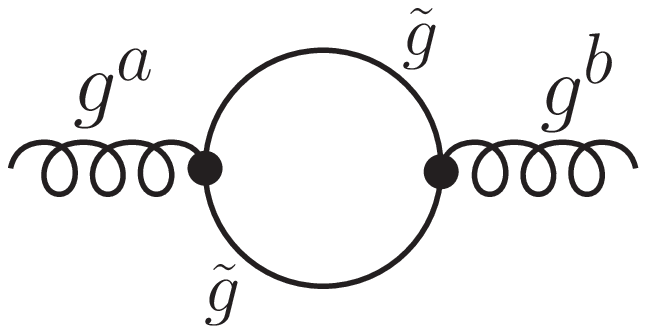}
&$\begin{array}{c} \mbox{\bf  +} \\[50pt]\end{array}$
&
\vspace*{-1.68cm}
\includegraphics[width=3.2cm,
  keepaspectratio]{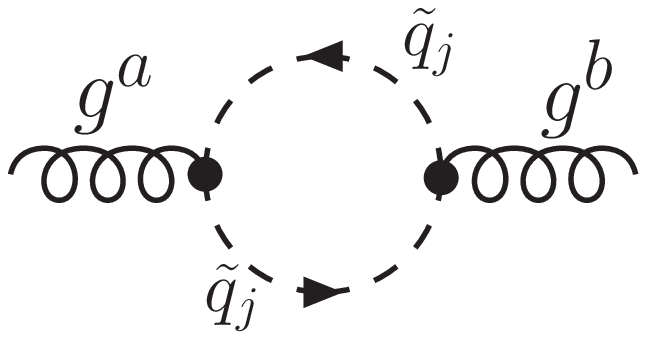}
& $\begin{array}{c} \mbox{\bf  +} \\[50pt]\end{array}$
& 
\vspace*{-2.1cm}
\includegraphics[width=2.9cm,
  keepaspectratio]{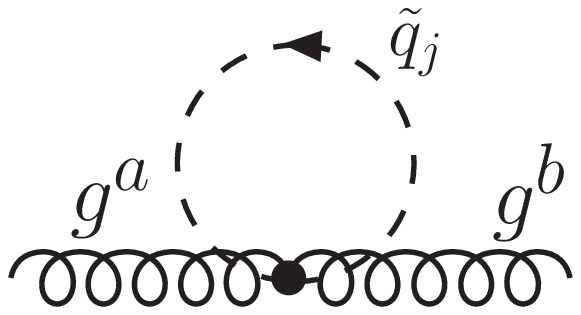}
&$\begin{array}{c} \mbox{\bf  +} \\[50pt]\end{array}$
&
\vspace*{-1.5cm}
\includegraphics[width=2.5cm,
  keepaspectratio]{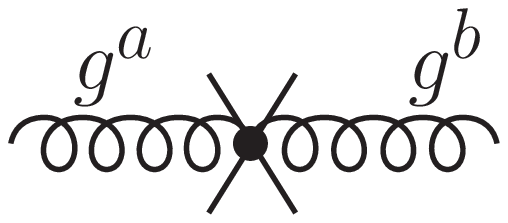}
\end{tabular}\\[-33pt]
\caption{\emph{Renormalized gluon self-energy, 
$-i \hat \Sigma_{\mu\nu}=i \delta_{ab} (k_\mu k_\nu \hat \Sigma_L-g_{\mu\nu} \hat \Sigma_T)$, at NLO SQCD.  Graphs
containing squarks are summed over the squark mass eigenstates j=L,R
(no mixing), j=1,2 (with mixing) and the quark flavors
q=\{u,d,s,c,b,t\}.}}
\label{fig:gself_renorm}
\end{figure}

\begin{figure}[h]
\vspace*{.9cm}
  \begin{tabular}{@{\hspace{-.06cm}}p{3.4cm}p{.5cm}p{3.4cm}p{.5cm}p{3.4cm}}
\vspace*{-1.57cm}
\includegraphics[width=3.4cm,
  keepaspectratio]{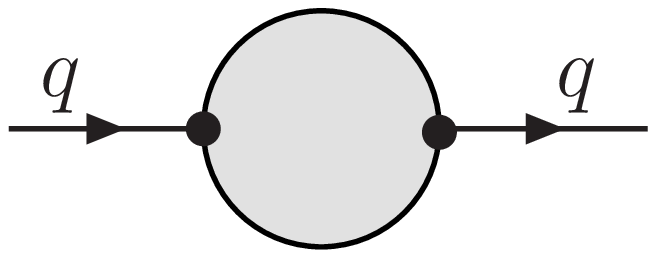}
&
$\begin{array}{c} \mbox{\bf  =} \\[50pt]\end{array}$
&
\vspace*{-1.74cm}
\includegraphics[width=3.4cm,
  keepaspectratio]{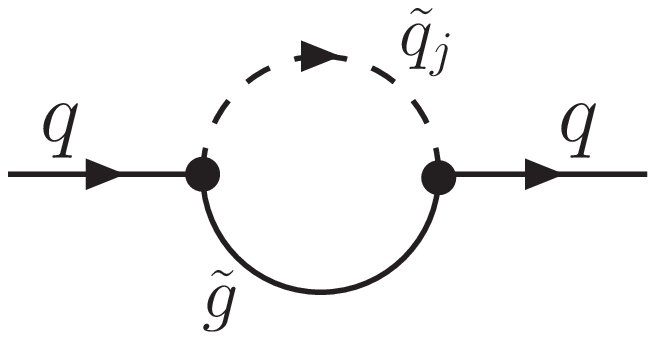}
&$\begin{array}{c} \mbox{\bf  +} \\[50pt]\end{array}$
&
\vspace*{-1.35cm}
\includegraphics[width=3.3cm,
  keepaspectratio]{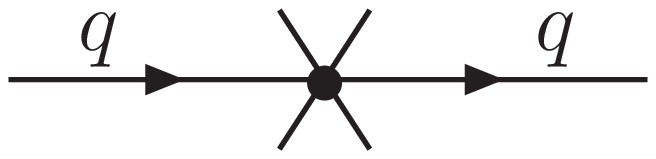}
\end{tabular}\\[-33pt]
\caption{\emph{Renormalized quark self-energy, 
$i\hat \Sigma=i (\not\! p [\hat\Sigma_V-\hat \Sigma_A \gamma_5]+m_q \hat \Sigma_S)$, at NLO SQCD.  Graphs
containing squarks are summed over the squark mass eigenstates j=L,R
(no mixing), j=1,2 (with mixing).}}
\label{fig:qself_renorm}
\end{figure}

The complete set of SQCD one-loop corrections to $\qqa$ and $\ggf$ is
gauge invariant and IR finite. However, the self-energy and vertex
corrections shown in Figs.~\ref{fig:qself_renorm},\ref{fig:ggg_renorm}
exhibit UV divergences that arise in form of
$\Delta=2/\epsilon-\gamma_E+\log(4\pi)$ terms when using dimensional
regularization in $d=4- \epsilon$ dimensions.  These singularities are
removed by introducing a suitable set of counterterms that are fixed
by a set of renormalization conditions.  We employ multiplicative
renormalization and perform the following replacements of the left and
right-handed quark fields, $\Psi_{L,R}=\frac{1}{2} (1 \mp \gamma_5)\, \Psi$, the
top-quark mass, $m_t$, the gluon field, $G_\mu^a$, and the strong
coupling constant, $g_s$, in the QCD Lagrangian:
\[ \Psi_{L,R} \to \sqrt{Z_{L,R}} \Psi_{L,R} \; \; , \; \; m_t \to m_t-\delta m_t \; \; , 
\; \; G_\mu^a \to \sqrt{Z_3} G_\mu^a \; \; , \; \; g_s \to Z_g g_s \;,\] 
With $Z_i=1+\delta Z_i$, this yields the counterterms to the quark
and gluon self-energies, $q\bar qg$ and
$ggg$ vertices as follows (with $\delta_{qt}=0$ for $q\ne t$ and $i,j; a,b,c$ denoting
respectively color degrees of freedom of quarks and gluons):\\[-4pt]
\begin{eqnarray}
\begin{array}{c}\mbox{}\\[-14pt]
\includegraphics[width=2.9cm,
  keepaspectratio]{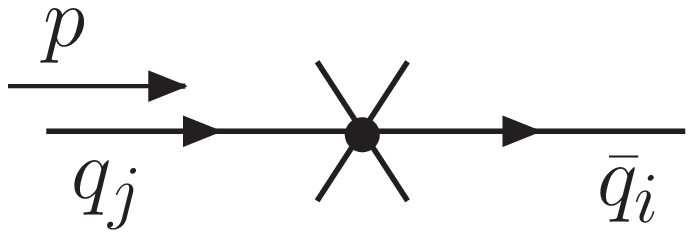}
\end{array}
& \qquad :\ & i \delta_{ij} [\not\! p (\delta Z_V-\delta Z_A \gamma_5)-\delta_{qt}(m_q \delta Z_V-\delta m_q)] 
\label{tt_counter}\\
\begin{array}{c}\mbox{}\\[-14pt]
\includegraphics[width=2.7cm,
  keepaspectratio]{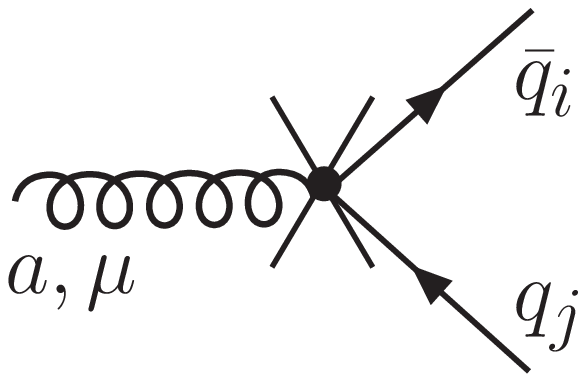}
\end{array}
& \qquad :\ & -i g_s T^a_{ij} \gamma_\mu (\delta Z_V+\delta Z_g+\frac{1}{2} \delta Z_3
-\delta Z_A \gamma_5) \label{gtt_counter}\\
\begin{array}{c}\mbox{}\\[-14pt]
\includegraphics[width=2.5cm,
  keepaspectratio]{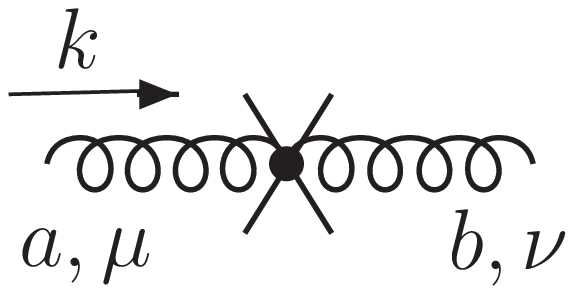}\hspace*{.2cm}
\end{array}
& \qquad :\ & i \delta_{ab} (k_\mu k_\nu -k^2 g_{\mu\nu}) \, \delta Z_3
\label{gg_counter}\\
\begin{array}{c}\mbox{}\\[-14pt]
\includegraphics[width=3.cm,
  keepaspectratio]{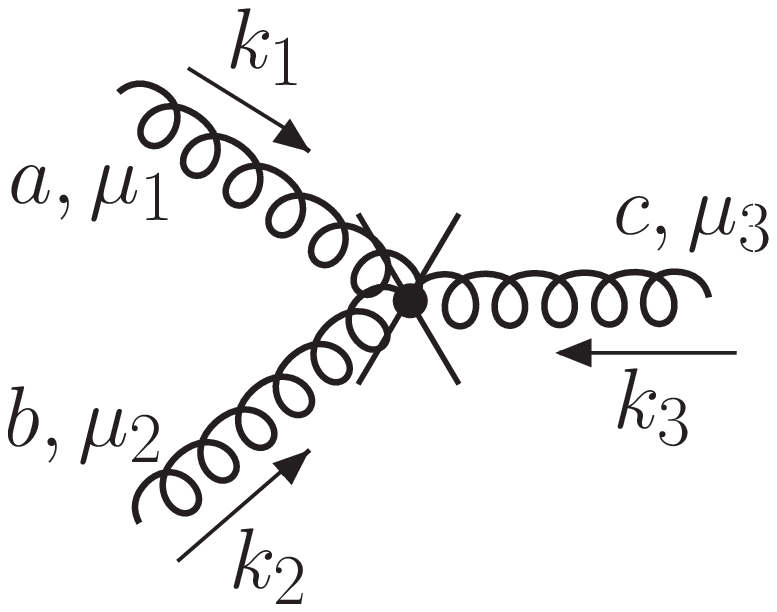}\hspace*{-.15cm}
\end{array}
& \qquad :\ & - g_s \, f_{abc} \,[(k_1-k_2)_{\mu_3} g_{\mu_1\mu_2} 
\nonumber\\[-20pt]    &  & \hspace{1.6cm}
+(k_2-k_3)_{\mu_1} g_{\mu_2\mu_3}
+(k_3-k_1)_{\mu_2} g_{\mu_3\mu_1}] \, \delta Z_1\; \label{ggg_counter}
\end{eqnarray}\\[-4pt]
with $\delta Z_1=\delta Z_g + 3 \delta Z_3/2$ and
$\delta Z_{V,A}=(\delta Z_L \pm \delta Z_R)/2$.
For the renormalization of the strong coupling constant 
and the gluon field, we use the
$\overline{MS}$ scheme, modified to decouple the heavy SUSY
particles
\cite{Collins:1978wz,Nason:1989zy}, i.e.~divergences
associated with the squark and gluino loops are subtracted at zero momentum:
\begin{itemize}
\item
\begin{equation}
\label{eq:deltaz3}
\delta Z_3=-\frac{\partial \Sigma_T(k^2)}{\partial k^2}|_{\epsilon,modified}
\equiv -\Pi(0)\; ,
\end{equation} 
where $\Sigma_T|_{\epsilon,modified}$ denotes 
the terms proportional to $\Delta_m=\Delta-\ln(m^2/\mu_R^2)$ 
(with $m=m_{\tilde q},m_{\tilde g}$)
of the transverse part of the gluon self-energy, yielding
\begin{equation}
\Pi(0)=\frac{\alpha_s}{4\pi} \left\{ \frac{1}{6} \sum_{q=u,c,t \atop d,s,b} 
\sum_j \Delta_{m_{\tilde q_j}}+2 \Delta_{m_{\tilde g}}\right\} \; .
\end{equation}
Only the transverse part of the gluon self-energy 
contributes to the NLO SQCD cross section for
$t\bar t$ production and the SQCD one-loop contribution to the
renormalized gluon self-energy shown in Fig.~\ref{fig:gself_renorm}
enters the partonic differential cross sections of 
Eqs.~(\ref{eq:qqannihilation}),~(\ref{eq:ggfusion}) as
follows:
\begin{eqnarray}
\label{eq:subgluon}
\frac{\alpha_s}{4\pi} \hat \Pi(k^2)&=&\frac{\Sigma_T(k^2)}{k^2}+\delta Z_3
\nonumber \\
&=& \frac{\alpha_s}{4\pi} \left\{ 
\sum_{q=u,c,t \atop d,s,b} \sum_j  
\frac{1}{6} \left[ 
\frac{(k^2-4 m_{\tilde q}^2)}{k^2} B_0(k^2,m_{\tilde q},m_{\tilde q})
+\frac{4 m_{\tilde q}^2}{k^2} B_0(0,m_{\tilde q_j},m_{\tilde q_j})+\frac{2}{3}\right] \right.
\nonumber\\
&-& \left. \frac{2}{3} +2 \frac{(2 m_{\tilde g}^2+k^2)}{k^2} B_0(k^2,m_{\tilde g},m_{\tilde g})- \frac{4m_{\tilde g}^2}{k^2} B_0(0,m_{\tilde g},m_{\tilde g})\right\}-\Pi(0) \; . 
\end{eqnarray}
where $j=L,R(1,2)$ (no(with) mixing) sums over the two squark mass
eigenstates and $B_0$ denotes the scalar two-point function (see,
e.g., Ref.~\cite{Hollik:1988ii}).
\item
\begin{equation}
\label{eq:deltaz1}
\delta Z_1=-\Gamma|_{\epsilon,modified}\equiv -\Pi(0)\; ,
\end{equation} 
where $\Gamma|_{\epsilon,modified}$ denotes the term proportional to
$\Delta_m$ (with $m=m_{\tilde q},m_{\tilde g}$) of the one-loop $ggg$
vertex correction shown in Fig.~\ref{fig:ggg_renorm} that multiplies
the tree level $ggg$ vertex.
\end{itemize}
Using Eqs.~(\ref{eq:deltaz3}),(\ref{eq:deltaz1}), one can verify that
$\delta Z_1=\delta Z_3$ and, thus, $\delta Z_g=-\delta Z_3/2$. 
It is interesting to note that in this modified version of the $\overline{MS}$ 
scheme the counter term for the $q\bar qg$ vertex coincides with the one of
Ref.~\cite{Beenakker:1993yr}.  Moreover, there is no
contribution from self-energy insertions into external gluon legs. 

To fix the renormalization constants for the external quarks, $\delta
Z_{V,A}(m_q)$, and the top-quark mass renormalization constant,
$\delta m_t$, we closely follow Ref.~\cite{Beenakker:1993yr} and use
{\it on-shell} renormalization conditions.
The renormalization constants $\delta Z_{V,A}(m_q), \delta \mt$ are
determined by the vector, axialvector and scalar parts of the quark
self-energy, evaluated at the on-shell quark mass,
$\Sigma_{V,A,S}(p^2=\mt^2)$ for the top quark and
$\Sigma_{V,A}(p^2=m_q^2=0)$ for the incoming quarks in $\qqa$.  The
SQCD one-loop corrections modify these quark self-energy contributions
as shown in Fig.~\ref{fig:qself_renorm} as follows:
\begin{eqnarray}
\label{eq:topself}
\Sigma_V(p^2) & = &- \frac{\alphas}{4\pi}\sum_j \lambda_j^+ \frac{2}{3} B_1(p^2,m_{\tilde g},m_{\tilde q_j})
\nonumber\\
\Sigma_S(p^2) & = & \frac{\alphas}{4\pi} \sum_j \lambda_j^- \frac{2}{3} 
\frac{m_{\tilde g}}{m_q} B_0(p^2,m_{\tilde g},m_{\tilde q_j}) 
\nonumber\\
\Sigma_A(p^2) & = & \frac{\alphas}{4\pi}\sum_j \lambda_j^A \frac{2}{3} B_1(p^2,m_{\tilde g},
m_{\tilde q_j}) \;  ,
\end{eqnarray}
yielding the following {\em on-shell} renormalization constants: 
\begin{eqnarray}
\label{eq:renorm}
\delta Z_V(m_q) & = & -\Sigma_V(p^2=m_q^2)-2m_q^2 \frac{\partial}{\partial p^2} (\Sigma_V+\Sigma_S)|_{p^2=m_q^2} \nonumber \\
&=& \frac{\alphas}{4\pi} \sum_j \frac{2}{3} \left[ \lambda_j^+ 
(B_1+2m_q^2 B_1')-2 m_q m_{\tilde g} \lambda_j^- B_0'\right](m_q^2,m_{\tilde g},m_{\tilde q_j})  
\\
\delta Z_A(m_q) & = & -\Sigma_A(p^2=m_q^2) 
\nonumber \\
&=& -\frac{\alphas}{4\pi}\sum_j \lambda_j^A \frac{2}{3} B_1(m_q^2,m_{\tilde g},m_{\tilde q_j})
\nonumber \\
\frac{\delta m_t}{m_t}&=&-\Sigma_V(p^2=m_q^2)-\Sigma_S(p^2=m_q^2)\\
&=&-\frac{\alphas}{4\pi} \sum_j \frac{2}{3} \left[- \lambda_j^+ 
B_1+\lambda_j^- \frac{m_{\tilde g}}{\mt} B_0\right](m_t^2,m_{\tilde g},m_{\tilde t_j})\; ,
\end{eqnarray}
where $j=L,R(1,2)$ (no(with) mixing) sums over the two squark mass
eigenstates, and the coupling constant factors, $\lambda^\pm_j,
\lambda_j^A$, are given in Appendix~\ref{sec:feynman_rules}.  The
scalar and vector two-point functions, $B_{0,1}$, and their
derivatives, $B_{0,1}'(p^2=m^2) \equiv \partial B_{0,1} /\partial
p^2|_{p^2=m^2}$, can be found in Ref.~\cite{Hollik:1988ii}, for
instance. In the {\em on-shell} renormalization scheme, the derivative
of the renormalized quark self-energy of Fig.~\ref{fig:qself_renorm}
is zero when evaluated at $p^2=m_q^2$, and thus there is no
contribution from the self-energy insertions in the external quark
lines. Finally, it is interesting to note that for our choice of
renormalization scheme, the renormalized SQCD one-loop corrections
only depend on $\mu_R$ through the strong coupling constant,
$\alpha_s(\mu_R)$.
%
%
%
%

\subsection{Scalar quark sector of the MSSM}
\label{sec:mssminput}

The superpartners of the left and right-handed top quarks are the left
and right-handed scalar top squarks ('stops'), $\tilde{t}_L$ and
$\tilde{t}_R$.  Assuming all supersymmetric breaking parameters are
real, the part of the MSSM Lagrangian that contains the stop mass
terms reads
\begin{eqnarray}
\label{eq:lagrangestlr}
{\cal L} &=& {}- (\begin{array}{cc}\tilde t_L^{\,*}\,, & \str^{\,*} \end{array}) \;
 {\cal M} 
\left(\begin{array}{c} \stl \\ \str \end{array}\right)
\hspace*{1.6cm}  {\rm with}\hspace*{1.6cm}
{\cal M} = \left(\begin{array}{cc}
\mstl^2 & m_t X_t \\
m_t X_t  & \mstr^2 \end{array}\right), 
\end{eqnarray}
where the diagonal entries $\mstl^2$ and $\mstr^2$ of the stop mass
matrix ${\cal M}$ as well as $X_t$ are defined as
\begin{eqnarray}
\label{eq:mtrlXt}
\mstl^2 &=& M_{\tilde{t}_L}^2+m_t^2+M_Z^2\cos{2\beta}\,(I_3^t-Q_t s_{\rm w}^2)~,
\hspace*{1.3cm} X_t = (A_t-\mu/\tanb)~,\nonumber\\
\mstr^2 &=& M_{\tilde{t}_R}^2+m_t^2+M_Z^2\cos{2\beta}\,Q_t s_{\rm w}^2
\end{eqnarray}
with $I_3^t = 1/2$, $Q_t = 2/3$ and $s_{\rm w}^2 = 1 -M^2_W/M_Z^2$.
Furthermore, $\tan\beta = v_2/v_1$ (also $v_u/v_d$ in the literature)
denotes the ratio of the two Higgs vacuum expectation values,
$A_{\tilde{t}}$ the trilinear Higgs-stop-stop coupling and $\mu$ is the
Higgs-mixing parameter. $M_{\tilde{t}_L}$ and $M_{\tilde{t}_R}$ are
the soft supersymmetry-breaking squark-mass parameters. As can be seen
from Eq.~(\ref{eq:lagrangestlr}), $\tilde{t}_L$ and $\tilde{t}_R$ are
not necessarily mass eigenstates, since ${\cal M}$ is of non-diagonal
form.  Thus, $\stl,\str$ can mix, so that the physical mass
eigenstates $\ste,\stz$ are model-dependent linear combinations of
these states. The latter are obtained by diagonalizing the mass matrix
performing the transformation
\begin{eqnarray}
\label{eq:trafo}
\left(\begin{array}{c} \ste \\ \stz \end{array}\right)
& = & 
\left(\begin{array}{rr} 
   \cos{\theta_{\tilde{t}}} & \quad \sin{\theta_{\tilde{t}}}  \\
   {}-\sin{\theta_{\tilde{t}}} & \cos{\theta_{\tilde{t}}}
\end{array}\ \right)
\left(\begin{array}{c} \stl \\ \str \end{array}\right)~.
\end{eqnarray}
The mass term of the supersymmetric Lagrangian will then transform to the 
diagonal form 
\begin{eqnarray}
\label{eq:Lagrangetop12}
{\cal L} &=& {}- (\begin{array}{cc}\ste^{\,*}\,, & \stz^{\,*} \end{array}) \;
   \left(\ \begin{array}{cc}
       \mste^2 & 0 \\
       0 & \quad \mstz^2 
   \end{array}\ \right)
\left(\begin{array}{c} \ste \\ \stz \end{array}\right)~.
\end{eqnarray}
Inserting Eq.~(\ref{eq:trafo}) and its adjoint form into 
Eq.~(\ref{eq:Lagrangetop12}) and comparing with Eq.~(\ref{eq:lagrangestlr})
the mass matrix ${\cal M}$  can be written as 
\begin{eqnarray}
{\cal M} = 
  \left(\begin{array}{cc}
    \cos^2\thetast\, \mste^2 +  \sin^2\thetast\, \mstz^2  & \quad
    \sin\thetast\cos\thetast\, (\mste^2-\mstz^2) \  \\
    \ \sin\thetast\cos\thetast\, (\mste^2-\mstz^2)  &
    \sin^2\thetast\,  \mste^2 +  \cos^2\thetast\, \mstz^2  
  \end{array}\right).
\end{eqnarray}
Comparing the two different descriptions of the mass matrix ${\cal M}$
we find
\begin{eqnarray}
\label{eq:comparemassmatrix}
\mstl^2\ \,\,&=& \cos^2\thetast\, \mste^2 +  \sin^2\thetast\, \mstz^2~,\nonumber\\
\mstr^2\ \,\,&=& \sin^2\thetast\,  \mste^2 +  \cos^2\thetast\, \mstz^2~,\nonumber\\
m_t X_t &=& \sin\thetast\cos\thetast\, (\mste^2-\mstz^2)~.
\end{eqnarray}
As input parameters we choose the physical measurable quantities
$\mste $, $\mstz$ and the stop mixing angle $\theta_{\tilde{t}}$.
Studying the properties of Eq.~(\ref{eq:comparemassmatrix}) we find
that the three equations, and therefore the cross sections of
Eq.~(\ref{eq:parton}), are invariant under the following two sets of
transformations:
\begin{eqnarray}
\theta_{\tilde{t}} &\rightarrow& \theta_{\tilde{t}}+n\cdot \pi 
\; \; (n \in {\cal I}) \\  
\mste^2 \leftrightarrow \mstz^2 \; \; &\mbox{and}& \; \; 
\theta_{\tilde{t}} \rightarrow \theta_{\tilde{t}}+ \pi/2 \; .
\end{eqnarray}
Therefore, for a complete exploration of the size of the SQCD one-loop
corrections, it is equivalent to either vary the masses $\mste^2$ and
$\mstz^2$ completely independently and vary the mixing angle between $\pm
\pi/4$, or to choose one of the squark masses to be always the lighter
one and vary $\theta_{\tilde{t}}$ between $\pm \pi/2$.  We choose the
latter option and take $\mstz^2$ always to be the lighter top
squark and vary the stop mixing angle in the range $-\pi/2 \le
\theta_{\tilde{t}} \le \pi/2$.

Taking $\mste $, $\mstz$ and $\theta_{\tilde{t}}$ as input parameters
and using Eq.~(\ref{eq:comparemassmatrix}), we can calculate
$\mstl^2,\, \mstr^2$ and $X_t$. The soft supersymmetry-breaking
parameters $M_{\tilde{t}_L}$ and $M_{\tilde{t}_R}$ can be determined
from Eq.~(\ref{eq:mtrlXt}) in dependence of $\tan\beta$.  Since the
SQCD NLO cross sections do not directly depend on $\tan\beta$ and the
dependence of $\mstl^2$ and $\mstr^2$ on $\tan\beta$ in the range
$1\le\tan\beta\le 50$ is rather weak, we choose in this paper an
arbitrary value for $\tan\beta$, i.e.~$\tan\beta=10$.

The equations~(\ref{eq:lagrangestlr})-(\ref{eq:comparemassmatrix}) can be
formulated in a similar way for the other squark flavors, one just has
to replace the index ``t'' by the appropriate squark index, as well as
$X_t \rightarrow X_d = (A_d - \mu \tan\beta)$ for the down-type
squarks. Since the off-diagonal elements of the squark mass matrices,
$m_q X_q$, are proportional to the quark masses $m_q$, and 
$X_q$ cannot be arbitrarily large ($X_q <$~\cite{Frere:1983ag}), 
mixing can be neglected for the first
two generations of squarks and may be only important in the sbottom
sector for large values of $\tan\beta$.  However, since the influence of the
sbottom sector on the hadronic $t\bar t$ cross sections is very small,
we also neglect mixing for the bottom squarks and 
in addition assume $m_{\tilde b_L}=m_{\tilde b_R}$.
Because gauge invariance requires
$M_{\tilde{t}_{L}} = M_{\tilde{b}_{L}}$, it follows that with
\begin{equation}
m_{\tilde{b}_L}^2 = 
M_{\tilde{t}_{L}}^2+m_b^2+M_Z^2\cos{2\beta}\,(I_3^b-Q_b s_{\rm w}^2)~,
\end{equation}
$I_3^b = -1/2$ and $Q_b = -1/3$, the bottom squark masses
are already fixed by the input parameters of the top
squark sector. 
We further assume universal squark masses for the first two
generations of squarks:
\begin{equation}
m_{\tilde u_L}=m_{\tilde u_R} = m_{\tilde d_L}=m_{\tilde d_R} =
m_{\tilde c_L}=m_{\tilde c_R} = m_{\tilde s_L}=m_{\tilde s_R} \ := m_{\tilde q}~.
\end{equation}
To summarize, in this paper we will study the dependence of the NLO
SQCD predictions on the following MSSM parameters: $m_{\tilde g}, \,
m_{\tilde t_1},\, m_{\tilde t_2},\, \theta_{\tilde{t}}$ and $m_{\tilde
q}$.


%
%
%
%
\section{Numerical Results}\label{sec:results}

In this section, we examine the numerical impact of the SQCD one-loop
corrections on the cross sections for unpolarized and polarized
top-pair production at the Tevatron~Run~II ($\sqrt{S}=1.96$ TeV) and
LHC ($\sqrt{S}=14$ TeV).  The details of the calculation are presented
in Section~\ref{sec:nlosqcd}, and the Feynman rules and explicit
analytic expressions are provided in Appendix~\ref{sec:feynman_rules}
and~\ref{sec:qqtt_analytic},~\ref{sec:ggtt_analytic}, respectively.

We performed a number of checks of our calculation: The results for
both the $q\bar q$ annihilation and gluon fusion matrix elements for
polarized and unpolarized top-pair production have been derived by at
least two independent analytic calculations based on the Feynman rules
of Ref.~\cite{Denner:1992vz}.  The numerical evaluation of the tensor
integrals was performed with an implementation described in
Ref.~\cite{Beenakker:1993yr} as well as by using the packages
LoopTools~\cite{Hahn:1998yk}/FF~\cite{vanOldenborgh:1989wn}.
Furthermore, one of the analytic calculations used the package
TRACER~\cite{Jamin:1991dp}.  The results of the analytic calculations
have been compared numerically with the results obtained with the
FeynArts~\cite{Hahn:2000kx, Kublbeck:1990xc} and
FormCalc~\cite{Hahn:2004rf,Hahn:1998yk} packages using the MSSM model
file of Ref.~\cite{Hahn:2001rv}. In all cases, we found perfect
numerical agreement.

In the computation of the $t \bar t$ observables, we use the following
values for the SM input parameters~\cite{Yao:2006px}:
\begin{eqnarray}
\mt = 175 \, \mbox{GeV} ,\quad m_b=4.7 \, \mbox{GeV} ,\quad M_W = 80.425 \, \mbox{GeV}
,\quad M_Z= 91.1876 \, \mbox{GeV}  \; .
\end{eqnarray}
As described in Section~\ref{sec:nlosqcd} (Eq.~(\ref{eq:hadwq})), we
use CTEQ6L1 PDFs~~\cite{Lai:1999wy} and the strong coupling constant
is evaluated at LO QCD with the factorization and renormalization
scales chosen to be equal to the top-quark mass.  Our choice of MSSM
input parameters is described in Section~\ref{sec:mssminput}.  As an
additional constraint on the MSSM parameter space, we calculate the
SUSY loop corrections to the $\rho$ parameter, $\Delta
\rho(m_{\tilde{t}},m_{\tilde{b}},\theta_{\tilde{t}})$ (for a review
see, e.g., Ref.~\cite{Heinemeyer:2004gx}), and only allow for MSSM
input parameters that yield $\Delta \rho \le
0.0035$~\cite{unknown:2005em}. If not stated otherwise, we choose the
masses of the supersymmetric partners of the light quarks to be
$m_{\tilde q}=2$~TeV. In varying the remaining MSSM input parameters
we observe the following experimentally motivated mass
limits~\cite{Abazov:2006bj,Abazov:2006wb,Affolder:2001tc,Affolder:2001nu}:
$m_{\tilde{g}} \ge 230~{\rm GeV}$ and $m_{\tilde{t}_2} \ge 100~{\rm
  GeV }$.  In the following sections, we study the numerical impact of
the SQCD one-loop corrections on the total $t \bar t$ production rate,
the invariant $t\bar t$ mass and top transverse momentum
distributions, and on polarization asymmetries.  We discuss the case
of unpolarized top quarks in Section~\ref{sec:res_unpolarized} and
consider polarized $t \bar t$ production in
Section~\ref{sec:res_polarized}.  Before we present results for the
observable hadronic cross sections at the Tevatron Run II and the LHC
in
Sections~\ref{sec:res_hadronic_unpolarized},~\ref{sec:res_hadronic_polarized},
we perform a detailed investigation of the impact of the SQCD one-loop
corrections on the parton-level cross sections in
Sections~\ref{sec:res_qqa_unpolarized},~\ref{sec:res_qqa_polarized}
($q\bar q$ annihilation) and
Sections~\ref{sec:res_gg_unpolarized},~\ref{sec:res_gg_polarized}
(gluon fusion).  Finally, in Section~\ref{sec:comparison}, we compare
our results with existing calculations in the literature where
available.

\subsection{Unpolarized top-pair production at NLO SQCD}
\label{sec:res_unpolarized}

We first consider the produced top quarks to be unpolarized, i.e.~the
corresponding partonic and hadronic cross sections are obtained from
the ones of Eqs.~(\ref{eq:parton}),(\ref{eq:hadwq}) by summing over
the top and antitop-quark helicity states, $\sum_{\lt,\ltb=\pm1/2}
d\hat \sigma^{LO,NLO}_{q\bar q,gg}(\hat t,\hat s,\lt,\ltb)$ and
$\sum_{\lt,\ltb=\pm1/2} d \sigma_{LO,NLO}(S,\lt,\ltb)$, respectively.

In order to reveal the numerical impact of the SQCD one-loop
corrections on observables to unpolarized top-pair production and to
study the dependences on the MSSM input parameters, the gluino mass
($m_{\tilde g}$), the two top-squark masses ($m_{\tilde t_{1,2}}$),
and the stop mixing angle ($\theta_{\tilde t}$), we introduce relative
corrections as follows:
\begin{itemize}
\item At the parton level, we use
\begin{eqnarray}
\label{eq:deltapart}
\hat{\Delta}_{q\bar q,gg}(\hat s)& = &\frac{\hat{\sigma}^{NLO}_{q\bar q,gg} - \hat{\sigma}_{q\bar q,gg}^{LO}}{\hat{\sigma}_{q\bar q,gg}^{LO}} \; ,
\end{eqnarray}
where $\hat{\sigma}^{LO,NLO}_{q\bar q,gg}(\hat s)$ denote the total
partonic production rates to unpolarized $t\bar t$ production via
$\qqa$ and $\ggf$ at LO QCD and NLO SQCD, respectively.
\item At the hadron level, we compute the total hadronic cross
  section, the invariant $t\bar t$ mass ($M_{t\bar t}$) and top
  transverse momentum ($p_T$) distributions at the Tevatron Run~II and
  the LHC, and use
\begin{eqnarray}\label{eq:haddeltas}
\Delta &=& \frac{\sigma_{NLO} - \sigma_{LO}}{\sigma_{LO}} \; ,\nonumber \\
\Delta(M_{t\bar t})& = &\frac{d\sigma_{NLO}/dM_{t\bar t} - d\sigma_{LO}/dM_{t\bar t}}{d\sigma_{LO}/dM_{t\bar t}} \; ,\nonumber \\
\Delta(p_T)& = &\frac{d\sigma_{NLO}/d p_T - d\sigma_{LO}/dp_T}{d\sigma_{LO}/dp_T} \; , 
\end{eqnarray}
where $d\sigma_{LO,NLO}$ describe the hadronic cross sections to
unpolarized $t\bar t$ production at LO QCD and NLO SQCD, respectively.
All hadron level results are obtained by including both processes
$q\bar q\to t\bar t$ and $gg\to t\bar t$.
\end{itemize}

%
%
\subsubsection{Effects of NLO SQCD corrections in $\qqa$ for unpolarized top quarks}
\label{sec:res_qqa_unpolarized}

At the Tevatron, top-quark pairs are mainly produced via $\qqa$ and,
thus, some of the characteristics of the corrections discussed here at
the parton level will manifest themselves again in the hadron-level
results for the Tevatron. In the following we study the impact of the
NLO SQCD corrections on the $u\bar u \to t\bar t$ subprocess,
representatively for all $q\bar q$ initiated processes.

\begin{figure}[h]
\begin{center}
  \includegraphics[width=8.1cm,
  keepaspectratio]{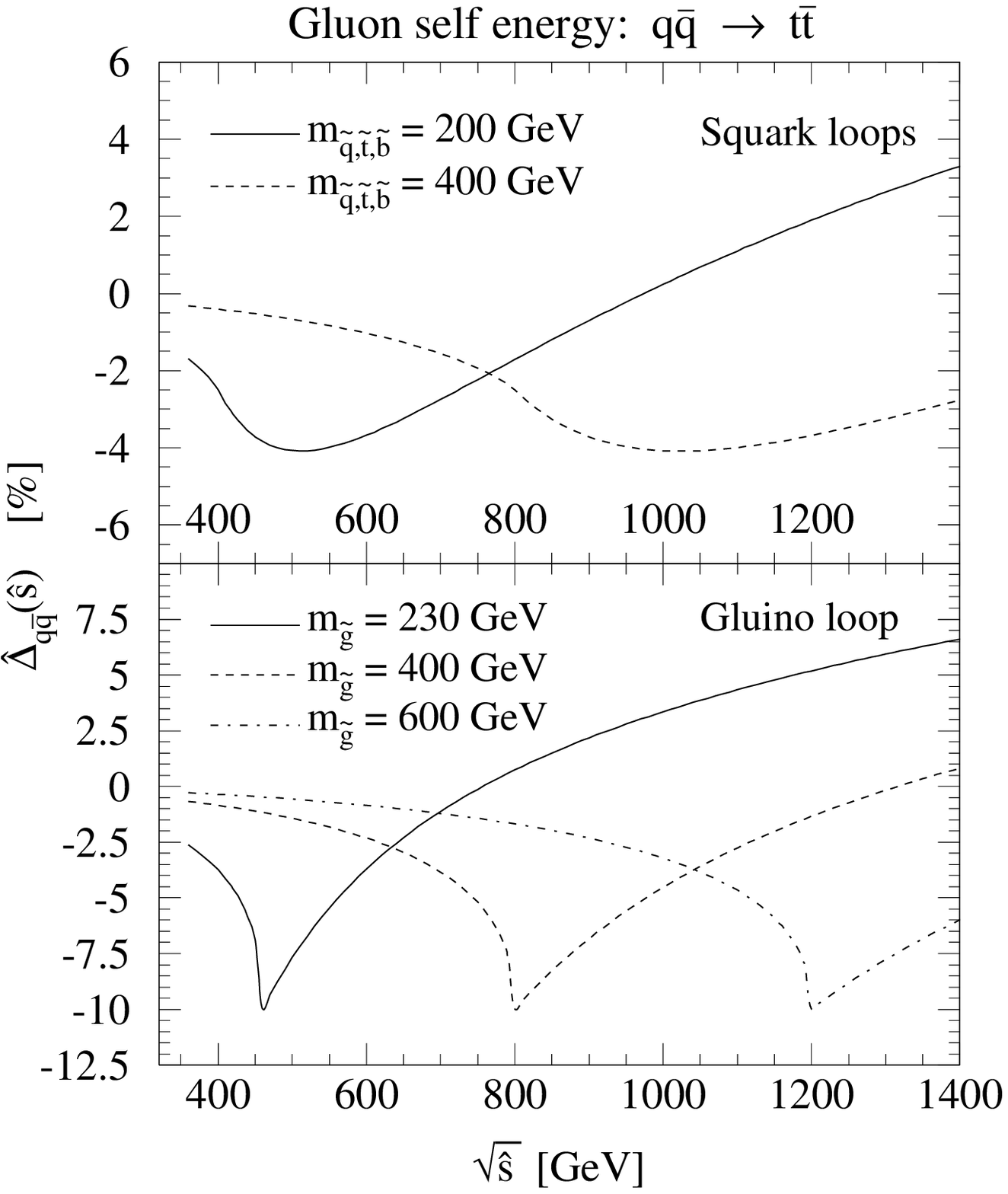}
\hspace*{-.0cm}
  \includegraphics[width=8.1cm,
  keepaspectratio]{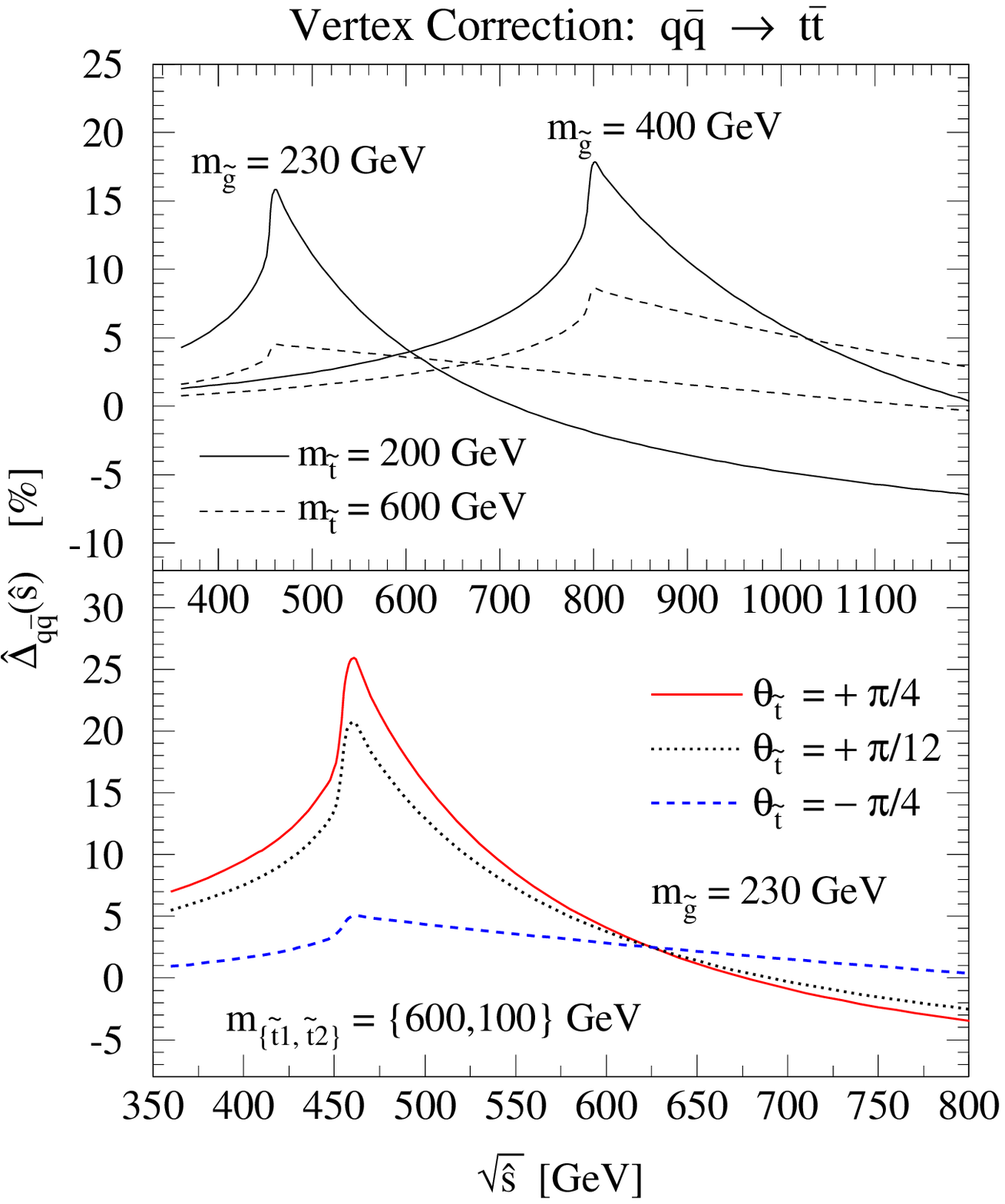}
\hspace*{-.8cm}\\[-30pt]
\hspace*{.6cm}(a)\hspace*{8cm}(b)\\
\vspace*{-25pt}
\end{center}
\caption{\emph{The relative correction $\hat{\Delta}_{q\bar q}(\hat s)$ for
$q\overline{q}$ annihilation due to (a) gluon self-energy and (b) vertex corrections (upper plot: $m_{\tilde t}\equiv m_{\tilde t_1}=m_{\tilde t_2}$).}}\label{fig:qqttselfvertex}
\end{figure}

Fig.~\ref{fig:qqttselfvertex}(a) illustrates the impact of the
(renormalized) gluon self-energy, shown in Fig.~\ref{fig:gself_renorm}
(see also Eq.~(\ref{eq:subgluon})), in dependence of the partonic CMS
energy, $\sqrt{\sd}$.  The upper plot shows the relative correction
$\hat \Delta_{q\bar q}$ of Eq.~(\ref{eq:deltapart}) due to the squark
loops, summed over all quark flavors and with squarks degenerate in
mass, and the lower plot the correction due to the gluino loop. The
gluino loop correction exhibits a characteristic ``dip'' at the
production threshold of a pair of gluinos, $\sqrt{\hat{s}}=2\,
m_{\tilde{g}}$, reducing the LO cross section by about $-10\%$. The
corrections then increase with $\sqrt{\sd}$ due to a
$\ln(\hat{s}/m_{\tilde{g}}^2)$ dependence.  A similar behavior is
observed in the squark loop correction, which can reduce $\hat
\sigma^{LO}_{q\bar q}$ by up to about $-4\%$, again due to a
threshold at $\sqrt{\hat{s}}=2\, m_{\tilde{q}}$, and then increases
with $\sqrt{\sd}$.  Both the gluino and the squark loop corrections do
not depend on the squark mixing angles.

Fig.~\ref{fig:qqttselfvertex}(b) shows the influence of the vertex
corrections of Fig.~\ref{fig:qqtt_gen_vertself}(b),(c), where in the
upper plot it is assumed that $m_{\tilde{t}_1} = m_{\tilde{t}_2}$. The
corrections are then independent of the top-squark mixing angle,
because the term that is proportional to $\theta_{\tilde t}$ cancels
in Eq.~(\ref{eq:comparemassmatrix}).  The corrections show again a
characteristic threshold behavior at $\sqrt{\hat{s}}=2\,
m_{\tilde{g}}$. This threshold is introduced by the vertex correction
containing the $\tilde{g}\tilde{g}\tilde{t}_{1,2}$ loop, shown in
Fig.~\ref{fig:gtt_renorm}. This loop correction is the dominant
contribution to the vertex corrections because of a large overall
color factor of $C_A^2 = 9$, arising due to the $g\tilde{g}\tilde{g}$
coupling. If the top squarks are non-degenerate in mass, as it is the
case in the lower plot of Fig.~\ref{fig:qqttselfvertex}(b), the vertex
correction depends very strongly on the stop mixing angle. Shown are
corrections for $m_{\tilde{t}_1} = 600$~GeV and $m_{\tilde{t}_2} =
100$~GeV for different values of the stop mixing angle. The largest
correction occurs always at the gluino-pair threshold, $\sqrt{\sd}=2\,
m_{\tilde g}$, for $\theta_{\tilde{t}} = \pi/4$ and the smallest
corrections are obtained for $\theta_{\tilde{t}} = -\pi/4$. For other
mixing angles the corrections lie between these two curves.  The
corrections are furthermore the larger the more pronounced the mass
splitting between $m_{\tilde{t}_1}$ and $m_{\tilde{t}_2}$. The amount
of allowed mass splitting, however, is restricted by the $\Delta \rho$
parameter.

\begin{figure}[h]
\begin{center}
  \includegraphics[width=8.1cm,
  keepaspectratio]{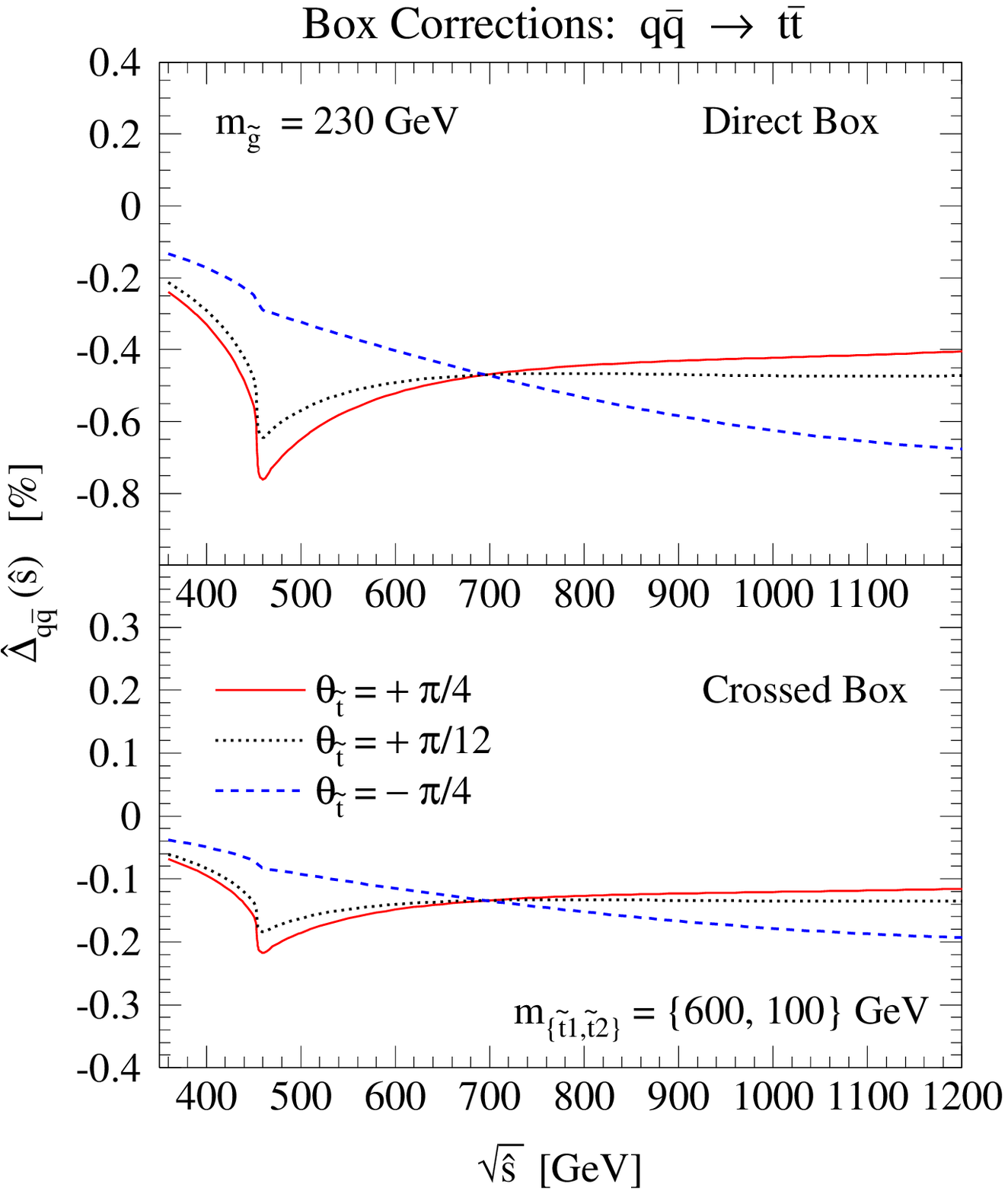}
 \hspace*{-.0cm}
  \includegraphics[width=8.1cm,
  keepaspectratio]{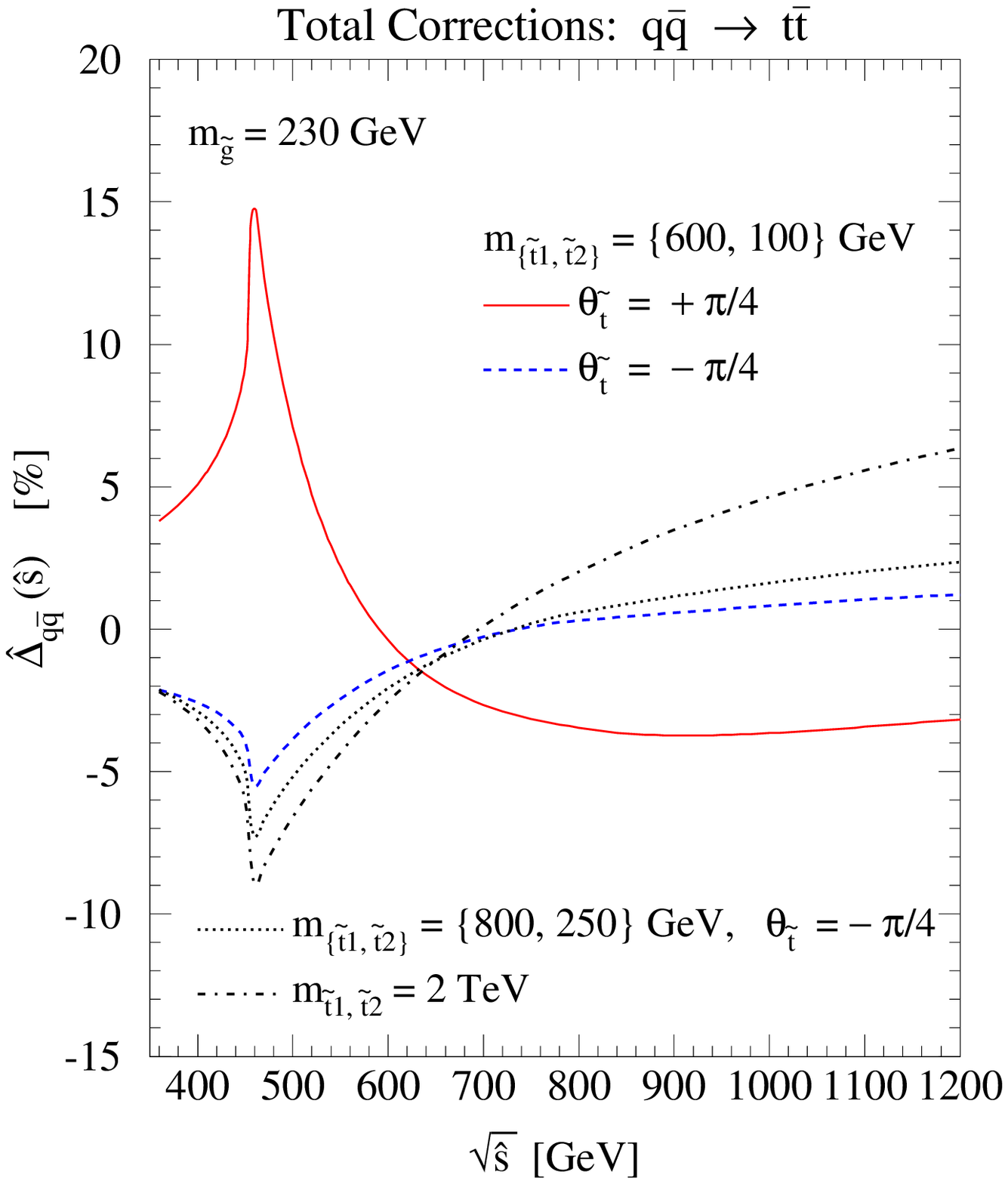}
\hspace*{-.8cm}\\[-30pt]
\hspace*{.6cm}(a)\hspace*{8cm}(b)\\
\vspace*{-25pt}
\end{center}
\caption{\emph{The relative correction $\hat{\Delta}_{q\bar q}(\hat s)$ for
$q\overline{q}$ annihilation 
due to (a) box corrections and (b) when including 
the complete SQCD one-loop corrections for $m_{\tilde{g}} = 230$~GeV.   
}}\label{fig:qqttboxcorr}
\end{figure}

The size of the box corrections is illustrated in
Fig.~\ref{fig:qqttboxcorr}(a) for $m_{\tilde{t}_1}=600$ GeV and
$m_{\tilde{t}_2}=100$ GeV and different values of the stop mixing
angle.  The upper plot shows the direct box contribution of
Fig.~\ref{fig:qqttboxes}(a) and the lower plot the crossed box
contribution of Fig.~\ref{fig:qqttboxes}(b).  The threshold peak at
$\sqrt{\hat{s}}=2\, m_{\tilde{g}}$ originates from the two gluinos
occurring in the Feynman diagrams of Fig.~\ref{fig:qqttboxes}.  The
corrections strongly depend on the stop mixing angle and in the
threshold region are largest for $\theta_{\tilde{t}} = \pi/4$, and can
be suppressed very effectively for $\theta_{\tilde{t}} = -\pi/4$.  The
direct and crossed box corrections have the same relative sign
compared to the LO cross section, as discussed in
Appendix~\ref{sec:qqtt_analytic} and are always negative in the
threshold region. The direct box correction is usually three to four
times larger than the crossed box contribution. Both box corrections
depend on the masses of the spartners of the initial-state quarks.
The corrections are small in this scenario, because of our choice of
the squark masses, $m_{\tilde{q}} = 2$~TeV (for $m_{\tilde{q}} \gtrsim
1$~TeV the corrections hardly change anymore).  The direct box
correction increases for smaller squark masses, e.g., for
$m_{\tilde{q}} = 200$~GeV at the gluino-pair threshold of
$\sqrt{\hat{s}} = 460$~GeV to $\hat\Delta_{q\bar q}=-22\%$ (for
$m_{\tilde{g}} = 230$~GeV, $m_{\tilde{t}_{1,2}} = 600,100~{\rm GeV},
\theta_{\tilde{t}} = +\pi/4$).  However, the positive vertex
correction also grows by the same amount and cancels the effect of the
box corrections.  We found that for every configuration with large box
corrections (and negative as always), the vertex corrections are
positive and also large, so that both largely cancel, resulting into a
small positive contribution to $\hat\Delta_{q\bar q}$.  As already
observed in the case of the vertex corrections, for top squarks
degenerate in mass, $m_{\tilde{t}_1}= m_{\tilde{t}_2}$, the direct and
crossed box corrections are independent of the top-squark mixing
angle, and are largest for the smallest possible choice of their
masses.

One can now deduce from the detailed discussion above, for which MSSM
scenarios the complete SQCD one-loop corrections, consisting of the
sum of self-energy, vertex and box corrections (see
Eq.~(\ref{eq:qqannihilation})), affect the $q\bar q \to t \bar t$
total cross section the most.  In Fig.~\ref{fig:qqttboxcorr}(b), we
show four representative scenarios, where we choose a light gluino
mass of $m_{\tilde{g}} = 230$~GeV.  If the top squarks are degenerate
in mass, $m_{\tilde{t}_1} = m_{\tilde{t}_2}$, the corrections are
independent of the squark mixing angle. If furthermore the stop masses
are relatively light, i.e.~smaller than $500$~GeV, no large
corrections occur because of a cancellation between the positive
vertex correction and the negative gluon self-energy. Only if the
stops are very heavy, i.e.~about 2~TeV (dot-dashed line in
Fig.~\ref{fig:qqttboxcorr}(b)), the vertex and box corrections are
strongly suppressed, and the negative gluon self-energy determines the
NLO SQCD $u \bar u \to t\bar t$ cross section.

If the stops are non-degenerate in mass, the corrections strongly
depend on~$\theta_{\tilde{t}}$~(red solid line and blue dashed line in
Fig.~\ref{fig:qqttboxcorr}(b)).  Large positive corrections can only
occur, if the positive vertex correction becomes large (and all other,
negative corrections are small) as is the case for
$\theta_{\tilde{t}}=\pi/4$.  Furthermore, the corrections are largest,
if the lighter stop mass, $m_{\tilde{t}_2}$, is as light as allowed by
the current mass limit, i.e.~$m_{\tilde{t}_2} \approx 100$~GeV, and for
the largest possible mass splitting.  The smallest correction for
top squarks non-degenerate in mass occurs for $\theta_{\tilde{t}}=-\pi/4$
because then the vertex correction is negligible and the total
correction is again dominated by the negative gluon self-energy. For
all other values of the stop mixing angle, the relative correction
$\hat \Delta_{q\bar q}$ will lie between the red solid and blue dashed
curves in Fig.~\ref{fig:qqttboxcorr}(b). 

%
%
\subsubsection{Effects of NLO SQCD corrections in gluon fusion for unpolarized top quarks}
\label{sec:res_gg_unpolarized}

Top-pair production at the LHC is dominated by gluon fusion and, thus,
the discussion of the SQCD one-loop corrections to $gg \to t \bar t$
at parton level is a good indicator of what to expect at the LHC.

\begin{figure}[h]
\begin{center}
  \includegraphics[width=8.1cm,
  keepaspectratio]{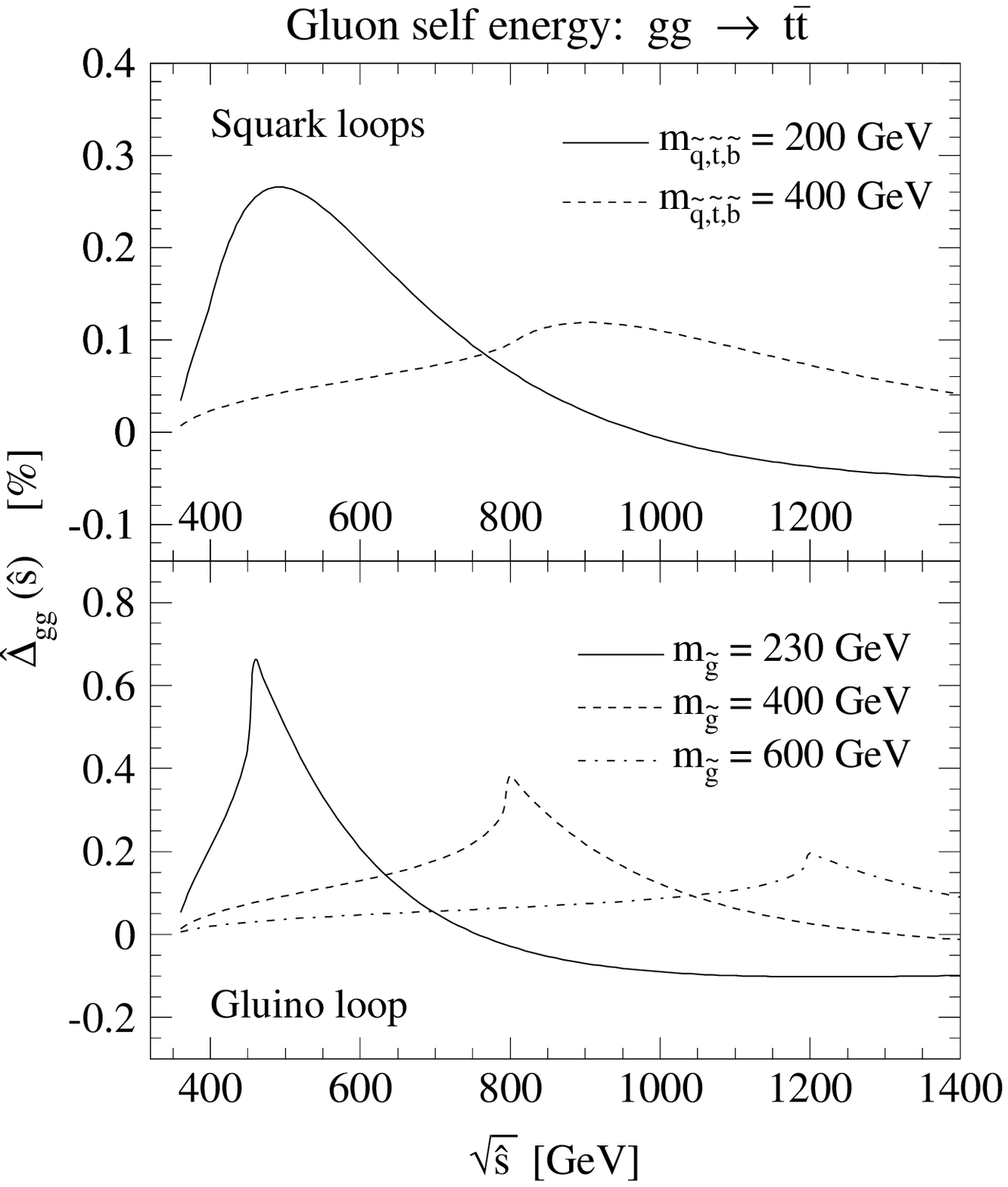}
\hspace*{-.0cm}
  \includegraphics[width=8.1cm,
  keepaspectratio]{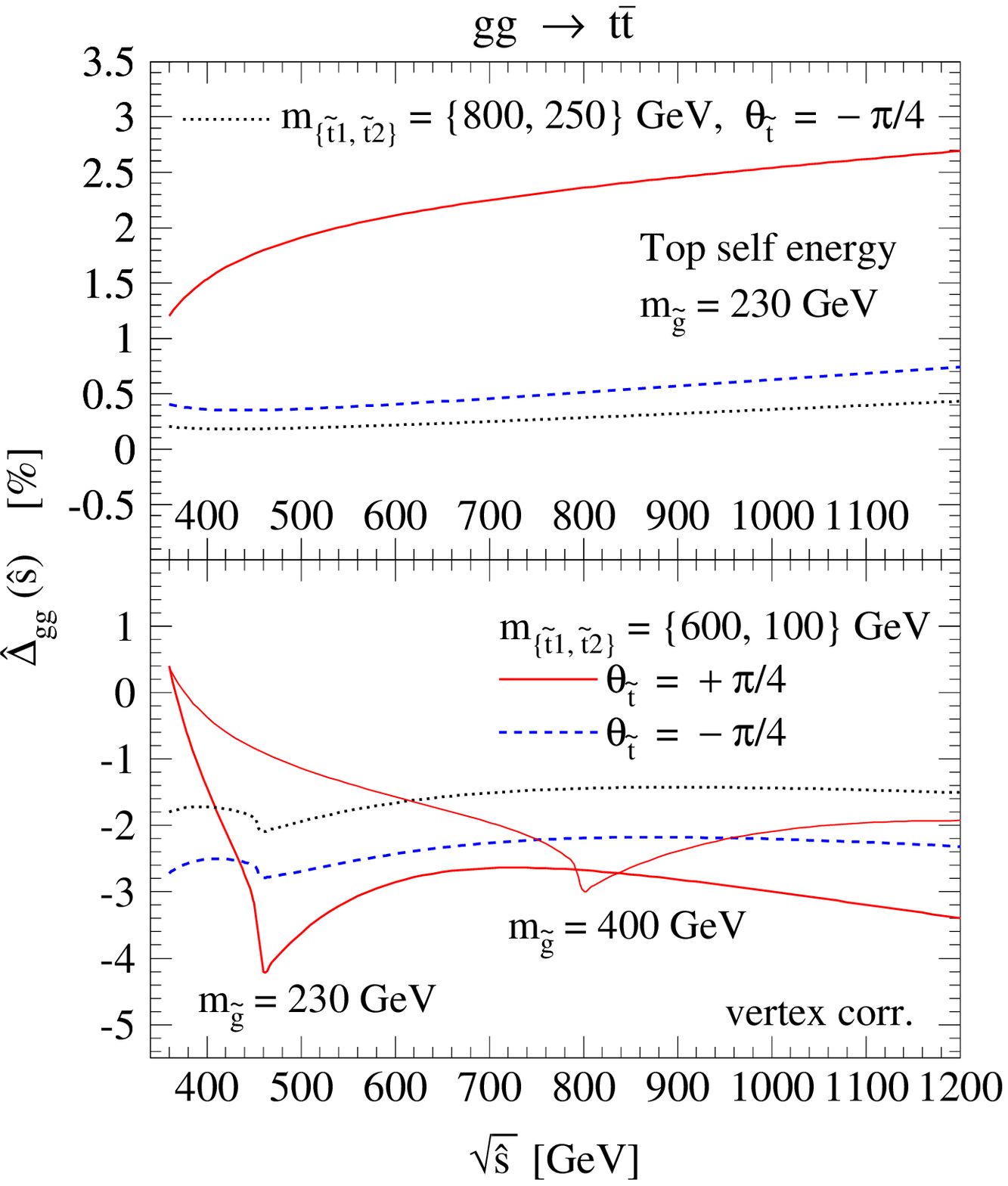}
\hspace*{-.8cm}\\[-30pt]
\hspace*{.6cm}(a)\hspace*{8cm}(b)\\
\vspace*{-25pt}
\end{center}
\caption{\emph{The relative correction $\hat{\Delta}_{gg}(\hat s)$ for
gluon fusion due to (a)~gluon self-energy, (b)(upper plot)~off-shell top quark self-energy, and (b)(lower plot) vertex corrections.}}\label{fig:ggtt_gtop_self}
\end{figure}

The impact of the gluon self-energy, top self-energy, and vertex
corrections to the total partonic cross section in gluon fusion is
shown in Fig.~\ref{fig:ggtt_gtop_self}, in form of the relative
correction $\hat\Delta_{gg}$ of Eq.~(\ref{eq:deltapart}). The gluon
self-energy correction to $gg\to t\bar t$ is displayed in
Fig.~\ref{fig:ggtt_gtop_self}(a), where the upper and lower plot shows
respectively the squark-loop (summed over all quark flavors and with
squarks degenerate in mass) and gluino-loop contributions.  Both
corrections are positive and much less pronounced than in $q\bar{q}$
annihilation.  This is because the $s$ channel contribution to the
gluon fusion cross section (see Fig.~\ref{fig:ggtt_gen_vertself})
reduces the total LO cross section by only 7\%, interfering
destructively with the $t$ and $u$ channels.  For the same reason, the
corrections are practically independent of the masses of the light
squarks, since diagrams that contain
$\tilde{u},\,\tilde{d},\,\tilde{c},\,\tilde{s}$ or $\tilde{b}$ quarks
occur only in the $s$ channel correction.

The upper plot of Fig.~\ref{fig:ggtt_gtop_self}(b) shows the effects
of the off-shell top quark self-energy contribution in the $t$ and $u$
channels of gluon fusion (see Figs.~\ref{fig:ggtt_gen_vertself}(b)).
Again the SQCD one-loop corrections are the larger the lighter the
gluino and the stops.  They are positive, increase with rising
$\sqrt{\hat{s}}$ up to about $\hat\Delta_{gg}=+3\%$ at
$\sqrt{\hat{s}}=1200$~GeV.  They strongly depend on the stop mixing
angle, if the squarks are non-degenerate in mass, and become largest
for $\theta_{\tilde{t}} =\pi/4$~(solid red line).  A similar behavior
can be observed, if $m_{\tilde{t}_1} = m_{\tilde{t}_2}$ and small,
i.e.~$\approx 200$~GeV.  With increasing stop masses this correction
quickly decreases.

The effects of the vertex corrections are displayed in the lower plot
of Fig.~\ref{fig:ggtt_gtop_self}(b). The corrections are negative and
exhibit the gluon-pair threshold behavior at $\sqrt{\hat{s}} = 2\,
m_{\tilde{g}}$ due to the vertex diagrams including the
$\tilde{g}\tilde{g}\tilde{t}_i$ loop (see Fig.~\ref{fig:gtt_renorm}).
The largest corrections arise again for light gluino masses, a large
mass splitting of the top squarks and small values of
$m_{\tilde{t}_2}$.  However, the influence of the mixing angle is
small and changes the correction in the threshold region by maximally
1.2\%.  Again, if both top squarks become heavier the size of the
corrections quickly decreases.

\begin{figure}[h]
\begin{center}
  \includegraphics[width=8.1cm,
  keepaspectratio]{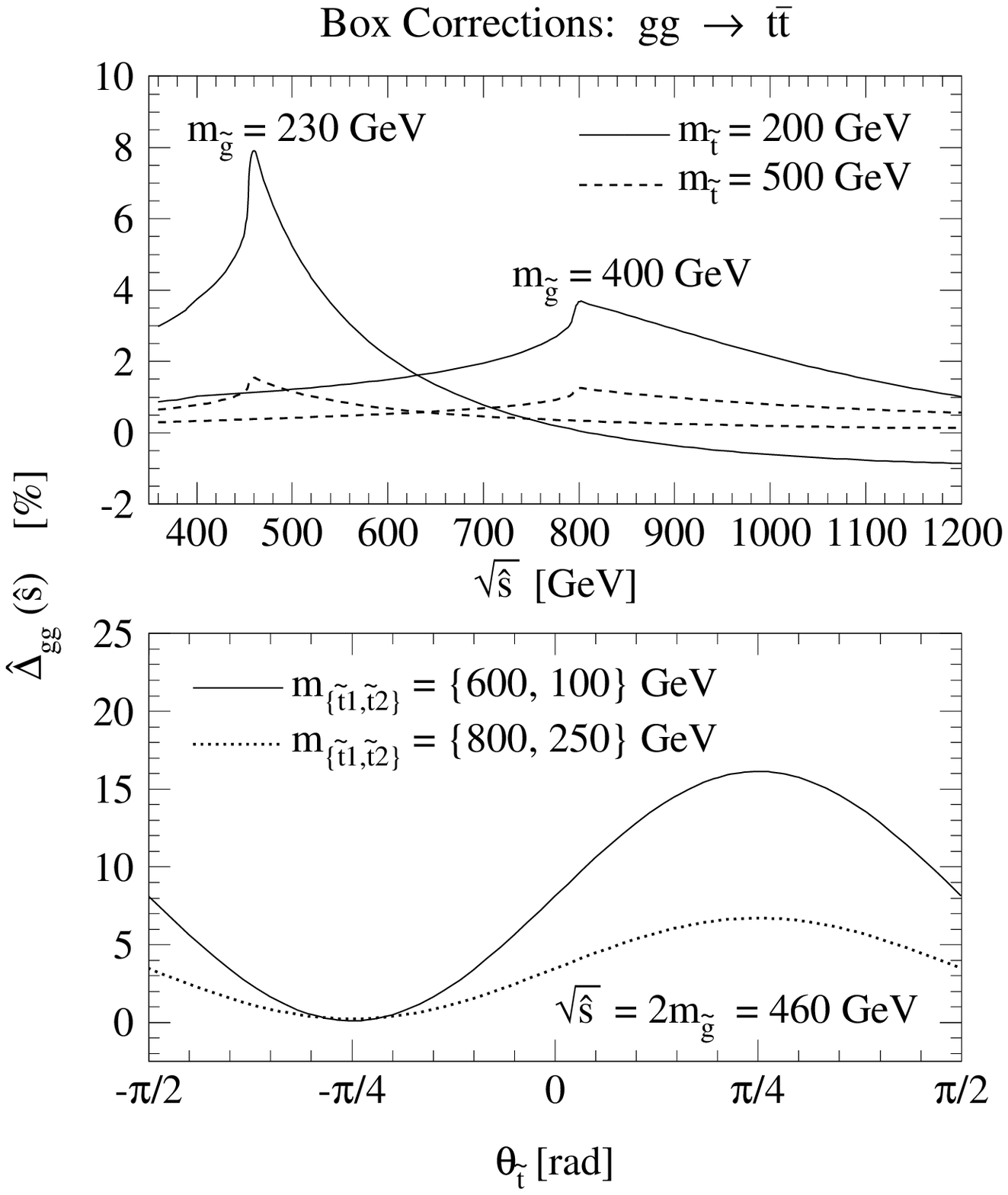}
\hspace*{-.0cm}
  \includegraphics[width=8.1cm,
  keepaspectratio]{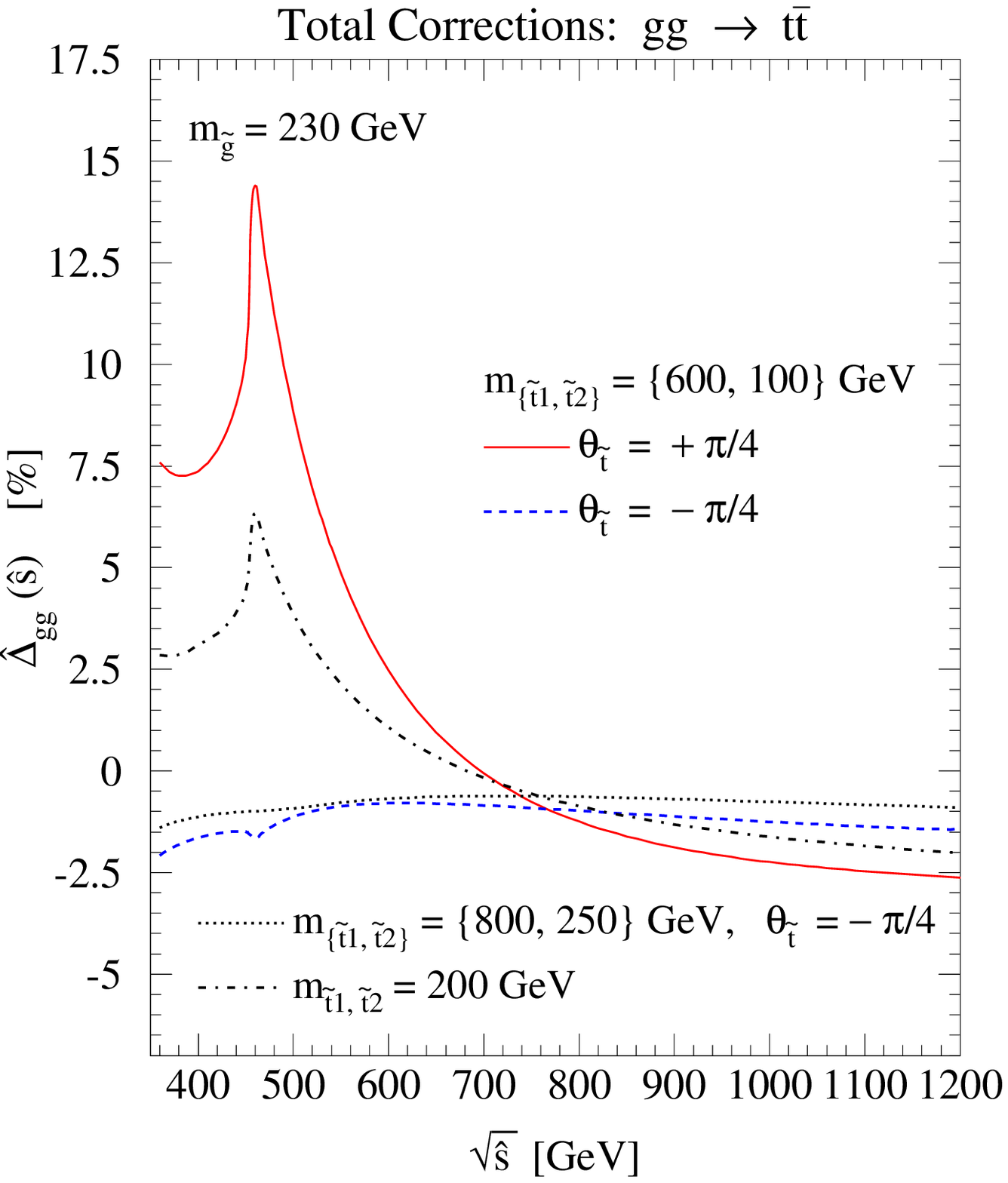}
\hspace*{-.8cm}\\[-30pt]
\hspace*{.6cm}(a)\hspace*{8cm}(b)\\
\vspace*{-25pt}
\end{center}
\caption{\emph{The relative correction $\hat{\Delta}_{gg}(\hat s)$ for
gluon fusion
due to (a)~box corrections (upper plot: $m_{\tilde t}\equiv m_{\tilde t_1}=m_{\tilde t_2}$) and (b) when including 
the complete SQCD one-loop corrections.}}\label{fig:ggtt_box_total}
\end{figure}

The largest corrections in gluon fusion arise due to the box diagrams
of Fig.~\ref{fig:ggtt_boxes}, as can be seen in
Fig.~\ref{fig:ggtt_box_total}(a).  In the gluino-pair threshold region
the corrections can reach up to $\hat\Delta_{gg}=+16\%$ for light
gluino masses of $230$~GeV, large mass splitting of the top squarks
and $\theta_{\tilde{t}} = \pi/4$. The corrections depend again
strongly on the stop mixing angle, if the stop masses are not equal,
as displayed in the lower plot of Fig.~\ref{fig:ggtt_box_total}(a). If
$m_{\tilde{t}_1} = m_{\tilde{t}_2}$, the box corrections at the
gluino-pair threshold reach $\hat\Delta_{gg}=+8\%$ for stop masses of
$200$~GeV and only $\hat\Delta_{gg}=+2\%$ for masses of $500$~GeV.

Since the top self-energy and vertex correction add destructively and
the gluon self-energy is negligible, the complete SQCD one-loop
correction to $\ggf$ is dominated by the box corrections and exhibits
the same characteristics. This is illustrated in
Fig.~\ref{fig:ggtt_box_total}(b), where we show the impact of the
complete SQCD one-loop correction to the partonic total cross section
of gluon fusion for four different MSSM scenarios, which differ by the
choices of top-squark masses and mixing angles.  The corrections
are practically independent of the light squark masses, $m_{\tilde
  q}$, because they contribute only to the gluon self-energy.  The
corrections are mostly positive, with a size of maximally
$\hat\Delta_{gg}=+15\%$ at the gluino-pair threshold. They can become
negative, reaching maximally $\hat\Delta_{gg}=-2.5\%$ for $\sqrt{\hat
  s} \approx 1$~TeV.

%
%
%
\subsubsection{Hadronic cross sections to unpolarized $pp,p\bar p \to t\bar t$ at NLO SQCD}
\label{sec:res_hadronic_unpolarized}

The effects of the NLO SQCD corrections to the hadronic cross sections
result from the combination of the already discussed effects to the
partonic cross sections and the PDFs (see Eq.~(\ref{eq:hadwq})).  The
hadronic $t\bar t$ cross sections at the Tevatron Run II and the LHC
are dominated respectively by the quark and gluon PDFs, which both
emphasize parton-level effects in the vicinity of the $t\bar t$
threshold, $\sqrt{\hat s}=2 m_t=350$ GeV, but are rapidly decreasing
for increasing values of $\sqrt{\hat s}$. The choice of the values for
the MSSM input parameters, $m_{\tilde{g}}$, $m_{\tilde{t}_1}$,
$m_{\tilde{t}_2}$ and $\theta_{\tilde{t}}$, is guided by the
discussion of the SQCD effects at the parton level of
Sections~\ref{sec:res_qqa_unpolarized},~\ref{sec:res_gg_unpolarized}.
We first study the dependence of the relative correction $\Delta$ of
Eq.~(\ref{eq:haddeltas}) on the gluino and heavier stop masses for four
different choices of the light stop mass and the stop mixing angle.
These choices are representative for the possible choices that affect
the $t\bar t$ cross sections the most.

\begin{figure}[h]
\begin{center}
  \includegraphics[width=8.1cm,
  keepaspectratio]{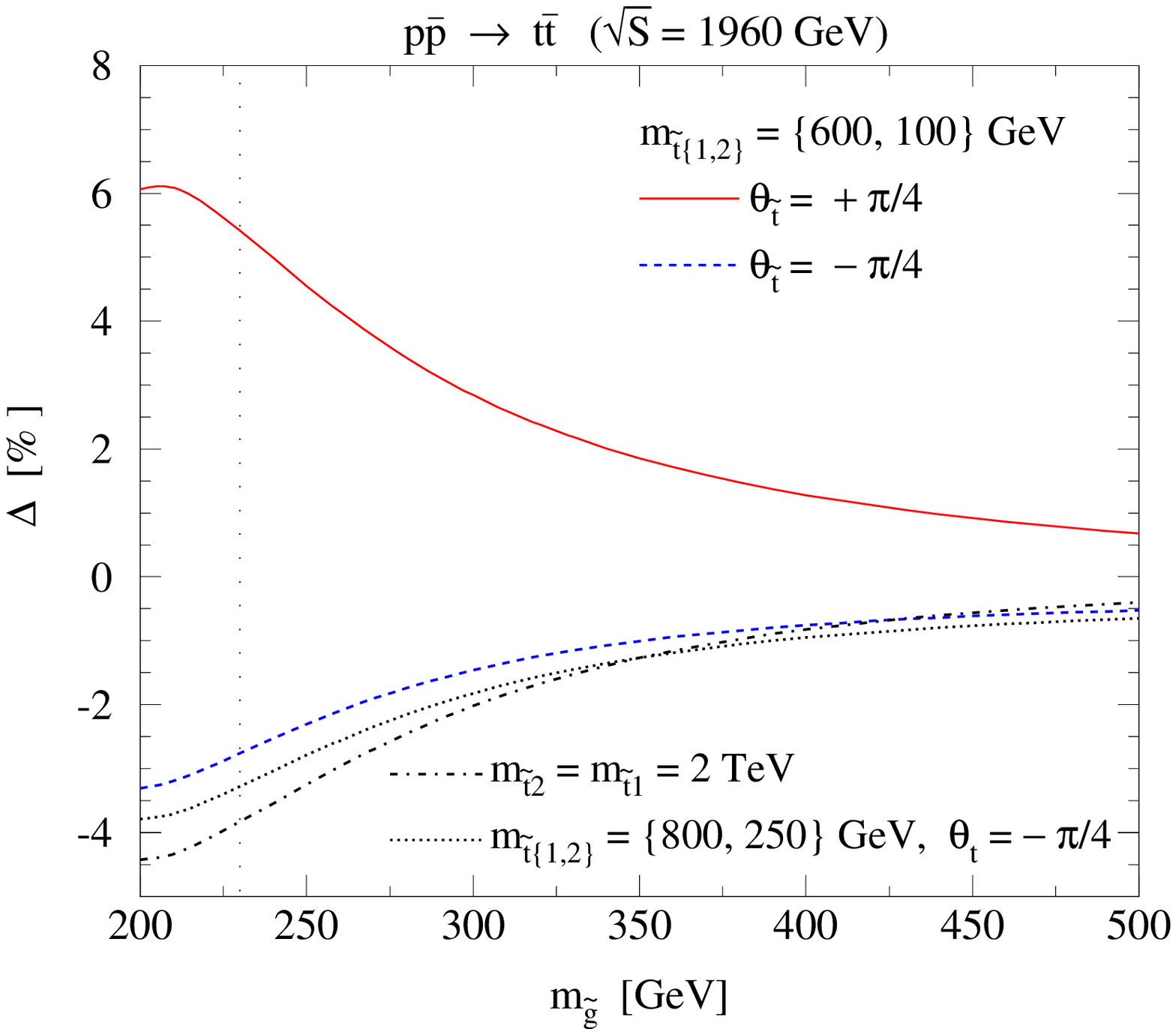}
\hspace*{-.0cm}
  \includegraphics[width=8.1cm,
  keepaspectratio]{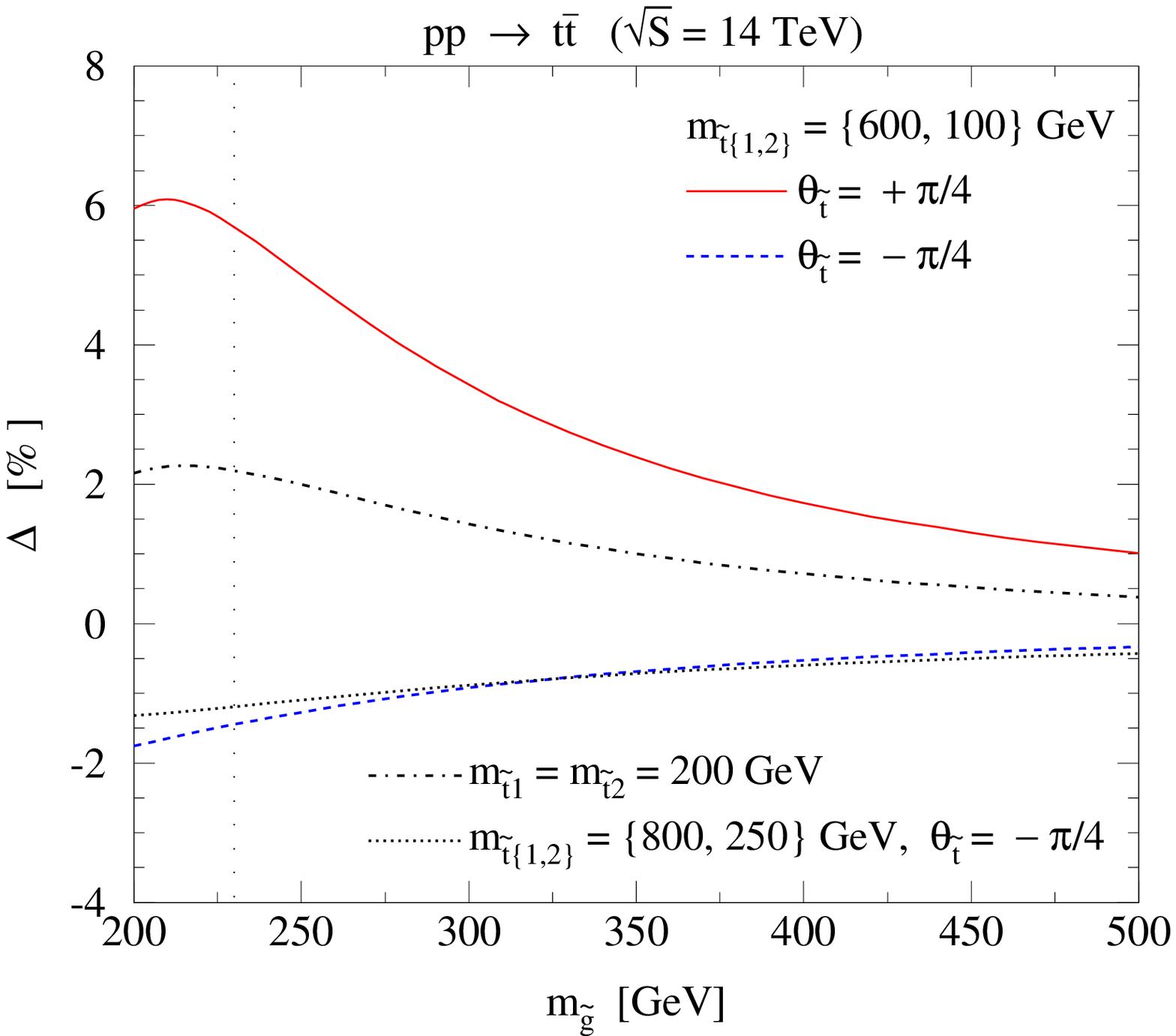}
\hspace*{-.8cm}\\[-30pt]
\hspace*{.6cm}(a)\hspace*{8cm}(b)\\
\vspace*{-25pt}
\end{center}
\caption{\emph{The relative correction $\Delta$ due to SQCD one-loop corrections
in dependence of the gluino mass, $m_{\tilde{g}}$, at 
(a)~the Tevatron Run~II and (b)~the LHC.}}\label{fig:Had_mg}
\end{figure}

The impact of the SQCD one-loop corrections on the total hadronic
cross section in dependence of the gluino mass is shown in
Fig.~\ref{fig:Had_mg}(a) for the Tevatron Run~II and in
Fig.~\ref{fig:Had_mg}(b) for the LHC.  The corrections are largest for
gluino masses of $200-250$~GeV, since for these masses the gluino-pair
threshold lies in the vicinity of the $t\bar t$~threshold. For
$m_{\tilde{g}} = 230$~GeV the SQCD one-loop corrections vary between
$\Delta=-4\%$ and $\Delta=+5\%$ at the Tevatron Run II, and between
$\Delta=-1.5\%$ and $\Delta=+5.5\%$ at the LHC.  They decrease to
approximately $\Delta=-0.5\%$ (Tevatron Run II) and $\Delta=+1\%$
(LHC) for gluino masses of $500$~GeV.  Since large corrections always
originate from the gluino-pair threshold, the relative corrections
decrease for heavier gluinos.

%
%

\begin{figure}[h]
\begin{center}
  \includegraphics[width=8.1cm,
  keepaspectratio]{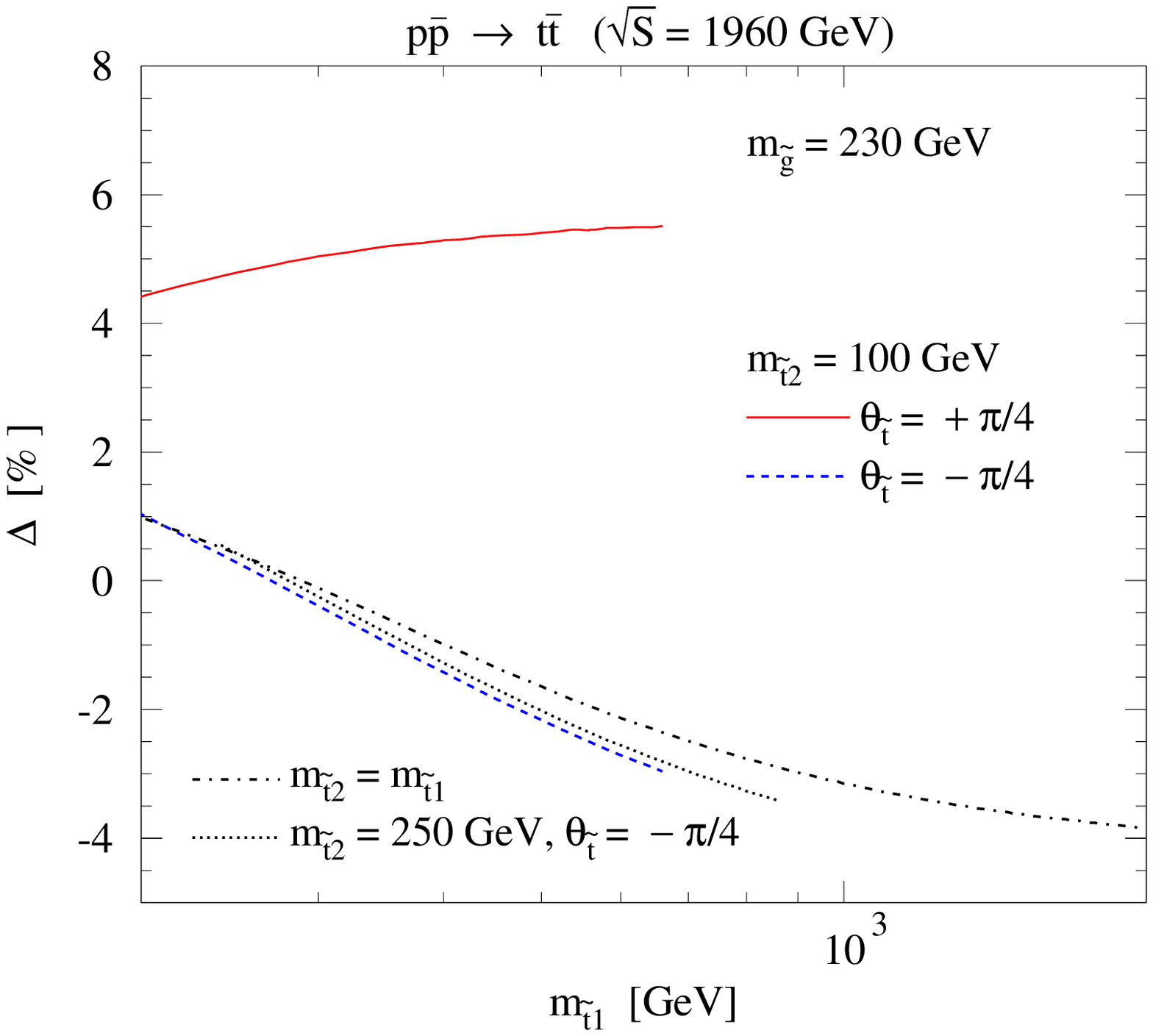}
\hspace*{-.0cm}
  \includegraphics[width=8.1cm,
  keepaspectratio]{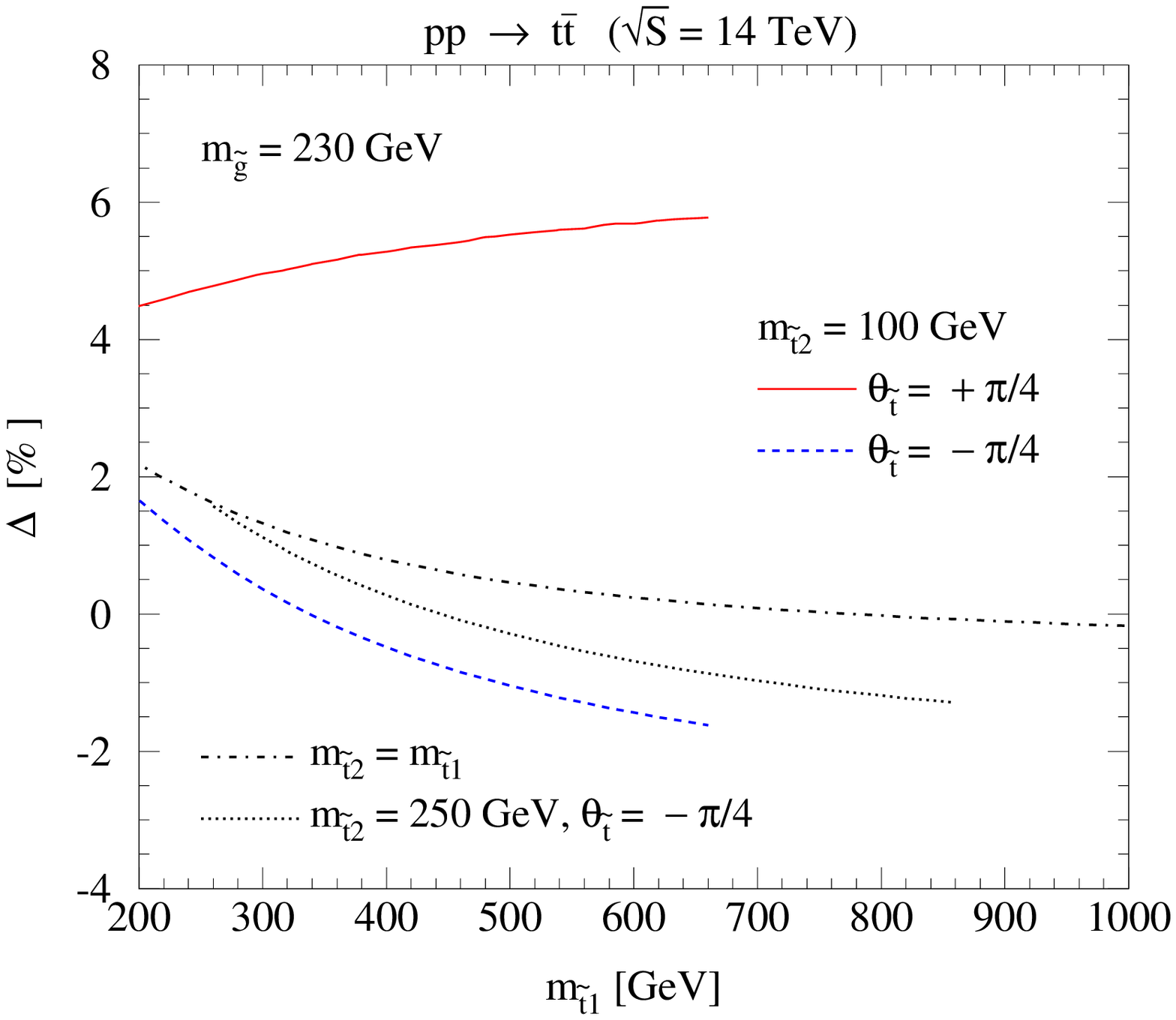}
\hspace*{-.8cm}\\[-30pt]
\hspace*{.6cm}(a)\hspace*{8cm}(b)\\
\vspace*{-25pt}
\end{center}
\caption{\emph{The relative correction $\Delta$ due to SQCD one-loop corrections 
in dependence of the heavier top-squark mass, $m_{\tilde t_1}$, at 
(a)~the Tevatron Run~II and (b)~the LHC.}}\label{fig:Had_mst}
\end{figure}

The influence of the top-squark masses on the SQCD one-loop
corrections is shown in Fig.~\ref{fig:Had_mst}(a) for the Tevatron
Run~II and in Fig.~\ref{fig:Had_mst}(b) for the LHC. Varied for these
plots is either the heavier top-squark mass alone ($m_{\tilde t_1}$),
or, in case of the dot-dashed line, both stop masses are varied
together.  The solid red, dashed blue and dotted black lines show that
the corrections increase with increasing difference in the top-squark
masses, as discussed at the parton level in
Section~\ref{sec:res_qqa_unpolarized},~\ref{sec:res_gg_unpolarized}.
The solid red and dashed blue lines end at $m_{\tilde t_1}\approx
650$~GeV and the dotted line at $m_{\tilde t_1}\approx 850$~GeV, because
for larger values of $m_{\tilde t_1}$, the parameter $\Delta \rho$
exceeds the current limit of $0.0035$~\cite{unknown:2005em}.  At the
Tevatron Run~II, if the top squarks are degenerate in mass, the
largest correction arises for very large top-squark masses because
then the correction is determined by the gluino loop in the gluon
self-energy, which amounts to $\Delta\approx -4\%$ at hadron level.
For $m_{\tilde{t}} > 2$~TeV the correction does not increase much
anymore and stays at about $\Delta=-4\%$. At the LHC, the gluon
self-energy plays a very minor rule, and all other corrections depend
on the top squark mass. Therefore, the corrections decouple and
approach zero if both top squarks are heavy.

\begin{figure}[h]
\begin{center}
  \includegraphics[width=8.1cm,
  keepaspectratio]{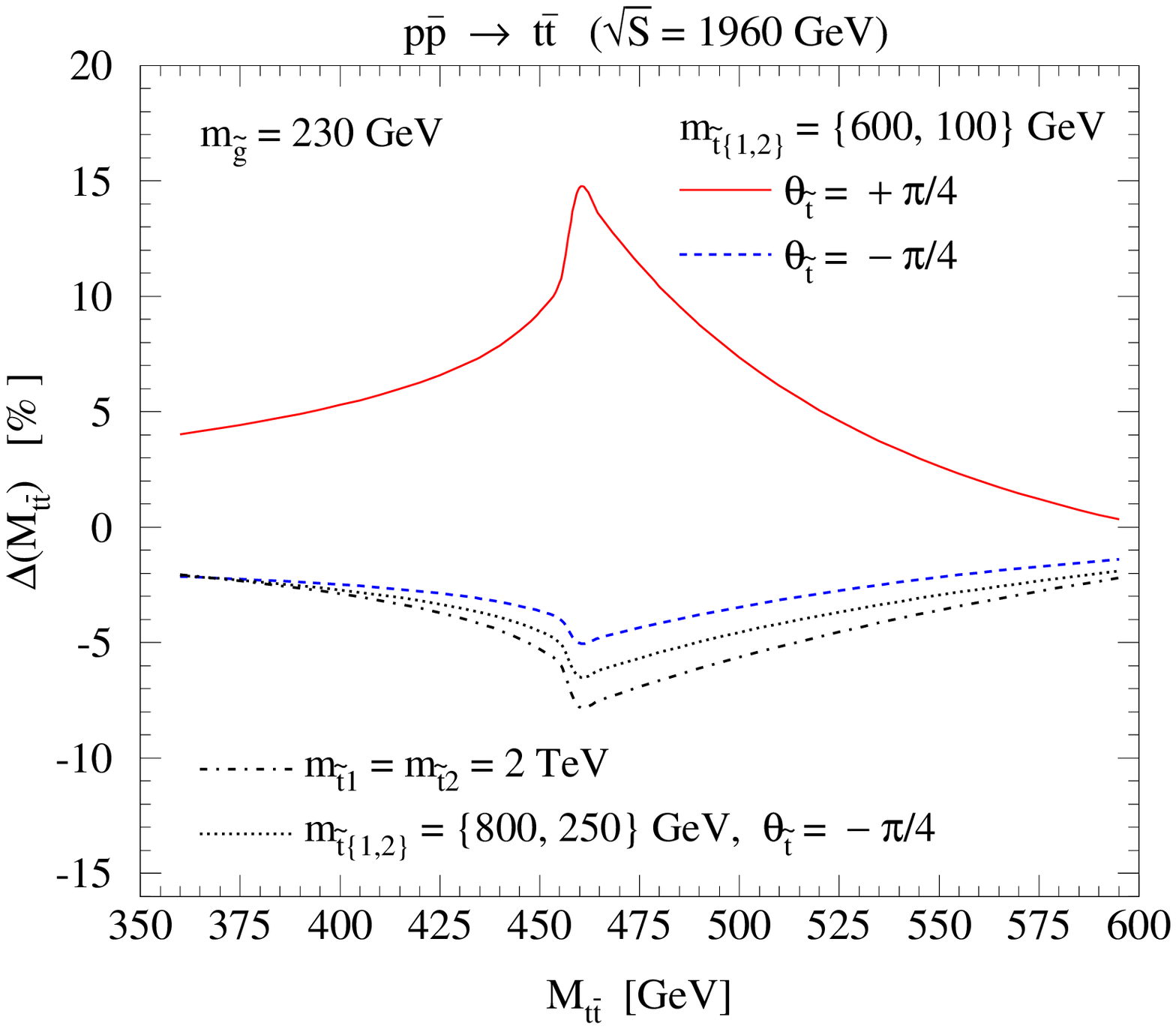}
\hspace*{-.0cm}
  \includegraphics[width=8.1cm,
  keepaspectratio]{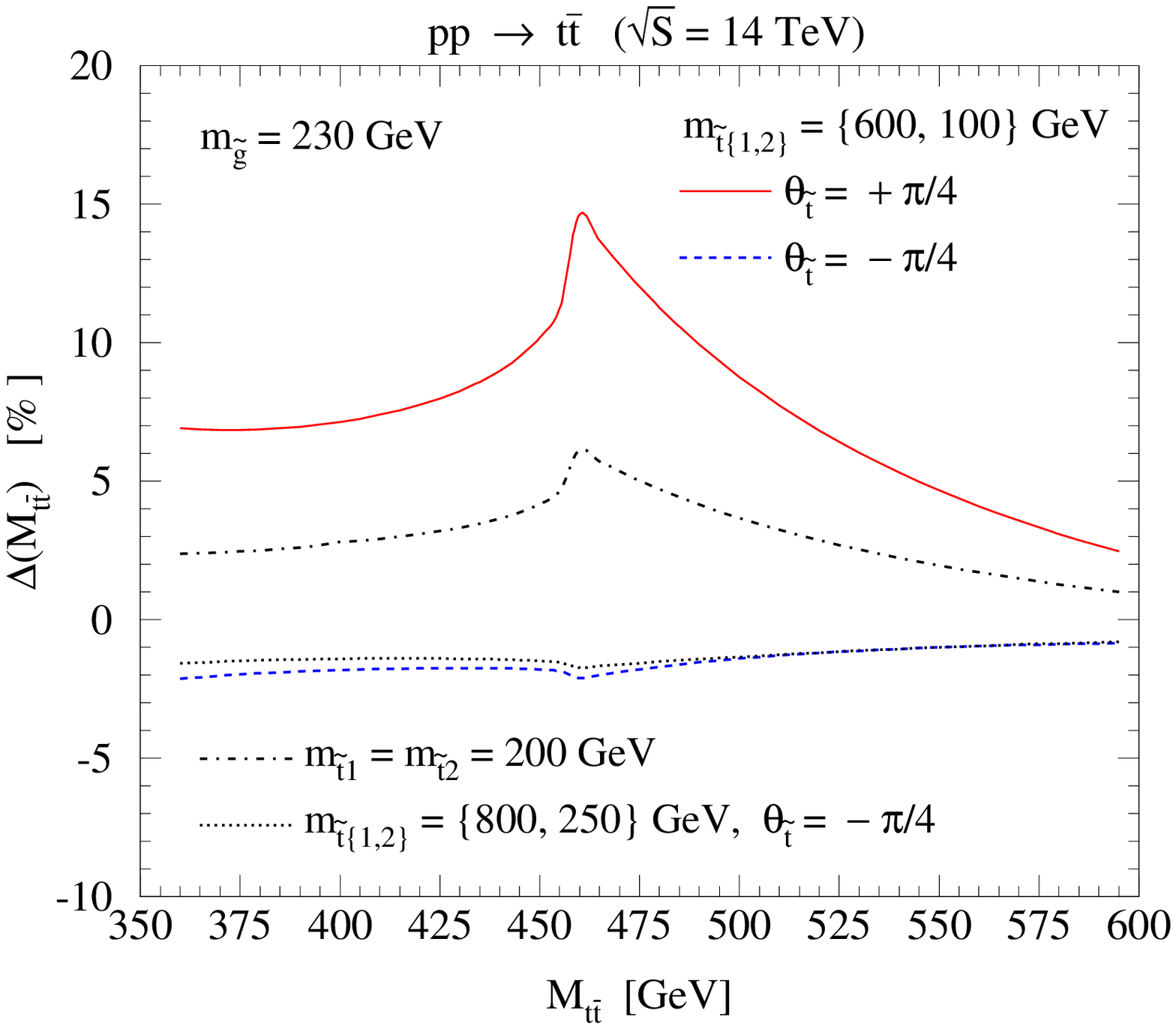}
\hspace*{-.8cm}\\[-30pt]
\hspace*{.6cm}(a)\hspace*{8cm}(b)\\
\vspace*{-25pt}
\end{center}
\caption{\emph{The relative correction $\Delta(M_{t\bar t})$ 
due to SQCD one-loop corrections at 
(a)~the Tevatron Run~II and (b)~the LHC.}}\label{fig:Had_mtt}
\end{figure}

\begin{figure}[h]
\begin{center}
  \includegraphics[width=8.1cm,
  keepaspectratio]{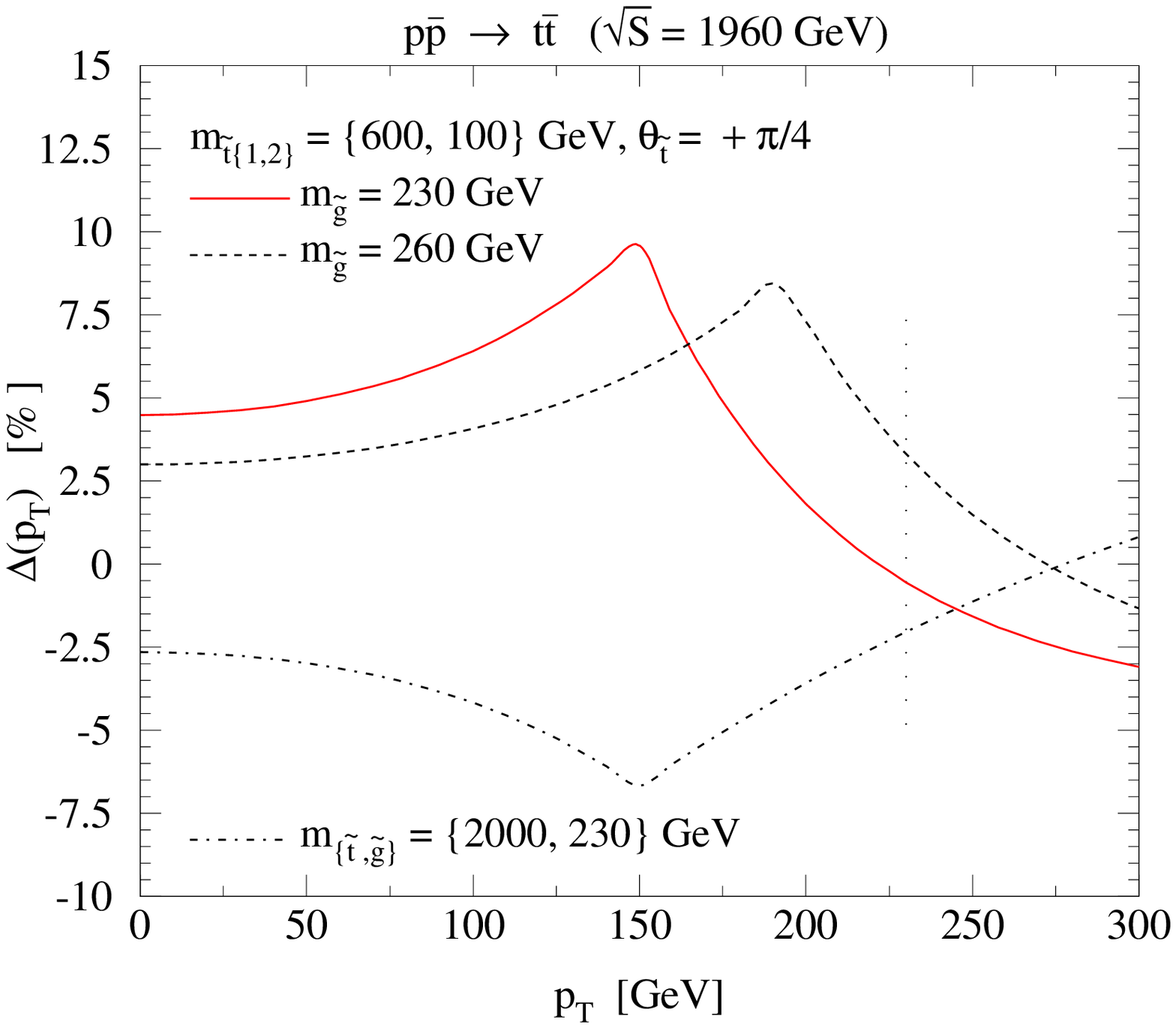}
\hspace*{-.0cm}
  \includegraphics[width=8.1cm,
  keepaspectratio]{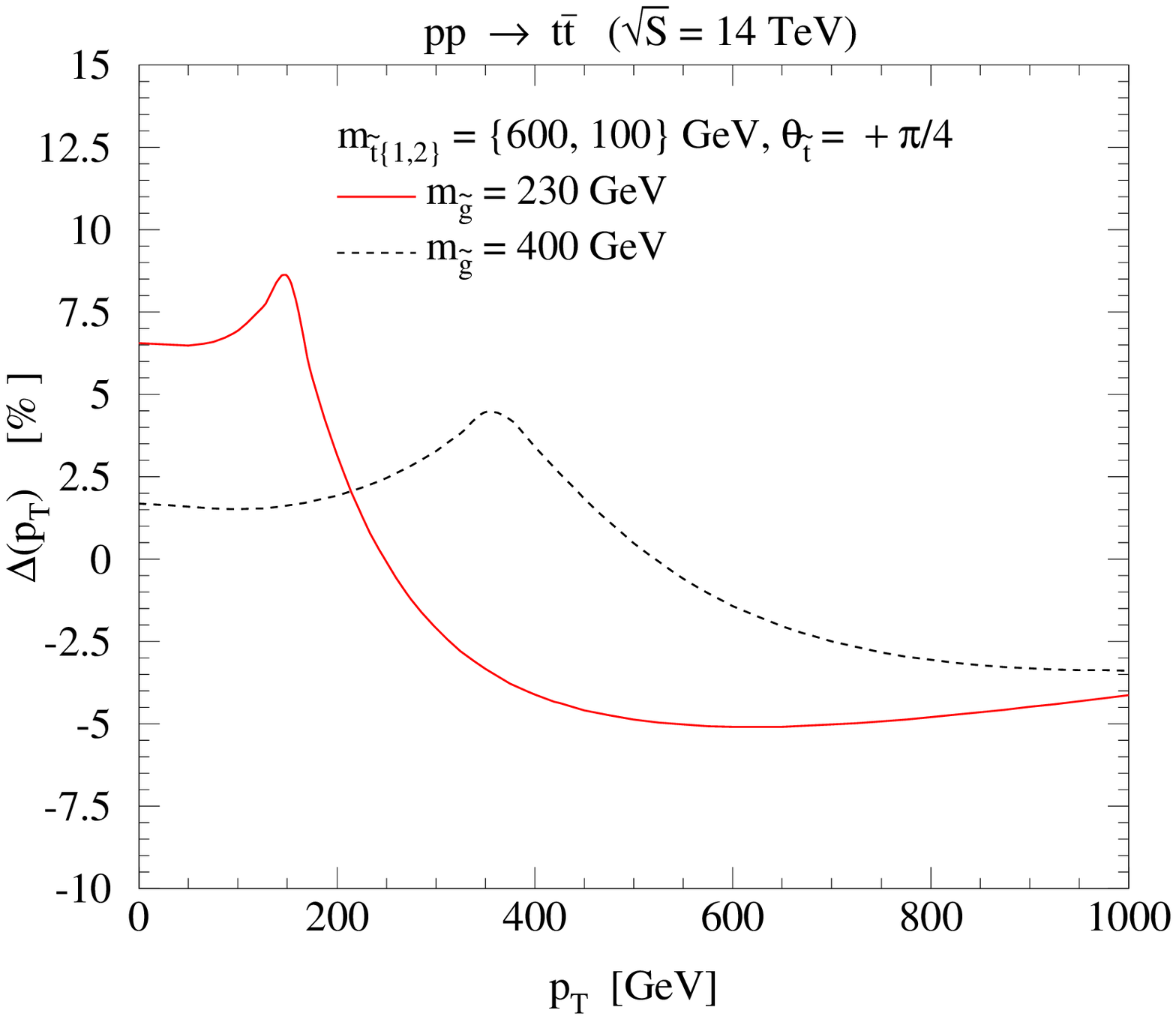}
\hspace*{-.8cm}\\[-30pt]
\hspace*{.6cm}(a)\hspace*{8cm}(b)\\
\vspace*{-25pt}
\end{center}
\caption{\emph{The relative correction $\Delta(p_T)$
due to SQCD one-loop corrections at 
(a)~the Tevatron Run~II and (b)~the LHC.}}\label{fig:Had_pt}
\end{figure}

The relative corrections to the $M_{t\bar t}$ and $p_T$ distributions
of Eq.~(\ref{eq:haddeltas}) are shown in Fig.~\ref{fig:Had_mtt} and
Fig.~\ref{fig:Had_pt}, respectively.
Both the $M_{t\bar{t}}$ and the $p_T$ distribution exhibit the
characteristic effects of the gluino-pair threshold, which can lead to
a significant distortion of the shape of these distributions. At the
Tevatron Run~II, for instance, when choosing
$m_{\tilde{t}_1,\tilde{t}_2} = 600,100$~GeV,
$\theta_{\tilde{t}}=\pi/4$ and $m_{\tilde{g}} = 230$~GeV, the
corrections to the $M_{t\bar t}(p_T)$~distribution increase for
$M_{t\bar t}<460$~GeV$(p_T < 150$~GeV) and can enhance the LO
distribution by up to $\Delta(M_{t\bar t})(\Delta(p_T))=+15(+9.5)\%$ and
then decrease quickly for larger values of $M_{t\bar t}(p_T)$.  This
behavior suggests that the relative corrections to the total $t\bar t$
cross section can be enhanced by applying cuts on the $p_T$ of the top
quark.  For instance, when restricting the top quark $p_T$ to the
range $75$~GeV~$<p_T<170$~GeV, the relative correction at the Tevatron
Run II reaches $\Delta=+7.1\%$ (instead of $\Delta=+5.4\%$ without
cuts, see~Fig.~\ref{fig:Had_mg}).  Keeping the same top-squark
parameters but choosing $m_{\tilde{g}} = 260$~GeV and
$100$~GeV~$<p_T<210$~GeV, we find $\Delta=+5.5\%$ (instead of
$\Delta=+4.1\%$ without cuts).  At the LHC, choosing
$m_{\tilde{t}_1,\tilde{t}_2} =600,100$~GeV,
$\theta_{\tilde{t}}=\pi/4$, $m_{\tilde{g}} = 230$~GeV and
$100$~GeV~$<p_T<170$~GeV, we find $\Delta=+7.5\%$ (instead of
$\Delta=+5.7\%$ without cuts).  We also studied the effects of
rapidity cuts and found that they do not affect the relative
corrections much, because the NLO SQCD corrections are quite stable
with respect to the rapidity distribution of the top quark.

In view of anticipated experimental uncertainties of about $10\%$ and
$5\%$ in the total $t\bar t$ production rate at the Tevatron Run II
and the LHC, respectively, and a current theoretical uncertainty of
the QCD prediction of about
12\%~\cite{Kidonakis:2003qe,Cacciari:2003fi}, it will be difficult to
reach sensitivity on SQCD effects in this observable. The $M_{t\bar
  t}$ and $p_T$ distributions exhibit an interesting signature of SQCD
one-loop corrections, but they are strongly affected in only a few
bins and the final verdict on its observability must be left to an
analysis, that includes for instance top decays and the detector
response.  However, the inclusion of the known SUSY electroweak
one-loop effects~\cite{Hollik:1997hm} may help to enhance SUSY
loop-induced effects. First results of combined SUSY electroweak and
SQCD one-loop corrections have been presented at the Tevatron in
Ref.~\cite{Wackeroth:1998wm}, but a detailed study at the LHC and the
Tevatron Run~II still needs to be done.

%
%

%
%
%
\subsection{Polarized top-pair production at NLO SQCD}
\label{sec:res_polarized}

In this section, we consider the production of polarized top-quark
pairs at the Tevatron Run~II and the LHC. We are working in the
helicity basis of left- and right-handed top quarks, as discussed in
Section~\ref{sec:parton}.  The incoming quarks and gluons are still
considered to be unpolarized. In the following, we use the notation
that the indices 'L' and 'R' denote left ($\lambda_{t(\bar t)}=-1/2$)
and right-handed ($\lambda_{t(\bar t)}=+1/2$) top(antitop) quarks,
respectively.

Since the SQCD couplings of gluinos and squarks contain an
axial-vector part (see Appendix~\ref{sec:feynman_rules},
Fig.~\ref{fig:feyntwo}), the SQCD one-loop corrections may affect the
production of left and right-handed top quarks differently.  To study
these differences in detail, we first discuss in
Sections~\ref{sec:res_qqa_polarized} ($q\bar q$ annihilation) and
Section~\ref{sec:res_gg_polarized} (gluon fusion) the impact of the
SQCD one-loop corrections on the total partonic $t\bar t$ cross
sections of Eq.~(\ref{eq:parton}) for each top and antitop helicity
state separately, using the following relative corrections at the
parton level:
\begin{equation}
\label{eq:deltaspin}
\hat\Delta_{q\bar q,gg}^{ab}(\hat s)= \frac{\hat{\sigma}^{NLO}_{q\bar q,gg}(\lambda_t,\lambda_{\bar t}) - 
\hat{\sigma}_{q\bar q,gg}^{LO}(\lambda_t,\lambda_{\bar t})}{\hat{\sigma}_{q\bar q,gg}^{LO}(\lambda_t,\lambda_{\bar t})} \; ,
\end{equation}
with $ab=LL,RR$ denoting the relative correction for spin-like $t\bar
t$ production with $\lambda_t=-1/2,\lambda_{\bar t}=-1/2$ and
$\lambda_t=+1/2,\lambda_{\bar t}=+1/2$, respectively. Similarly,
$ab=LR,RL$ denotes the relative correction for spin-unlike $t\bar t$
production with $\lambda_t=-1/2,\lambda_{\bar t}=+1/2$ and
$\lambda_t=+1/2,\lambda_{\bar t}=-1/2$, respectively.  Naturally,
since QCD preserves parity, there is no difference in the spin-unlike
amplitudes at LO QCD, i.e.~$\hat{\sigma}_{q\bar
  q,gg}^{LO}(+1/2,-1/2)=\hat{\sigma}_{q\bar q,gg}^{LO}(-1/2,+1/2)$.
Since we are not considering any CP violating couplings, the spin-like
amplitudes are the same at both LO QCD and NLO SQCD, i.e.~
$\hat{\sigma}_{q\bar q,gg}^{LO,NLO}(-1/2,-1/2)=\hat{\sigma}_{q\bar
  q,gg}^{LO,NLO}(+1/2,+1/2)$.

Differences in the spin-unlike amplitudes at NLO SQCD manifest
themselves in parity-violating polarization asymmetries, which we
study in detail at hadron level in
Section~\ref{sec:res_hadronic_polarized}.  We consider the following
differential and integrated polarization
asymmetries~\cite{Kao:1999kj}:
\begin{itemize}
\item 
The left-right asymmetry in the $M_{t\bar t}$ distribution 
\begin{eqnarray}\label{eq:mttasymone}
\delta {\cal A}_{LR}(M_{t\bar t}) &= & \frac{d\sigma_{RL}/dM_{t\bar{t}}-d\sigma_{LR}/dM_{t\bar{t}}}{d\sigma_{RL}/dM_{t\bar{t}}+d\sigma_{LR}/dM_{t\bar{t}}} 
\end{eqnarray}
and in the total hadronic $t\bar t$ cross section
\begin{eqnarray}\label{eq:asymone}
{\cal A}_{LR} &= & \frac{\sigma_{RL}-\sigma_{LR}}{\sigma_{RL}+\sigma_{LR}} \; ,
\end{eqnarray}
where we introduced the notation
$d\sigma_{LR(RL)}\equiv d\sigma_{NLO}(\lambda_t=-1/2(+1/2),\lambda_{\bar t}=+1/2(-1/2))$
with the hadronic NLO SQCD cross sections of Eq.~(\ref{eq:hadwq}).
\item The left-right asymmetry in the $M_{t \bar t}$ distribution,
  when assuming that the polarization of the antitop quark is not
  measured in the experiment (denoted by $U$=unpolarized),
\begin{eqnarray}\label{eq:mttasymtwo}
\delta {\cal A}(M_{tt}) &= & 
\frac{(d\sigma_{RL}/dM_{t\bar{t}}+d\sigma_{RR}/dM_{t\bar{t}})-
(d\sigma_{LL}/dM_{t\bar{t}}+d\sigma_{LR}/dM_{t\bar{t}})}
{(d\sigma_{RL}/dM_{t\bar{t}}+d\sigma_{RR}/dM_{t\bar{t}})+(d\sigma_{LL}/dM_{t\bar{t}}+d\sigma_{LR}/dM_{t\bar{t}})}
\nonumber\\
&=& 
\frac{d\sigma_{RU}/dM_{t\bar{t}}-d\sigma_{LU}/dM_{t\bar{t}}}{d\sigma_{NLO}/dM_{t\bar{t}}} 
\end{eqnarray}
and the corresponding asymmetry in the total hadronic $t\bar t$ cross section
\begin{eqnarray}\label{eq:asymtwo}
{\cal A} &= & 
\frac{(\sigma_{RL}+\sigma_{RR})-
(\sigma_{LL}+\sigma_{LR})}
{\sigma_{RL}+\sigma_{RR}+\sigma_{LL}+\sigma_{LR}}
= 
\frac{\sigma_{RU}-\sigma_{LU}}{\sigma_{NLO}}  
\end{eqnarray}
with the unpolarized NLO SQCD cross sections, $d\sigma_{NLO} =
\sum_{\lambda_t,\lambda_{\bar t}=\pm 1/2}
d\sigma_{NLO}(S,\lambda_t,\lambda_{\bar t})$.  In this way, one
complicated polarization measurement is avoided.
\end{itemize}
These parity-violating asymmetries are zero for stop mixing angles
$\theta_{\tilde{t}}=\pm \frac{\pi}{4}$ (then $\lambda_j^A=0$, see
Appendix~\ref{sec:feynman_rules}) and top squarks degenerate in mass.
The latter is due a cancellation of the parity-violating terms in the
sum of the $\tilde t_1$ and $\tilde t_2$ contributions, since
$\lambda_1^A=-\lambda_2^A$ and for equal stop masses the terms
multiplying $\lambda_A^j$ are the same.

In addition to the parity-violating polarization asymmetries, we study
in Section~\ref{sec:res_hadronic_polarized} the spin correlation functions
of Ref.~\cite{Stelzer:1995gc}, which are parity-conserving asymmetries
in the spin-like and spin-unlike contributions to the $t\bar t$ cross
sections.  They are defined as follows:
\begin{eqnarray}\label{eq:corrone}
\overline{C}& = &
\frac{(d\sigma_{RR}/dM_{t\bar{t}}+d\sigma_{LL}/dM_{t\bar{t}})
-(d\sigma_{LR}/dM_{t\bar{t}}+d\sigma_{RL}/dM_{t\bar{t}})}{
(d\sigma_{RR}/dM_{t\bar{t}}+d\sigma_{LL}/dM_{t\bar{t}})
+(d\sigma_{LR}/dM_{t\bar{t}}+d\sigma_{RL}/dM_{t\bar{t}})} \nonumber\\
&=&
\frac{(d\sigma_{RR}/dM_{t\bar{t}}+d\sigma_{LL}/dM_{t\bar{t}})
-(d\sigma_{LR}/dM_{t\bar{t}}+d\sigma_{RL}/dM_{t\bar{t}})}{
d\sigma_{NLO}/dM_{t\bar{t}}}
\end{eqnarray}
and
\begin{eqnarray}\label{eq:corrtwo}
{C} &= & 
\frac{(\sigma_{RR}+\sigma_{LL})-(\sigma_{LR}+\sigma_{RL})}{
(\sigma_{RR}+\sigma_{LL})+(\sigma_{LR}+\sigma_{RL})} 
=
\frac{(\sigma_{RR}+\sigma_{LL})-(\sigma_{LR}+\sigma_{RL})}{\sigma_{NLO}} 
\; .
\end{eqnarray}

%
%
%
%
\subsubsection{Effects of NLO SQCD corrections in $q\bar q$ annihilation for polarized top quarks}
\label{sec:res_qqa_polarized}

\begin{figure}[h]
\begin{center}
  \includegraphics[width=8.1cm,
  keepaspectratio]{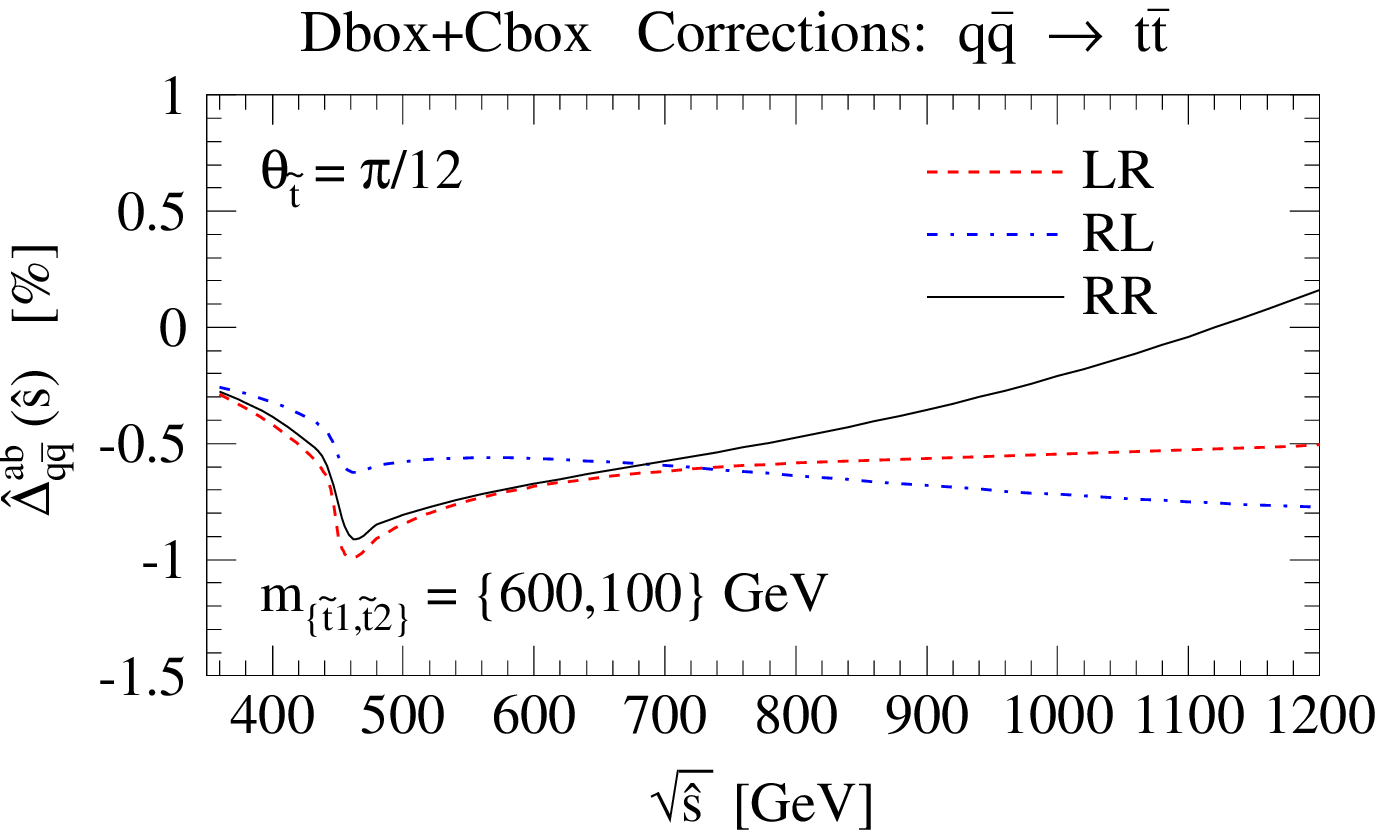}
\hspace*{-.0cm}
  \includegraphics[width=8.1cm,
  keepaspectratio]{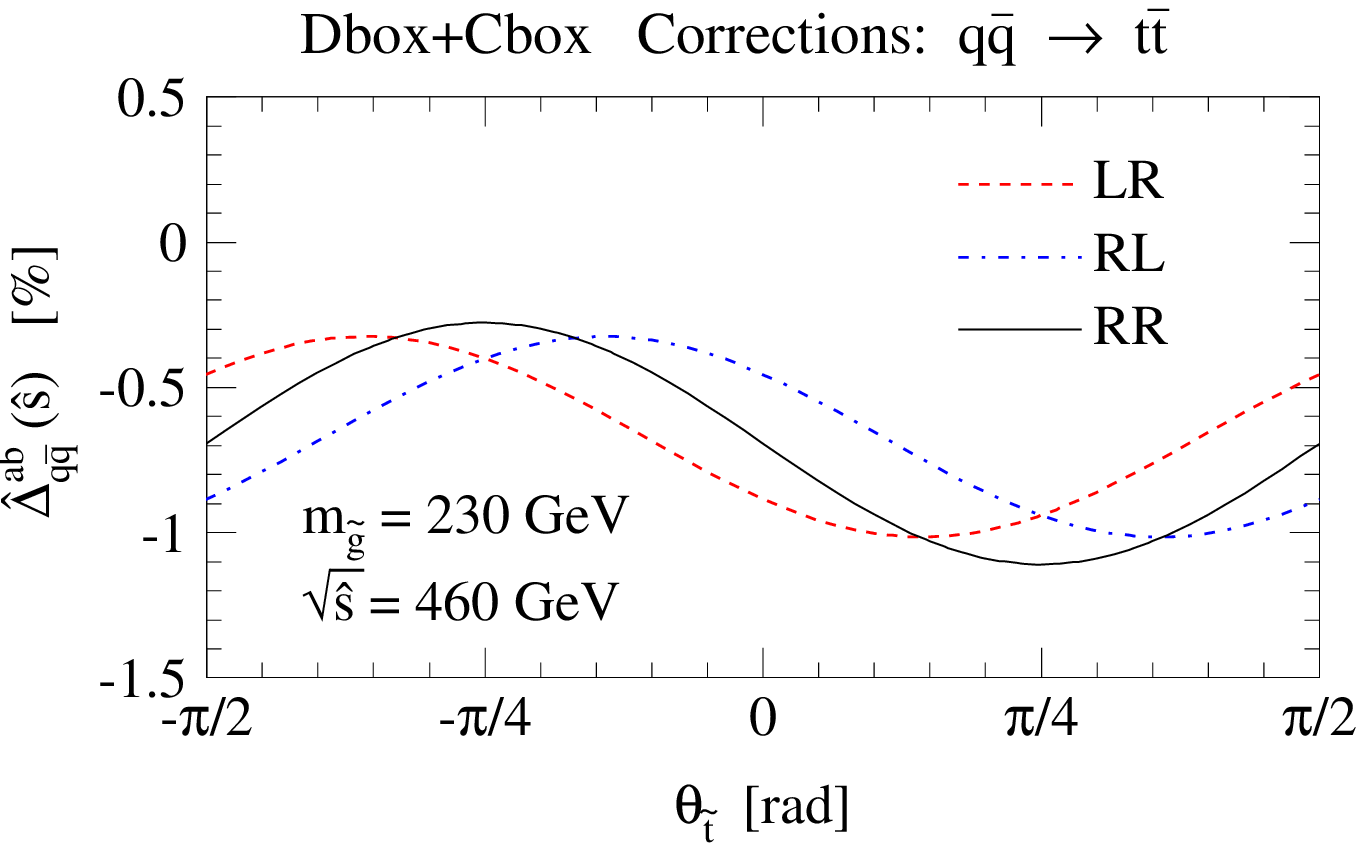}
\hspace*{-.8cm}\\[-30pt]
\hspace*{.6cm}(a)\hspace*{8cm}(b)\\
\vspace*{-25pt}
\end{center}
\caption{\emph{The relative corrections $\hat \Delta_{q\bar q}^{ab}(\hat s) \,(ab=LR,RL,RR=LL)$ due to the box correction 
for polarized top-quark pairs in $q\bar q$ annihilation in dependence of (a) the partonic CMS energy and (b)
the stop mixing angle.}}\label{fig:Pol_qqtt_cdbox}
\end{figure}

In this section, we study the effects of the SQCD one-loop corrections
on the partonic cross sections for polarized top quarks in $\qqa$. We
are especially interested in the differences between the corrections
to $t_L \bar t_R$ and $t_R \bar t_L$ production, since they will
determine the size of the polarization asymmetries at hadron level at
the Tevatron. As in the unpolarized case, we only show results for
$u\bar u \to t\bar t$, representatively for all $q\bar q$-initiated
processes. In Fig.~\ref{fig:Pol_qqtt_cdbox}, we show the relative
correction $\hat \Delta_{q\bar q}^{ab}$ of Eq.~(\ref{eq:deltaspin})
for the sum of the direct and crossed box corrections, separately for
each top(antitop) polarization state.  The characteristics of the box
corrections are similar to the ones observed in the unpolarized case,
which we discussed in Section~\ref{sec:res_qqa_unpolarized}.  In
particular, the polarized box correction also exhibits the threshold
behavior at $\sqrt{\hat{s}} = 2\, m_{\tilde{g}}$.
Fig.~\ref{fig:Pol_qqtt_cdbox}(a) shows the correction in dependence of
the partonic CMS energy for a small gluino mass of
$m_{\tilde{g}}=230$~GeV and for a choice of the stop mixing angle,
where we expect the left-right asymmetry to be large.  The dependence
of the box correction on the stop mixing angle is shown in
Fig.~\ref{fig:Pol_qqtt_cdbox}(b) with $\sqrt{\hat{s}}$ chosen to be at
the gluino-pair threshold. As can be seen, the differences between the
relative corrections to $t_L \bar t_R$ and $t_R \bar t_L$ production
are largest for $\theta_{\tilde t}=\pm\pi/2,0$.

\begin{figure}[h]
\begin{center}
  \includegraphics[width=8.1cm,
  keepaspectratio]{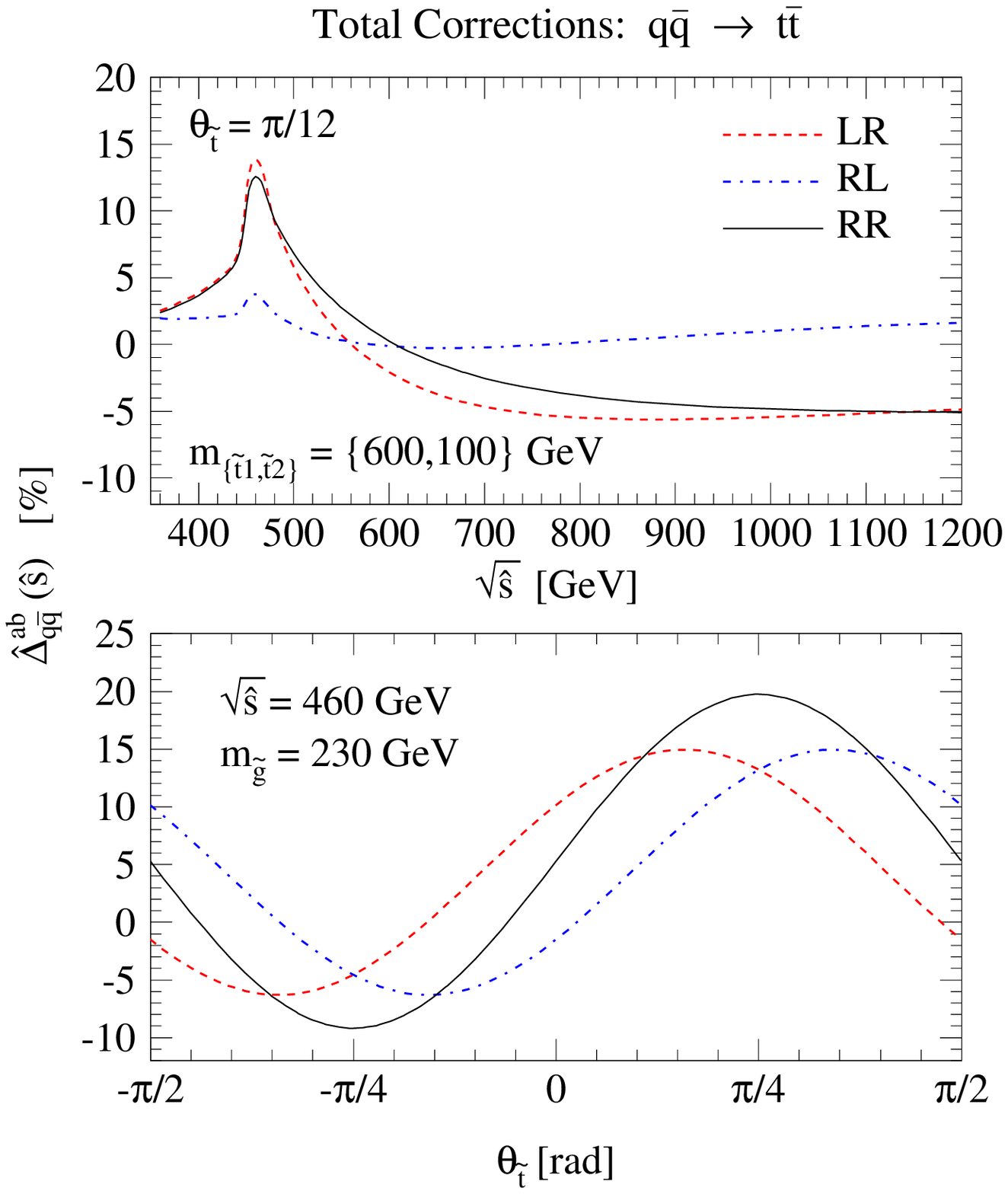}
\hspace*{-.0cm}
  \includegraphics[width=8.1cm,
  keepaspectratio]{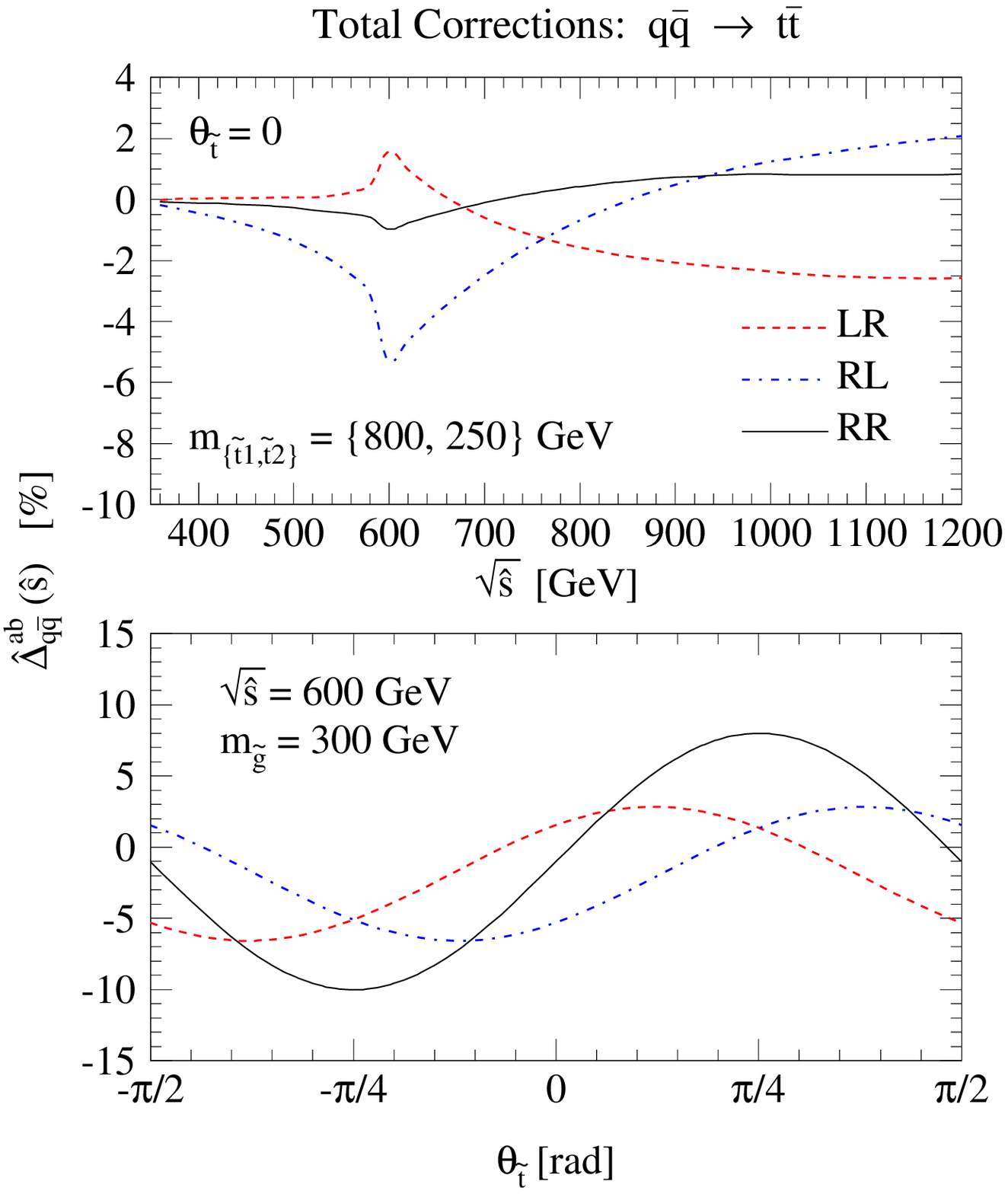}
\hspace*{-.8cm}\\[-30pt]
\hspace*{.6cm}(a)\hspace*{8cm}(b)\\
\vspace*{-25pt}
\end{center}
\caption{\emph{The relative corrections $\hat \Delta_{q\bar q}^{ab}(\hat s) \, (ab=LR,RL,RR=LL)$ due to the 
complete SQCD one-loop corrections for polarized top-quark pairs in $\qqa$ in dependence
of the partonic CMS energy (upper plots) and the stop mixing angle (lower plots) for two
sets of choices of the MSSM parameters: 
(a)~$m_{\tilde{g}} = 230$~GeV, $m_{\tilde{t}_1,\tilde{t}_2} = 600,100$~GeV and 
(b)~$m_{\tilde{g}} = 300$~GeV, $m_{\tilde{t}_1,\tilde{t}_2} = 800,250$~GeV.
}}\label{fig:Pol_qqtt_tot_mg300}
\end{figure}

In Fig.~\ref{fig:Pol_qqtt_tot_mg300} we show the relative correction
due to the complete SQCD one-loop correction in dependence of the
partonic CMS energy (upper plots) and the stop mixing angle (lower
plots).  Since the box correction is very small compared to the
complete correction, and the gluon self-energy is independent of the
top(antitop) polarization (and thus cancel in the polarization
asymmetries), the differences between the relative corrections to $t_L \bar
t_R$ and $t_R \bar t_L$ production observed in
Fig.~\ref{fig:Pol_qqtt_tot_mg300} are mainly due to the vertex
correction.  Fig.~\ref{fig:Pol_qqtt_tot_mg300}(a) shows the
corrections for $m_{\tilde{g}}=230$~GeV and $m_{\tilde{t}_2}=100$~GeV,
and Fig.~\ref{fig:Pol_qqtt_tot_mg300}(b) the corrections for somewhat
heavier particles, $m_{\tilde{g}}=300$~GeV and
$m_{\tilde{t}_2}=250$~GeV. The vertex correction introduces large
partonic left-right asymmetries in the gluino-pair threshold region
and at high $\sqrt{\hat s}$ for $\theta_{\tilde t}=\pm \pi/2,0$.
Furthermore, as illustrated in Fig.~\ref{fig:Pol_qqtt_tot_mg300}
(lower plots), the largest relative corrections occur in spin-like
top-pair production ($\hat\Delta_{q\bar q}^{RR}=\hat\Delta_{q\bar
  q}^{LL}$) at $\theta_{\tilde{t}} = \pm\frac{\pi}{4}$.  In this case,
the relative difference between the corrections to spin-like ($t_R \bar t_R,
t_L \bar t_L$) and spin-unlike ($t_R \bar t_L, t_L \bar t_R$) top-pair
production increases with increasing gluino mass.  We found that if
the stops are degenerate in mass, the corrections to spin-like and
spin-unlike $t\bar t$ production are of comparable size. If in
addition the top squarks are very heavy ($m_{\tilde t_1}=m_{\tilde
  t_2}\approx 2$~TeV), the corrections are completely independent of
the top polarization states, because then they are determined by the
gluon self-energy, as discussed in
Section~\ref{sec:res_qqa_unpolarized}.

%
%
\subsubsection{Effects of NLO SQCD corrections in gluon fusion for polarized top quarks}
\label{sec:res_gg_polarized}

\begin{figure}[h]
\begin{center}
  \includegraphics[width=8.1cm,
  keepaspectratio]{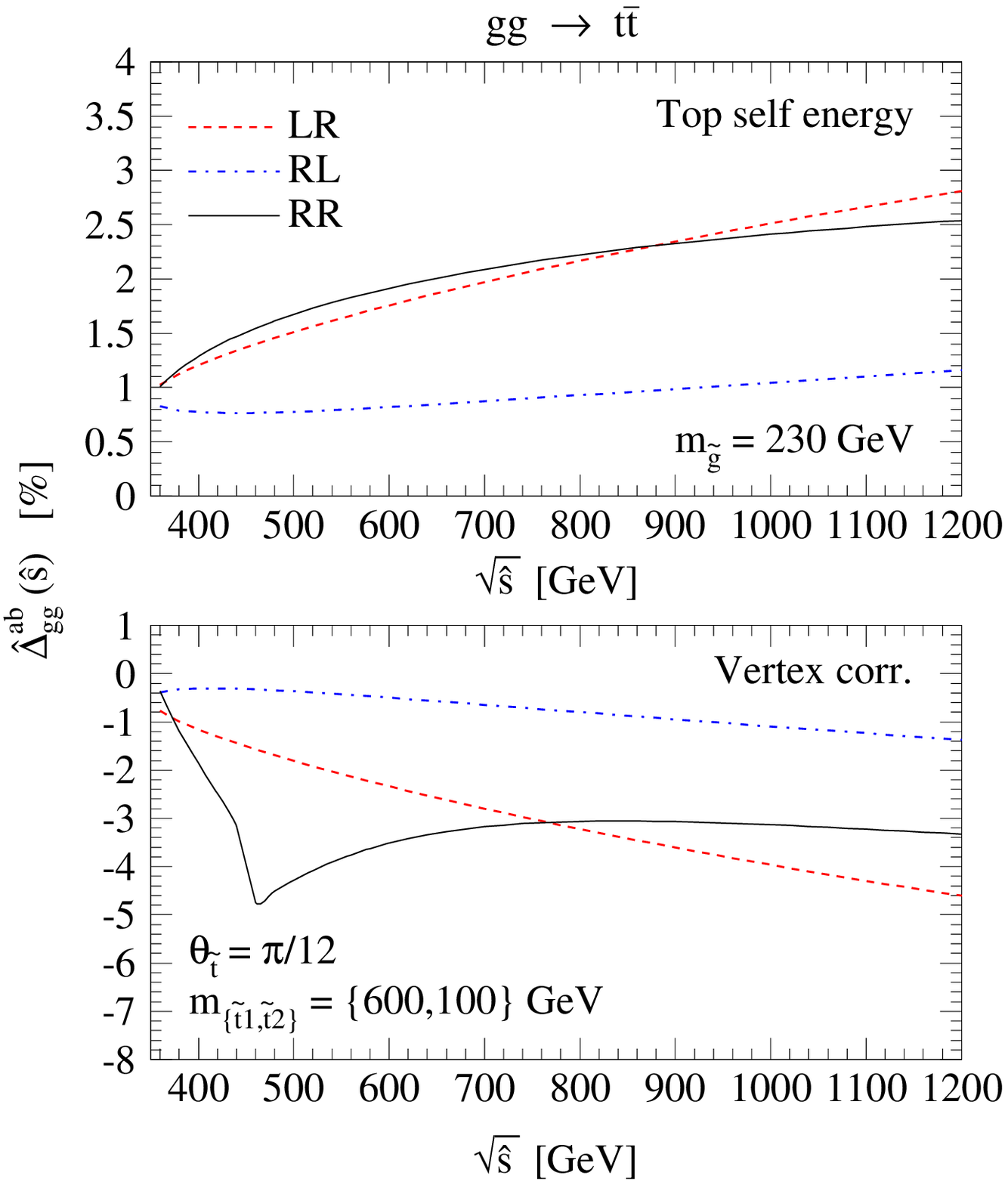}
\hspace*{-.0cm}
  \includegraphics[width=8.1cm,
  keepaspectratio]{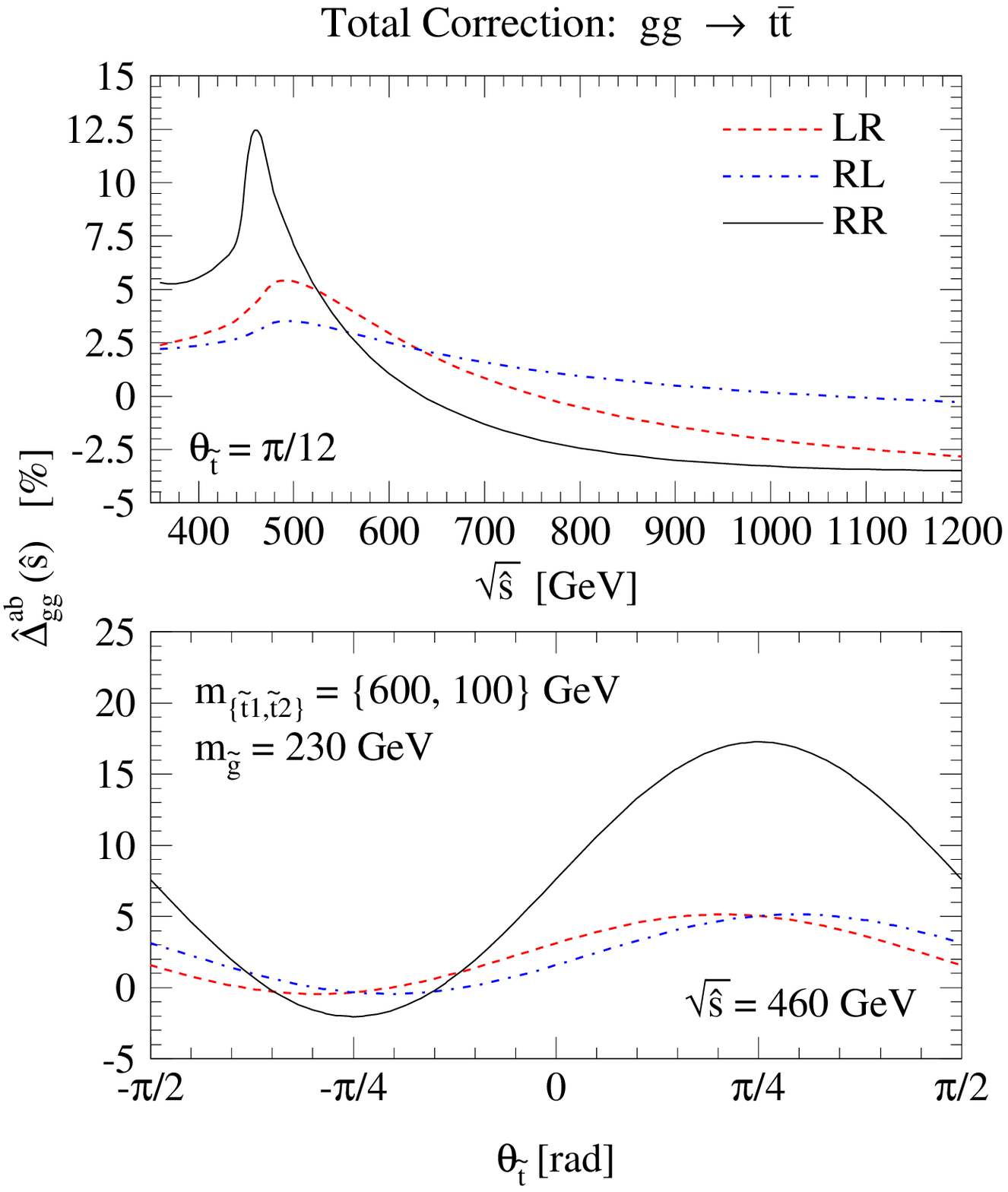}
\hspace*{-.8cm}\\[-30pt]
\hspace*{.6cm}(a)\hspace*{8cm}(b)\\
\vspace*{-25pt}
\end{center}
\caption{\emph{The relative corrections $\hat \Delta_{gg}^{ab}(\hat s)\, (ab=LR,RL,RR=LL)$ 
for polarized top-quark pairs in gluon fusion due to (a) top self-energy and vertex corrections
and (b) the complete SQCD one-loop correction in dependence of the partonic CMS energy. Also shown is the stop mixing angle dependence
of the complete SQCD one-loop correction ((b) lower plot).}}\label{fig:Pol_ggtt_partonic}
\end{figure}

When studying the effects of the SQCD one-loop corrections on the
partonic cross sections for polarized top quarks in gluon fusion, we
again are especially interested in the differences between $t_R \bar
t_L$ and $t_L \bar t_R$ production, since this will determine the size
of the polarization asymmetries at hadron level at the LHC.  As for
$q\bar{q}$ annihilation, also in gluon fusion the gluon self-energy
correction is independent of the top(antitop) polarization states. In
Fig.~\ref{fig:Pol_ggtt_partonic}(a) we therefore show the relative
corrections $\hat\Delta_{gg}^{ab}$ of Eq.~(\ref{eq:deltaspin})
separately only for the top self-energy and vertex corrections,
choosing $m_{\tilde{g}}=230$~GeV and $m_{\tilde t_{1,2}}=600,100$~GeV.
As can be seen, the top self-energy and the vertex corrections depend
strongly on the polarizations of the top and antitop quarks, inducing
differences between the corrections to $t_R \bar t_L$ and $t_L \bar
t_R$ production of several percent.  However, the partonic left-right
asymmetries due to these two corrections are of opposite sign and
therefore almost cancel in the complete SQCD one-loop correction. Also
the corrections to spin-like top-pair production have opposite signs
and interfere destructively.  Therefore, the relative correction due
to the complete SQCD one-loop correction, shown in
Fig.~\ref{fig:Pol_ggtt_partonic}(b), is largely determined by the box
correction. The relative correction for spin-like top-pair production
is much larger than the one for $t_R \bar t_L,t_L \bar t_R$
production, which is still the case for larger values of
$m_{\tilde{g}}$.  The difference in the box correction to $t_R \bar
t_L$ and $t_L \bar t_R$ production is relatively small and therefore
also the partonic left-right asymmetry induced by the complete SQCD
one-loop corrections is small.

%
%
\subsubsection{Hadronic cross sections to polarized $pp,p\bar p \to t\bar t$ at NLO SQCD}\label{sec:res_hadronic_polarized}

In the following discussion of parity violating effects in polarized
$t\bar t$ production at the Tevatron Run II and the LHC, we benefit
from the detailed study of these effects at the parton level in
Sections~\ref{sec:res_qqa_polarized},~\ref{sec:res_gg_polarized}, so
that we can restrict our discussion to the following two sets of
choices for the stop masses and mixing angle:
\begin{eqnarray}\label{Eq:StopMassSets_LR_Asym}
&(I)  & \ m_{\tilde{t}_1} = 600~{\rm GeV},\quad m_{\tilde{t}_2} = 100~{\rm GeV},
\quad \theta_{\tilde{t}} = \pi/12 \nonumber\\
&(II) & \ m_{\tilde{t}_1} = 800~{\rm GeV}, \quad m_{\tilde{t}_2} = 250~{\rm GeV},
\quad \theta_{\tilde{t}} = 0
\end{eqnarray}
and to a few values of the gluino mass.  These parameter sets are
representative for the choices that yield the largest numerical
impact.  The top-squark mixing angle of set~(I) is chosen to be
$\pi/12$ instead of zero, since in this case smaller values of
$\theta_{\tilde t}$ render the soft supersymmetry breaking mass
$M^2_{\tilde{t}_R}$ negative.

\begin{figure}[h]
\begin{center}
  \includegraphics[width=7.9cm,
  keepaspectratio]{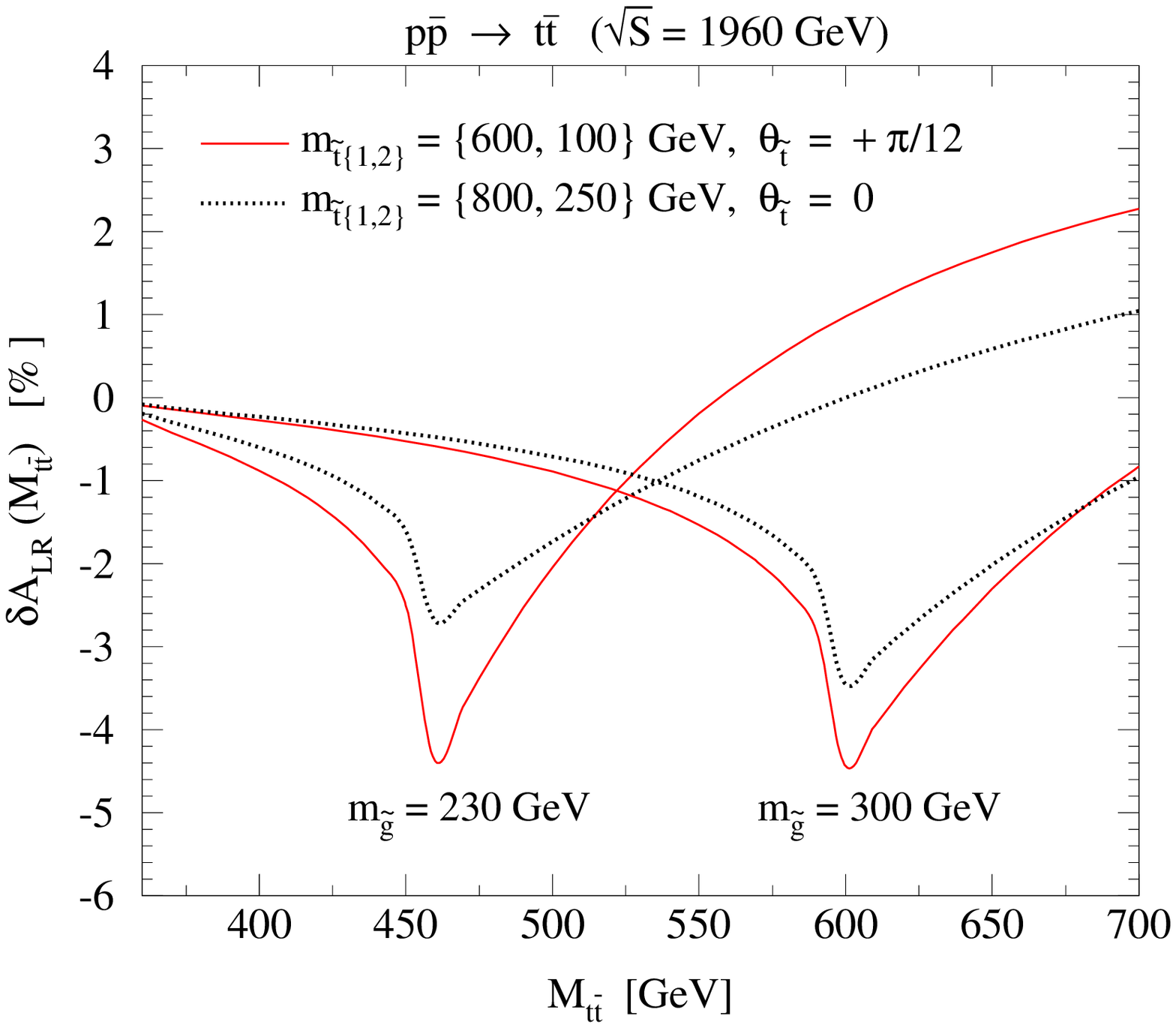}
\hspace*{-.0cm}
  \includegraphics[width=7.9cm,
  keepaspectratio]{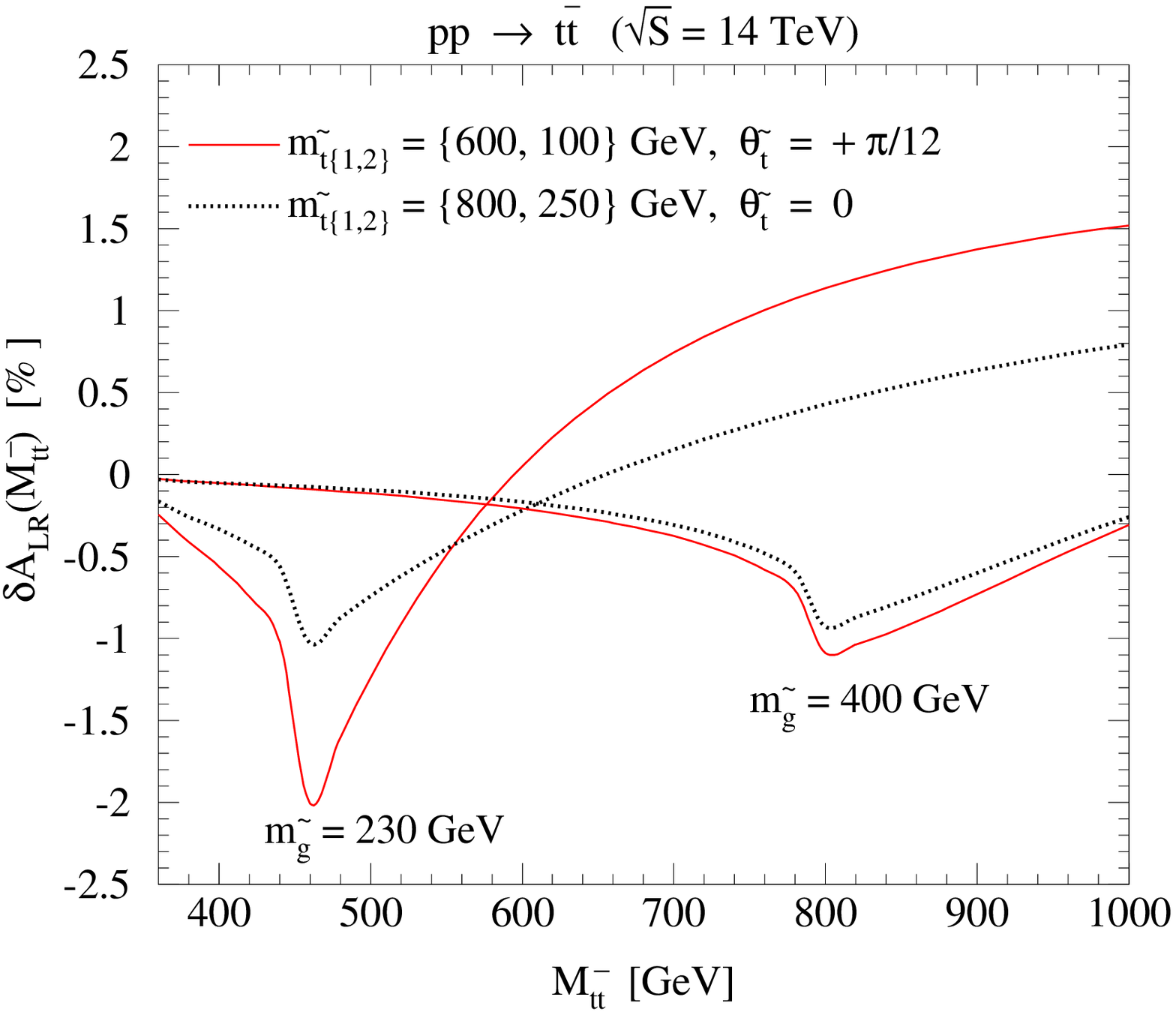}
\hspace*{-.0cm}
  \includegraphics[width=7.9cm,
  keepaspectratio]{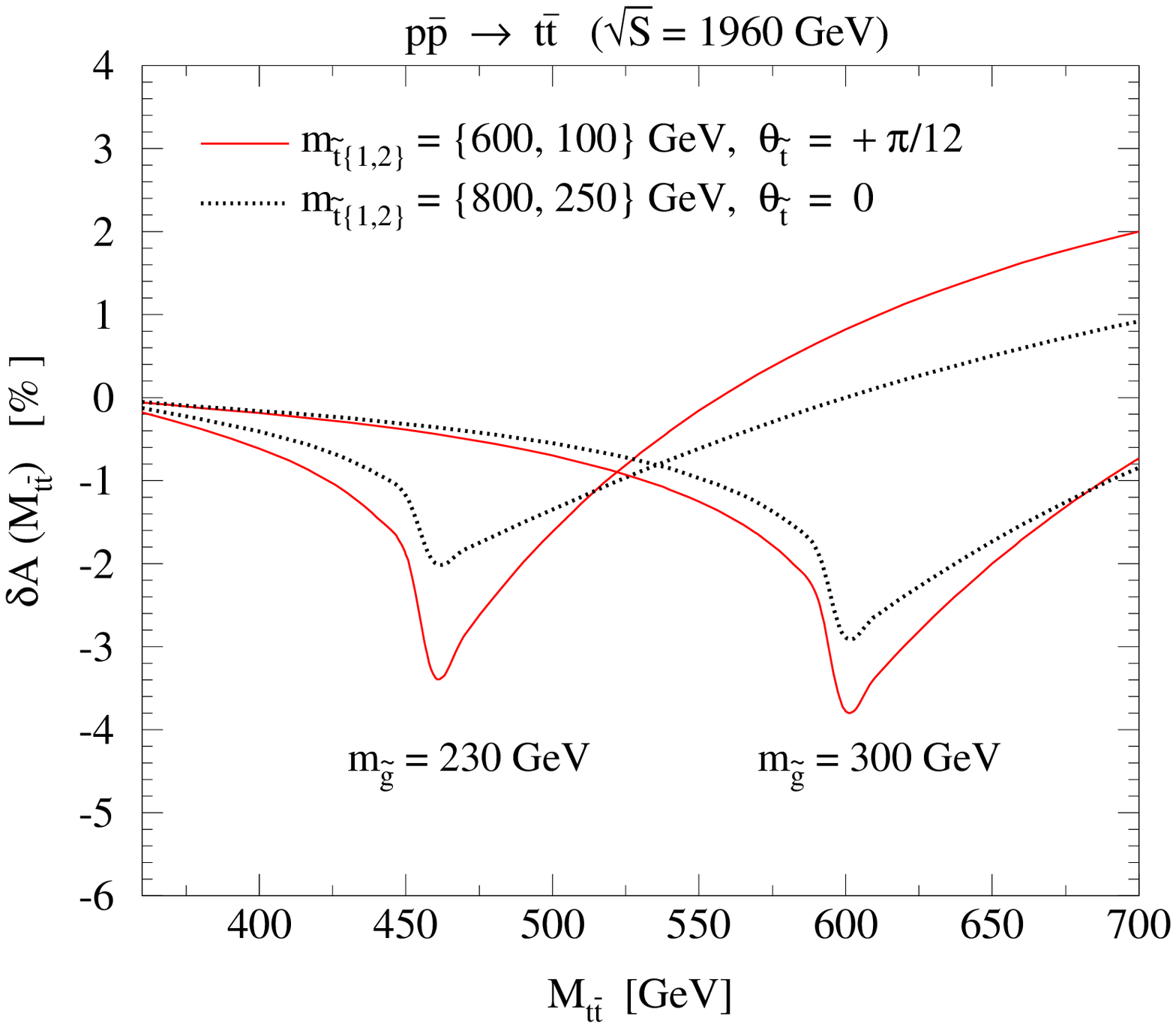}
\hspace*{-.0cm}
  \includegraphics[width=7.9cm,
  keepaspectratio]{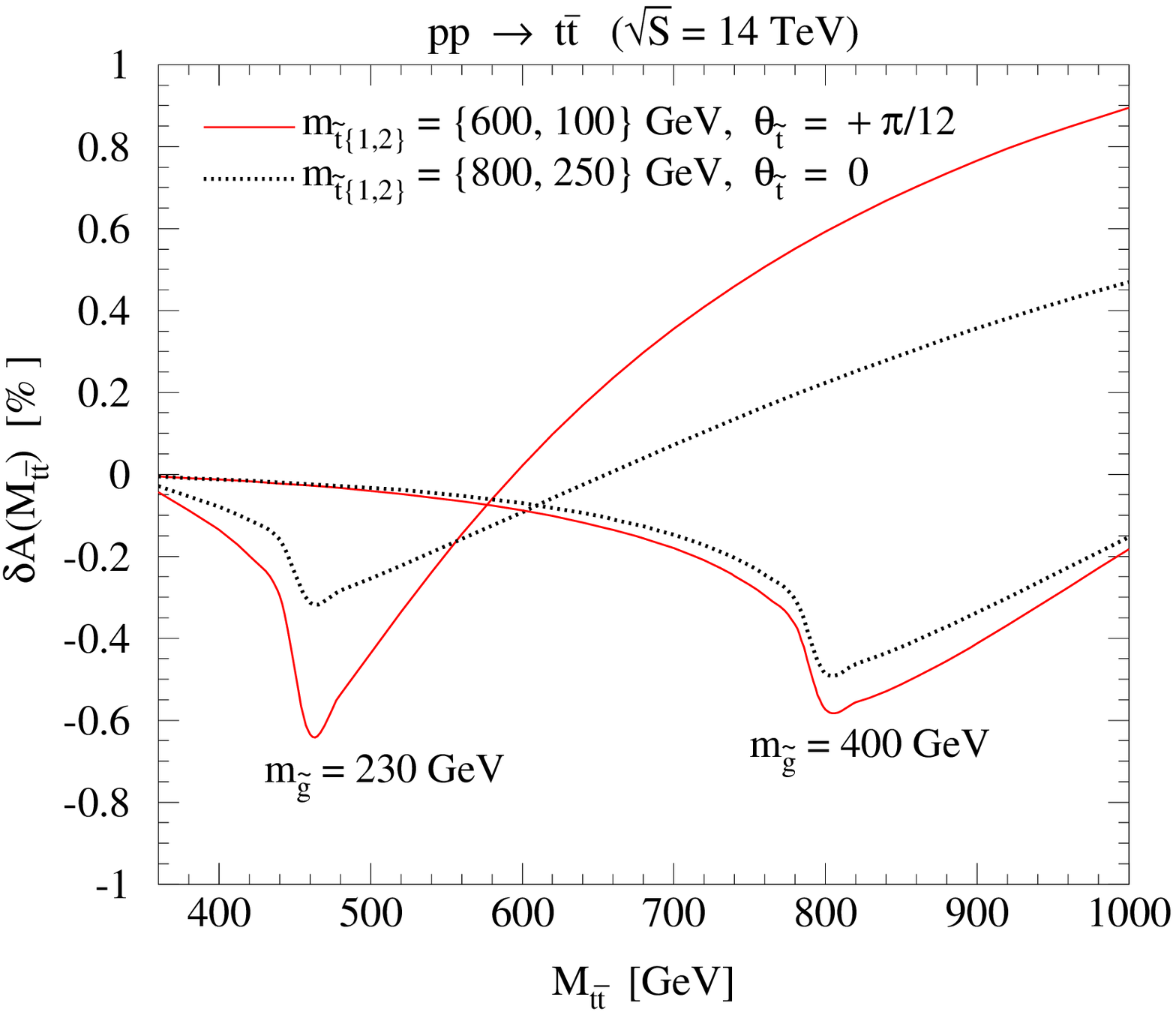}
\hspace*{-.8cm}\\[-10pt]
\hspace*{.6cm}(a)\hspace*{8cm}(b)\\
\vspace*{-25pt}
\end{center}
\caption{\emph{Parity-violating asymmetries in the $M_{t\bar t}$ distribution  
for polarized top-quark pairs (upper plots) and
for polarized top and unpolarized antitop quarks (lower plots) at (a) 
the Tevatron Run~II and (b)~the LHC.
}}\label{fig:Had_LR_Tev}
\end{figure}

At the Tevatron Run II, the integrated left-right asymmetries in the
total hadronic cross section of
Eqs.~(\ref{eq:asymone}),~(\ref{eq:asymtwo}) are largest for set (I): We
find ${\cal A}_{LR}=-1.08\%, {\cal A}=-0.79\%$ for $m_{\tilde{g}} =
230$~GeV, and ${\cal A}_{LR}=-0.72\%, {\cal A}=-0.53 \%$ for
$m_{\tilde{g}} = 300$~GeV. These small asymmetries are clearly not
observable at the Tevatron: For instance, the statistical significance
$N_S$ of Eq.~(19) in Ref.~\cite{Kao:1999kj} only amounts to
$N_S=2.0(1.7)$ for $|{\cal A}_{LR}|=1.08\%(|{\cal A}|=0.79\%)$ for an
integrated luminosity of ${\cal L}=8 \, \rm{fb}^{-1}$, while
conservatively $N_S>4$ is required to be statistically significant.
In Fig.~\ref{fig:Had_LR_Tev}(a) we show the parity-violating
asymmetries in the $M_{t \bar t}$ distribution, $\delta{\cal
  A}_{LR}(M_{t\bar t})$ of Eq.~(\ref{eq:mttasymone}) and $\delta{\cal
  A}(M_{t\bar t})$ of Eq.~(\ref{eq:mttasymtwo}), for sets (I) and (II)
and a gluino mass of $230$~GeV and $300$~GeV. The asymmetry
$\delta{\cal A}(M_{t\bar t})$ could be an interesting observable at
the Tevatron, since the cross sections for the spin-like polarization
states ($t_L \bar t_L,t_R \bar t_R$) are much smaller than the ones
for the spin-unlike polarization states ($t_L \bar t_R,t_R \bar t_L$)
and, thus, do not significantly decrease this asymmetry compared to
$\delta{\cal A}_{LR}(M_{t\bar t})$.  Due to characteristic peaks at
the gluino-pair threshold, $\delta{\cal A}(M_{t\bar t})$ can reach
$-3.4\%(-4.1\%)$ for set (I) and $m_{\tilde g}=230(300)$ GeV, and
$-2.1\%(-3.0\%)$ for the heavier top-squark mass set~(II).  If the
polarization of the antitop quark is also measured, the resulting
left-right asymmetry $\delta{\cal A}_{LR}(M_{t\bar t})$ of
Eq.~(\ref{eq:mttasymone}) could reach $-4.7\%$ (set (I) and
$m_{\tilde{g}} = 230$~GeV) and $-4.6 \%$ (set (I) and $m_{\tilde{g}} =
300$~GeV).

In Fig.~\ref{fig:Had_LR_Tev}(b), we show the parity-violating
asymmetries in the $M_{t\bar t}$ distribution of
Eqs.~(\ref{eq:mttasymone}),~(\ref{eq:mttasymtwo}) at the LHC.  Since at
the LHC the cross section of $t_R \bar t_R,t_L \bar t_L$ production is
much larger (roughly three times) than the one for $t_L \bar t_R,t_R
\bar t_L$ production, the asymmetries $\delta{\cal A}(M_{t \bar t}),
{\cal A}$ of Eqs.~(\ref{eq:mttasymtwo}),~(\ref{eq:asymtwo}) are
significantly smaller than $\delta{\cal A}_{LR}(M_{t \bar t}), {\cal
  A}_{LR}$ of Eqs.~(\ref{eq:mttasymone}),~(\ref{eq:asymone}).  At the
LHC, the left-right asymmetry $\delta{\cal A}_{LR}(M_{t\bar t})$
reaches a maximum for set (I) of $-2.0\%$ for $m_{\tilde{g}}=230$~GeV
and $-1.1\%$ for $m_{\tilde{g}} = 400$~GeV. In comparison, if the
polarization of the antitop quark is not measured, we find for the
same set of MSSM parameters that $\delta{\cal A}(M_{t\bar t})$ is
maximally $-0.64\%$ and $-0.58\%$.  The asymmetries in the total
hadronic cross section, ${\cal A}_{LR}$ of Eq.~(\ref{eq:asymone}) and
${\cal A}$ of Eq.~(\ref{eq:asymtwo}), reach for set~(I) a maximum
value of $-0.24\%$ and $-0.08\%$ for $m_{\tilde{g}} = 230$~GeV and
$-0.37\%$ and $-0.13\%$ for $m_{\tilde{g}} = 280$~GeV.  For set~(II)
the asymmetries are maximal at roughly $m_{\tilde{g}} = 300$~GeV with
${\cal A}_{LR}=-0.26\%$ and ${\cal A}=-0.09\%$. For both parameter
sets, ${\cal A}_{LR}$ and ${\cal A}$ decrease slowly for heavier
gluinos. At the LHC, integrated parity-violating asymmetries as small
as $0.1\%$ may be accessible: For instance, we find statistical
significances of $N_S=9.0(5.3)$ for $|{\cal A}_{LR}|=0.37\%(|{\cal
  A}|=0.13\%)$ (with ${\cal L}=30 \, \rm{fb}^{-1}$).

%
%

\begin{figure}[h]
\begin{center}
\hspace*{-.4cm}  \includegraphics[width=7.8cm,
  keepaspectratio]{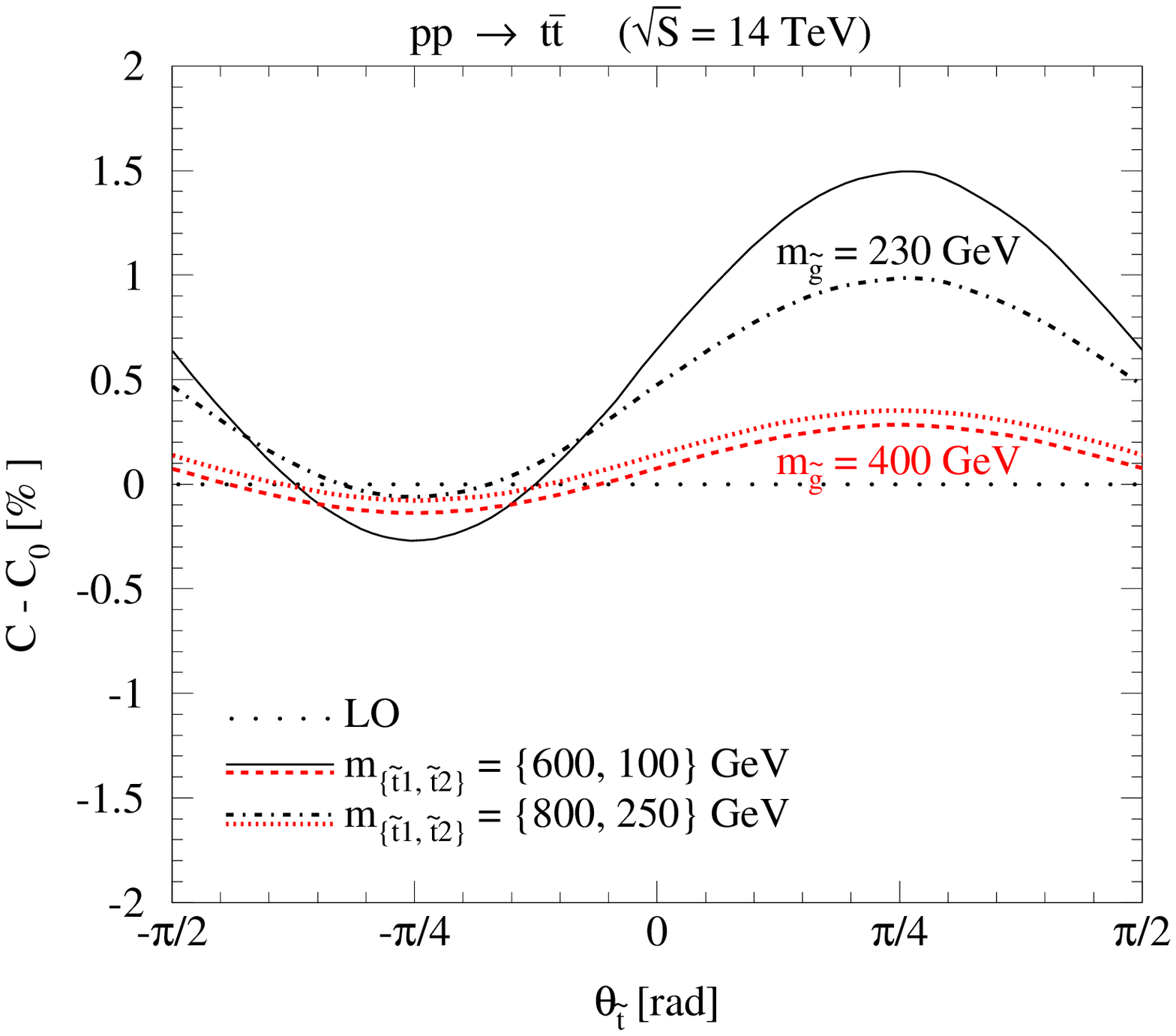}
\hspace*{-.4cm}
  \includegraphics[width=8.3cm,
  keepaspectratio]{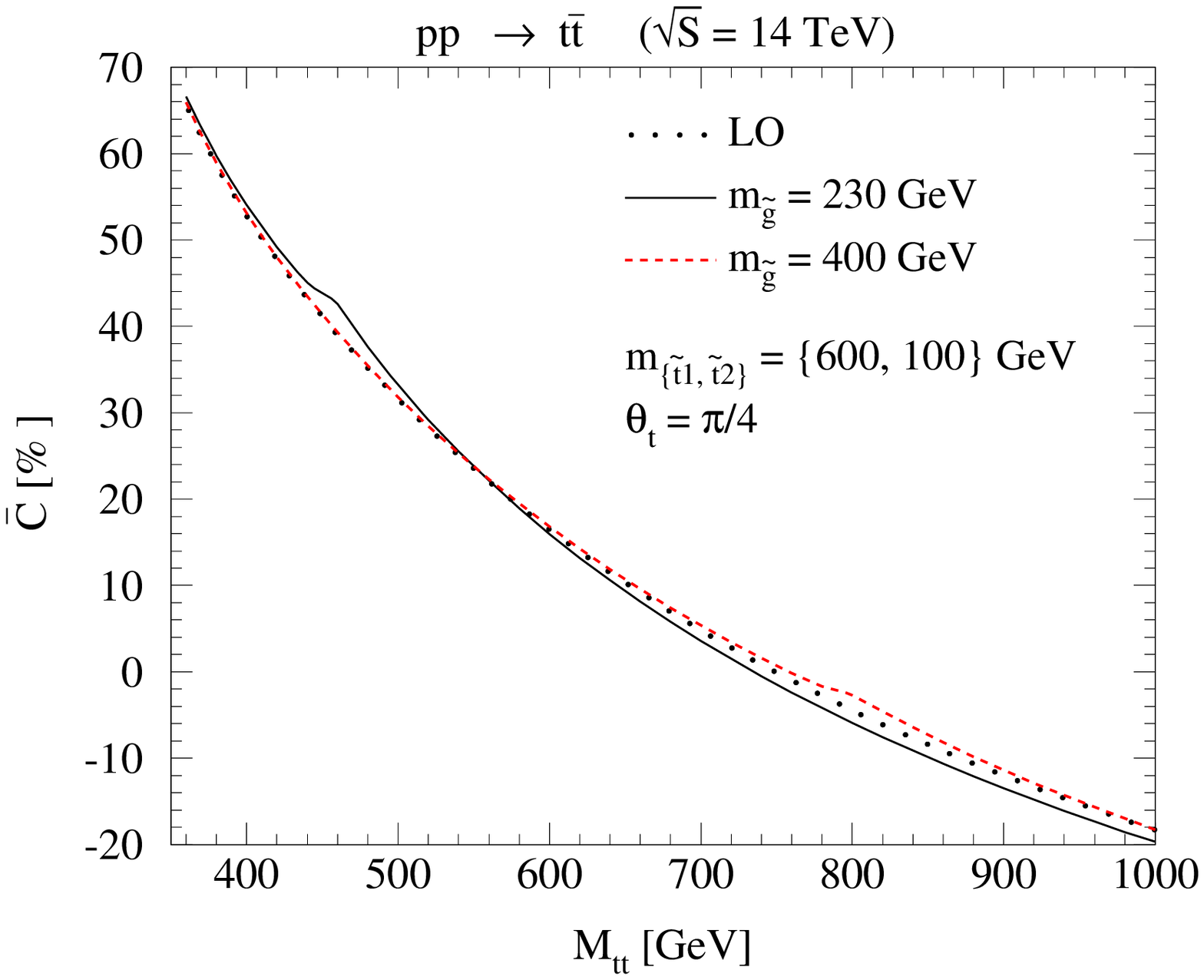}
\hspace*{-.8cm}\\[-6pt]
\hspace*{.6cm}(a)\hspace*{7.8cm}(b)\\
\vspace*{-24pt}
\vspace*{5pt}
\end{center}
\caption{\emph{Spin correlation functions in polarized $t\bar t$ production at the LHC 
at LO QCD and NLO SQCD. Shown is (a) the difference between the LO QCD and NLO SQCD 
spin correlation functions, $C_0$ and $C$, in dependence of the top-squark mixing angle $\theta_{\tilde{t}}$
and (b) $\overline{C}$ in dependence of the invariant $t\bar t$ mass. 
}}\label{fig:LHC_corr}
\end{figure}

\begin{figure}[h]
\begin{center}
  \includegraphics[width=8.1cm,
  keepaspectratio]{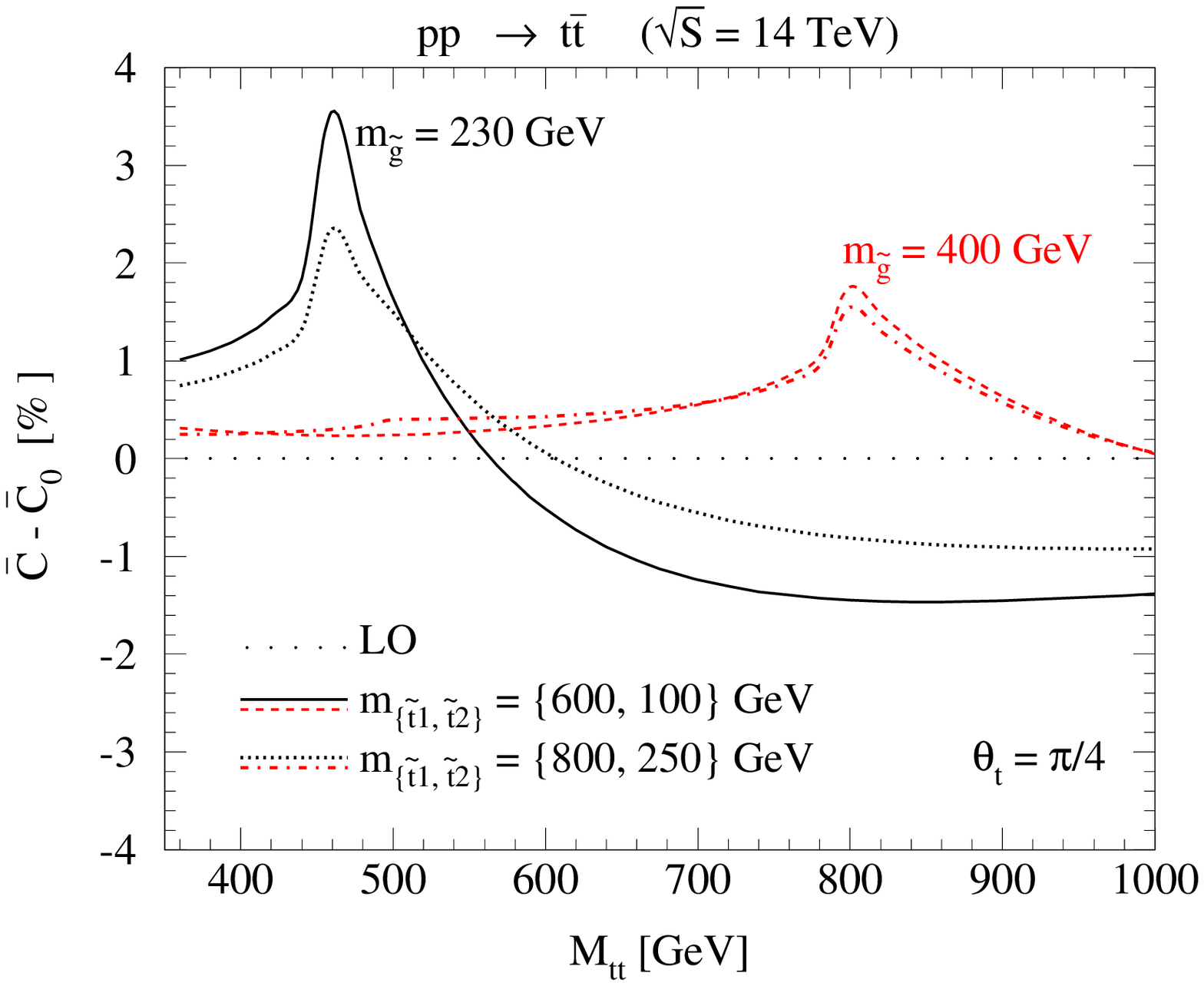}
\hspace*{-.0cm}
  \includegraphics[width=8.1cm,
  keepaspectratio]{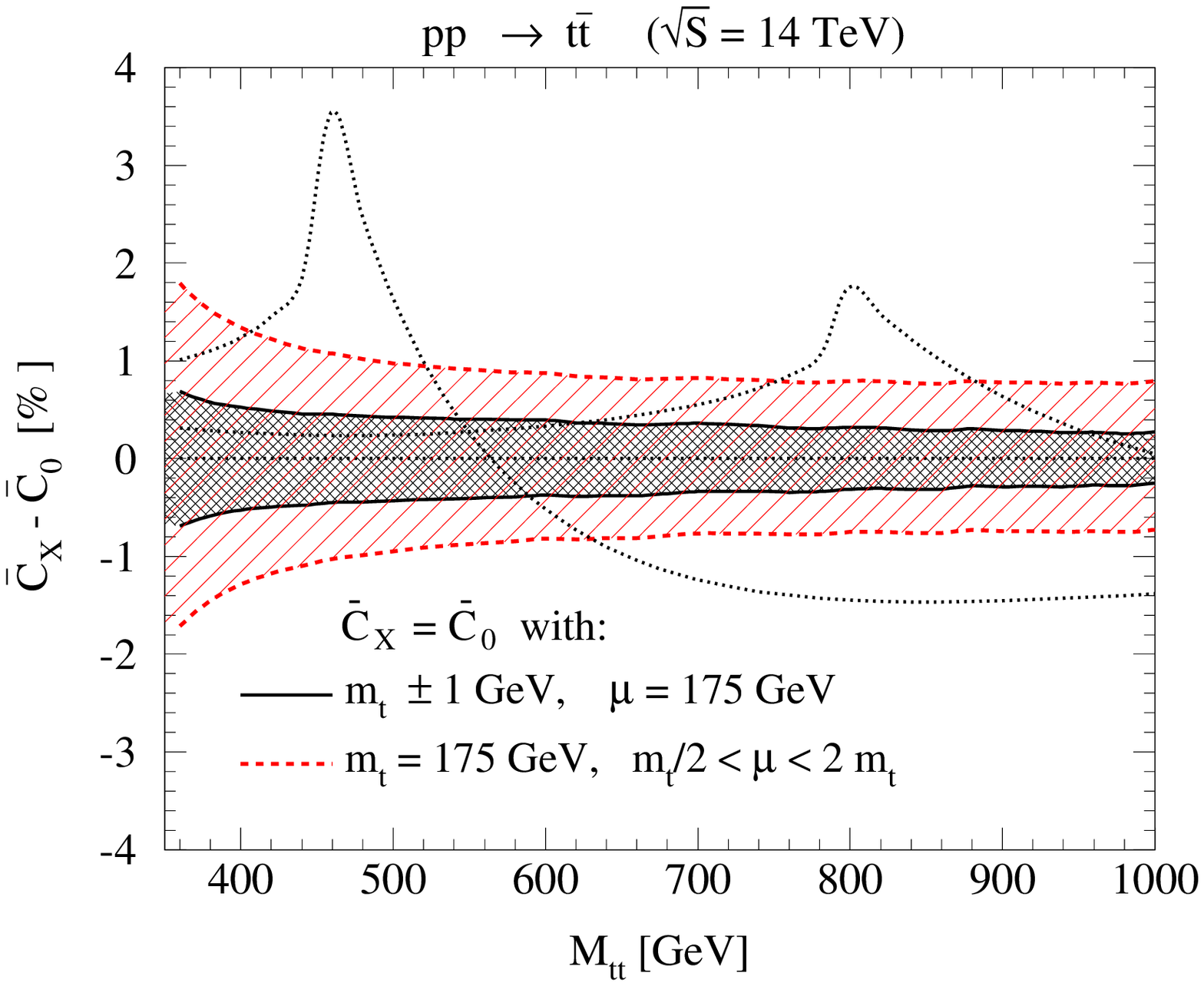}
\hspace*{-.8cm}\\[-30pt]
\hspace*{.6cm}(a)\hspace*{8cm}(b)\\
\vspace*{-25pt}
\vspace*{5pt}
\end{center}
\caption{
\emph{
Difference between the LO QCD and NLO SQCD spin correlation functions,
$\overline{C}_0$ and $\overline{C}$, at the LHC in dependence of the
invariant $t\bar t$ mass.
In (b) we compare the NLO SQCD prediction for this difference (dotted
lines with $\overline{C}_X=\overline{C}$ and $m_{\tilde{t}_1} =
600~{\rm GeV}, m_{\tilde{t}_2} = 100~{\rm GeV}, \theta_{\tilde{t}} =
\pi/4, m_{\tilde g}=230,400$ GeV) to the LO QCD prediction with
$\overline{C}_X$ describing the LO spin correlation function for
varying top mass and renormalization and factorization scales.  The
bands are obtained due to a variation of $m_t$ by $\pm 1$~GeV (grey,
cross-hatched band with the lower line corresponding to $m_t - 1$~GeV and
the upper line to $m_t +1$~GeV) and due to a variation of $\mu_F$ and
$\mu_R$ by $m_t/2 < \mu_F=\mu_R <2m_t$ (red, hatched band).
$\overline{C}_0$ in (a) and (b) is calculated for $m_t = \mu_F=\mu_R =
175$~GeV.
}}\label{fig:LHC_corrdiff}
\end{figure}

In the remaining part of this section we will provide numerical
results for the spin correlation functions of
Eqs.~(\ref{eq:corrone}),~(\ref{eq:corrtwo}) at the LHC.
Fig.~\ref{fig:LHC_corr}(a) shows the difference between the LO QCD and
NLO SQCD predictions for the spin correlation function~$C$ of
Eq.~(\ref{eq:corrone}) in dependence of the top-squark mixing angle.
A comparison of the LO QCD and NLO SQCD predictions for the spin
correlation function~$\overline{C}$ of Eq.~(\ref{eq:corrtwo}) is shown
in Fig.~\ref{fig:LHC_corr}(b).  The LO QCD predictions are obtained
from Eqs.~(\ref{eq:corrone}),~(\ref{eq:corrtwo}) by replacing the NLO
cross section with the LO cross sections.  As shown in
Fig.~\ref{fig:Pol_ggtt_partonic}(b), the SQCD one-loop corrections to
the $t_R \bar t_R,t_L \bar t_L$ and $t_L \bar t_R,t_R \bar t_L$
production can differ considerably, resulting in large deviations from
the LO QCD prediction for $C$, and thus can lead to potentially
observable effects at the LHC. Since this difference strongly depends
on the stop mixing angle, this spin correlation function may have the
potential for extracting information about $\theta_{\tilde{t}}$.
In Fig.~\ref{fig:LHC_corrdiff}, we discuss the difference between the
LO QCD ($\overline{C}_0$) and NLO SQCD~($\overline{C}$) predictions
for the spin correlation function of Eq.~(\ref{eq:corrtwo}) in more detail.
As can be seen in Fig.~\ref{fig:LHC_corrdiff}(a),
a deviation from the LO QCD prediction of maximal $3.7\%$ can be
achieved for light stop and gluino masses and a top-squark mixing
angle of $\theta_{\tilde{t}} = \pi/4$.  Since the spin correlation
functions $C,\,\overline{C}$ do not suffer from luminosity uncertainties
(they cancel in the cross section ratios), and due to the high $t\bar
t$ yield at the LHC, they may be interesting observables to search for
loop-induced SUSY effects in $t\bar t$ production. To see how these
effects compare to theoretical uncertainties in the predictions for
$\overline{C}$ we show again in Fig.~\ref{fig:LHC_corrdiff}(b) the
difference $\overline{C}_X-\overline{C}_0$ with
$\overline{C}_X=\overline{C}$ for 
$m_{\tilde{t}_1} = 600~{\rm GeV},\, m_{\tilde{t}_2} = 100~{\rm GeV},
\,\theta_{\tilde{t}} = \pi/4$
and a light and heavier
gluino, together with the uncertainty bands induced by the LO QCD
scale dependence, varied between $m_t/2 < \mu_F=\mu_R <2m_t$, and an
experimental top-mass uncertainty of $\Delta m_t=1$ GeV. The bands are
calculated with $\overline{C}_X$ being the LO spin correlation
function for different values of $m_t$ and $\mu_R,\mu_F$, and
$\overline{C}_0$ is calculated for $m_t = \mu_F=\mu_R = 175$~GeV.
Since the scale uncertainty is obtained at LO QCD, we expect the
corresponding band to be considerably less pronounced when including
the known QCD NLO
results~\cite{Nason:1987xz,Beenakker:1988bq,Altarelli:1988qr,Beenakker:1990ma,Mangano:1991jk,Frixione:1995fj}.
In view of an anticipated relative experimental error on
$\overline{C}$ of about $4\%$~\cite{Hubaut:2005er}, this observable
has the potential to be sensitive to loop-induced SUSY effects.

%
%
\subsection{Comparison with existing calculations}
\label{sec:comparison}

The NLO SQCD corrections to $q\bar{q} \to t\bar t$ for unpolarized top
quarks have also been calculated in
Refs.~\cite{Li:1995fj,Alam:1996mh,Sullivan:1996ry}, however in
Ref.~\cite{Li:1995fj} the contribution of the gluon self-energy and
the crossed box diagram (Fig.~\ref{fig:qqttboxes}(b)) is missing.  We
compared our results analytically and numerically with those presented
in Refs.~\cite{Alam:1996mh,Sullivan:1996ry} and found agreement, if we
adjust for a missing $(-1)$ sign for the direct box
(Fig.~\ref{fig:qqttboxes}(a)) contribution in Ref.~\cite{Alam:1996mh}
and missing $(-1)$ signs for the direct~(Fig.~\ref{fig:qqttboxes}(a))
and crossed box~(Fig.~\ref{fig:qqttboxes}(b)) contributions of
Ref.~\cite{Sullivan:1996ry}.  A detailed discussion of the box
contributions can be found in Appendix~\ref{sec:qqtt_analytic}. The
NLO SQCD corrections to gluon fusion for unpolarized top quarks have
also been calculated in Ref.~\cite{Zhou:1997fw}. When we use their
choices of the SM and MSSM input parameters and assume $\alpha_s =
0.120$ we were able to reproduce the numerical results presented in
the figures of Ref.~\cite{Zhou:1997fw}.


\section{Conclusion}
\label{sec:conclusion}

We studied in detail the effects of SQCD one-loop corrections to the
main production processes for polarized and unpolarized strong $t\bar
t$ production, $q \bar q\to t \bar t$ and $gg\to t \bar t$, at the
Tevatron Run II and the LHC. We presented numerical results for the
total $t\bar t$ production rate, the invariant $t\bar t$ mass and top
transverse momentum distributions in unpolarized $t \bar t$ production
and for a number of asymmetries in polarized $t \bar t$ production for
different choices of the MSSM input parameters, $m_{\tilde g}, \,
m_{\tilde t_1},\, m_{\tilde t_2}$ and $\theta_{\tilde{t}}$.  We found
that the largest corrections occur for the lightest gluino allowed by
current experimental limits and large stop-mass splittings. For
instance, for $m_{\tilde t_1},\, m_{\tilde t_2}=600,100$~GeV and
$\theta_{\tilde{t}}=\pi/4$, the SQCD one-loop corrections enhance the
LO total hadronic $t\bar t$ cross section by $7.1\%$ (when restricting
the top quark $p_T$ to $75\, {\rm GeV}<p_T<170$~GeV) at the Tevatron
Run II for $m_{\tilde g}=230$ GeV and by $7.5\%$ ($100\, {\rm
  GeV}<p_T<210$~GeV) at the LHC for $m_{\tilde g}=260$ GeV.  The NLO
SQCD $M_{t\bar t}$ and $p_T$ distributions can be significantly
distorted due to a gluino-pair production threshold at $\sqrt{\hat
  s}=2 m_{\tilde g}$. For $M_{t\bar t}$ and $p_T$ values in the
vicinity of this threshold, the SQCD one-loop corrections can enhance
the LO $M_{t\bar t}$ and $p_T$ distributions by $15 \%$ and $9.5\%$,
respectively.  When considering polarized $t\bar t$ production, we
studied the effects on parity-violating and parity-conserving
asymmetries in the total hadronic cross section and the $M_{t\bar t}$
distribution.  We found that in view of the anticipated top-quark
yield at the Tevatron Run~II, it will be difficult to observe
loop-induced SUSY effects in polarization asymmetries. At the LHC,
however, we find promising effects: The parity-violating asymmetries
in the production of left and right-handed top and antitop quarks can
reach up to $|\delta {\cal A}_{LR}(M_{t\bar t})|=2.0\%$ in the
$M_{t\bar t}$ distribution and $|{\cal A}_{LR}|=0.37\%$ in the total
hadronic cross section. The parity-conserving spin correlation function,
$\overline{C}$, that describes an asymmetry in the $M_{t\bar t}$
distributions of spin-like and spin-unlike $t\bar t$ production can
differ from the LO QCD prediction by up to $3.7\%$.  These effects are
promising enough to motivate a future study which includes top-quark
decays and the detector response, in order to determine whether they
are observable at the LHC. Such a study should also include SM and
SUSY electroweak one-loop corrections, since they can either enhance
or diminish the SQCD induced effects.


%
%

\section*{Acknowledgments}

We thank F.~I.~Olness, W.~Bernreuther, S.~Heinemeyer and B.~Kehoe for
discussion.  W.M.~M.~is grateful for the kind hospitality extended to
him by the Particle Physics Theory Group of the Paul-Scherrer
Institut, where part of this work was done.  D.~W.~is
particularly grateful to D.~Dicus for the opportunity to spend a
research semester at the Center for Particles and Fields of the University of
Texas at Austin, where a substantial part of this work was completed, supported in part by
the U.S.~Department of Energy under grant DE-FG03-93ER40757.  This
research is supported in part by the European Community's Marie-Curie
Research Training Network under contract MRTN-CT-2006-035505 {\em Tools
and Precision Calculations for Physics Discoveries at Colliders}
(HEPTOOLS).  The work of D.~W.~is supported in part by the National
Science Foundation under grants NSF-PHY-0244875, NSF-PHY-0456681 and
NSF-PHY-0547564.  The work of S.~B.~is supported in part by the
U.S.~Department of Energy under grant DE-FG03-95ER40908, Contract
W-31-109-ENG-38, by the Lightner-Sams Foundation, and by Deutsche
Forschungsgemeinschaft SFB/TR9.
%
%
\appendix
\section{Feynman rules}\label{sec:feynman_rules}

\begin{figure}[h]
\begin{center}
\setlength{\unitlength}{1cm}
\begin{picture}(15,2.3)
\put(0,0){\includegraphics[width=2.9cm,
  keepaspectratio]{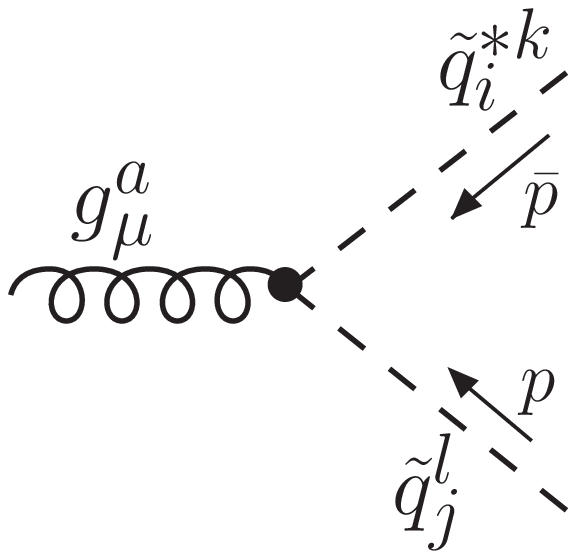}}
\put(5,1.2){\makebox(0,0){: $-i g_s T^a_{kl} (p-\overline p)_{\mu}
\delta_{ij}$}}
\put(8.4,0){\includegraphics[width=2.4cm,
  keepaspectratio]{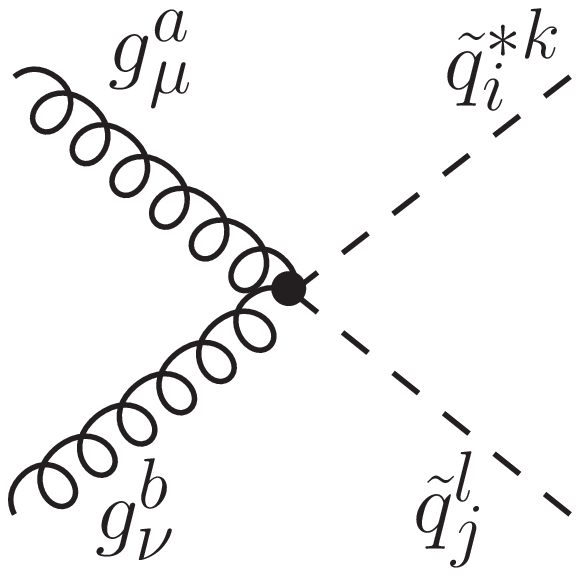}}
\put(13.4,1.2){\makebox(0,0){: 
$i g_s^2 (\frac{\delta_{ab}}{3}\delta_{kl}+d_{abc} T^c_{kl})
g_{\mu\nu}\delta_{ij}$}}
\end{picture}
\end{center}
\mbox{}\\[-45pt]
\caption{The Feynman rules for the triple and quartic squark-gluon
interactions.  $a,b,c=1\ldots 8$ and $k,l=1\ldots 3$ are color indices
and $i,j=L,R$ and $i,j=1,2(L,R)$ with(without) mixing are the squark indices.  
Here $d^{abc}$ is total symmetric, defined by
$d^{abc}  = 2 {\rm Tr}\,[\{T^a,T^b\}T^c]$.
%
}
\label{fig:feynone}
\end{figure}
\begin{figure}[h]
\begin{center}
\setlength{\unitlength}{1cm}
\begin{picture}(15,1.8)
\put(0,0){\includegraphics[width=2.9cm,
  keepaspectratio]{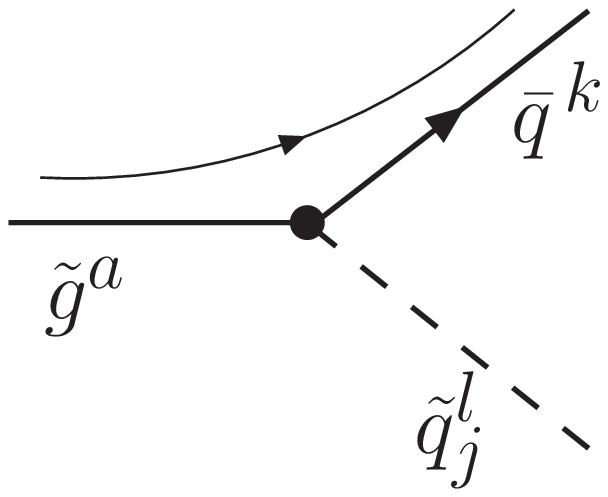}}
\put(5.01,1.15){\makebox(0,0){: $\Gamma_{1,kl}^{a,j} = -i \frac{g_s T^a_{kl}}{\sqrt{2}} (g_s^j+g_p^j \gamma_5)$}}
\put(8.2,0){\includegraphics[width=2.9cm,
  keepaspectratio]{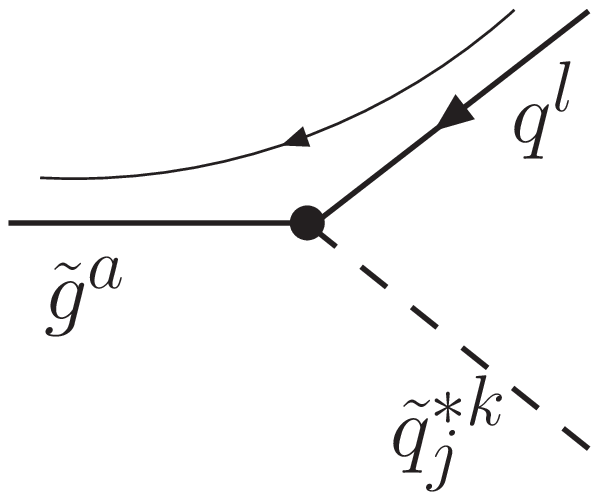}}
\put(12.8,1.15){\makebox(0,0){: $\Gamma_{2,kl}^{a,j} = -i \frac{g_s T^a_{kl}}{\sqrt{2}} (g_{s}^j - g_{p}^j \gamma_5)$}}
\end{picture}
\end{center}
\mbox{}\\[-45pt]
\caption{
The $\tilde g$-$\tilde q_j$-$q$ - vertex written in generalized form in
terms of scalar and pseudo scalar couplings, $g_{s,p}^{L,R}$, with
$g_{s}^{L,R}=\pm 1$ and $g_p^{L,R}=1$. The thin line with arrow represents 
the fermion flow of Ref.~\cite{Denner:1992vz}.}
\label{fig:feyntwo}
\end{figure}
\begin{figure}[h]
\begin{center}
\setlength{\unitlength}{1cm}
\begin{picture}(15,1.8)
\put(0,0){\includegraphics[width=2.4cm,
  keepaspectratio]{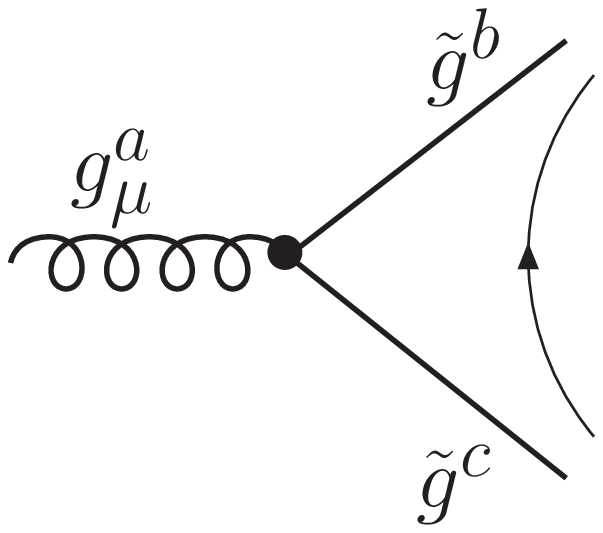}}
\put(4.,1.13){\makebox(0,0){: $-g_s f_{abc} \gamma_\mu$}}
\put(8.2,1.15){\includegraphics[width=2.5cm,
  keepaspectratio]{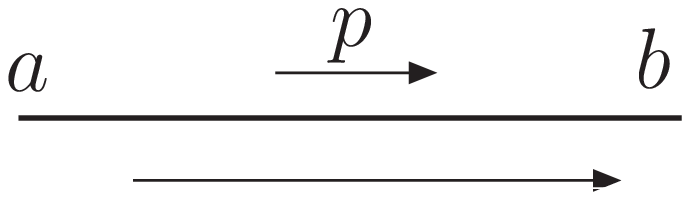}}
\put(12.4,1.55){\makebox(0,0){: 
{\large $ \frac{i \delta^{ab} }{{\footnotesize \pslash-m+i\varepsilon}}$}
}}
\put(8.2,0.3){\includegraphics[width=2.5cm,
  keepaspectratio]{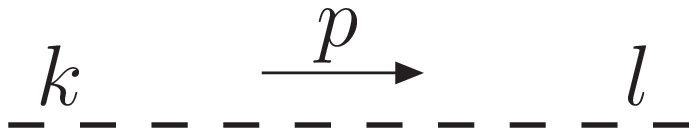}}
\put(12.53,.4){\makebox(0,0){: 
{\large $ \frac{i \delta_{kl} }{{\footnotesize p^2-m^2+i\varepsilon}}$}
}}
\end{picture}
\end{center}
\mbox{}\\[-45pt]
\caption{The $g$-$\tilde g$-$\tilde g$ - vertex with the SU(3) structure
constants $f_{abc}$, and the gluino and squark propagators.}
\label{fig:feynthree}
\end{figure}
\begin{figure}[h]
\begin{center}
\setlength{\unitlength}{1cm}
\begin{picture}(15,4.5)
\put(0.3,2.5){\includegraphics[width=2.4cm,
  keepaspectratio]{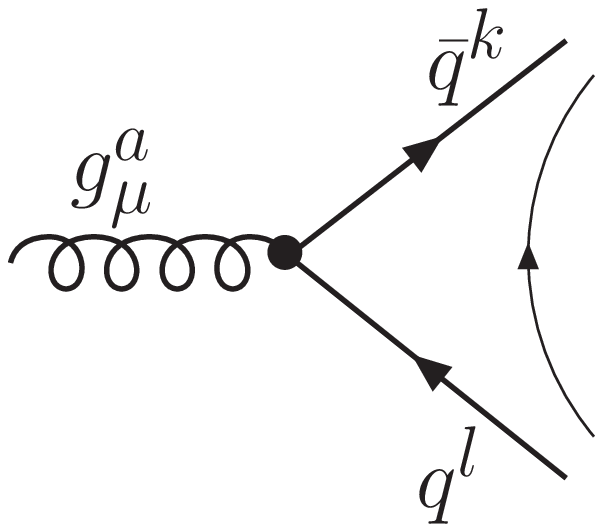}}
\put(4.3,3.6){\makebox(0,0){: $-ig_s T^a_{kl} \gamma_\mu$}}
\put(8.2,3.8){\includegraphics[width=2.7cm,
  keepaspectratio]{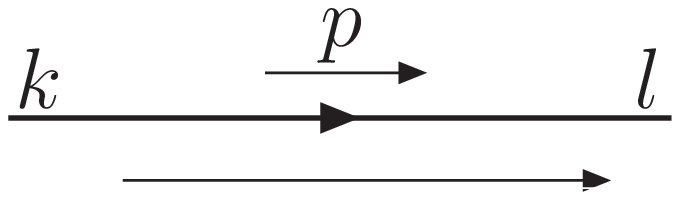}}
\put(12.4,4.1){\makebox(0,0){: 
{\large $ \frac{i \delta_{kl}}{{\footnotesize \pslash-m+i\varepsilon}}$}}}
\put(8.1,2.5){\includegraphics[width=2.7cm,
  keepaspectratio]{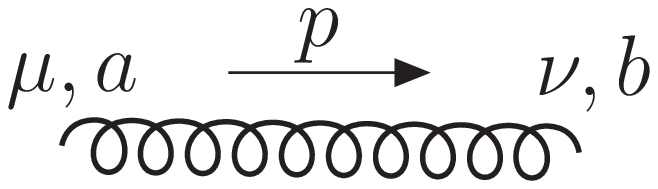}}
\put(12.4,2.7){\makebox(0,0){: 
{\large $ \frac{-i g_{\mu\nu}\delta^{ab}}{{\footnotesize p^2+i\varepsilon}}$}}}
\put(0,-.2){\includegraphics[width=3.2cm,
  keepaspectratio]{eps_feyndiags/3Gl-Vertex.epsi}}
\put(3.3,1.){\makebox(0,0){ :}}
\put(4.58,1.7){\makebox(0,0){\footnotesize $ -g_s f_{a_1a_2a_3} \times$}}
\put(5.2,1.2){\makebox(0,0){\footnotesize $   \{(k_1-k_2)_{\mu_3} g_{\mu_1 \mu_2} $}}
\put(5.31,0.7){\makebox(0,0){\footnotesize $   +(k_2-k_3)_{\mu_1} g_{\mu_2 \mu_3}  $}}
\put(5.38,0.2){\makebox(0,0){\footnotesize $   +(k_3-k_1)_{\mu_2} g_{\mu_3 \mu_1}\!\} $}}
\put(8.3,-0.2){\includegraphics[width=2.6cm,
  keepaspectratio]{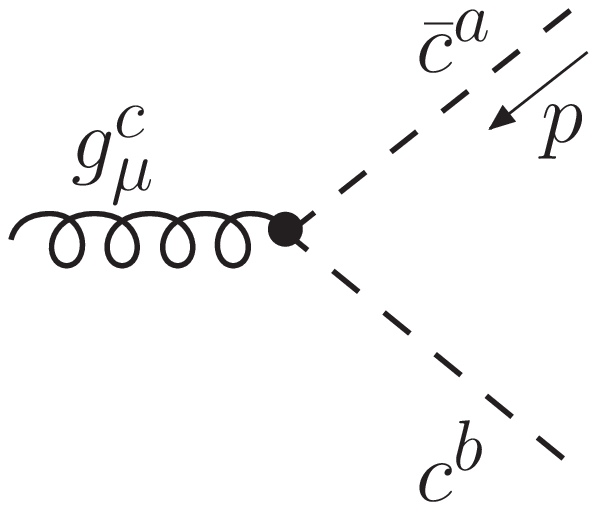}}
\put(12.45,0.9){\makebox(0,0){: 
{ $ { g_s  p_{\mu} f^{abc} }$}}}
\end{picture}
\end{center}
\mbox{}\\[-35pt]
\caption{SM QCD Feynman rules. The long thin lines with arrows denote the 
fermion flow of Ref.~\cite{Denner:1992vz}.}
\label{fig:sm_gqq}
\end{figure}

The relevant QCD Lagrangian of the MSSM is given in
Ref.~\cite{Haber:1984rc} with the convention of the covariant
derivative $D_{\mu} = \partial_{\mu} + i g_s G_{\mu}^{a} T^a$ with the
gluon field $G^a_{\mu}$, strong coupling parameter
$g_s=\sqrt{4\pi\alpha_s}$, and $T^a=\lambda^a/2$, where $\lambda^a$
denote the Gell-Mann-matrices.  Furthermore the methods of
Ref.~\cite{Denner:1992vz} are applied to address Feynman graphs
containing Majorana particles.  In~\cite{Denner:1992vz} a ``fermion
flow'', denoted by a thin line with arrow, is introduced in addition
to the standard ``fermion number flow'' to deal with diagrams
containing fermion number violating fermion chains.  The relevant SQCD
Feynman rules used in this paper are given in
Figs.~\ref{fig:feynone}-\ref{fig:feynthree}.
%
%
%
%
%
\\[-6pt]
When considering
$\tilde q_L$-$\tilde q_R$-mixing the interaction eigenstates are replaced by the
mass eigenstates 
in the interaction Lagrangian, as discussed in Section~\ref{sec:mssminput}, which has the following impact on the
squark couplings of Fig.~\ref{fig:feyntwo}:
\begin{eqnarray}
g_s^{1,2} & = & \cos{\theta_{\tilde q}} \; g_s^{L,R}\pm \sin{\theta_{\tilde q}} \; g_s^{R,L}
\nonumber\\
g_p^{1,2} & = & \cos{\theta_{\tilde q}} \; g_p^{L,R}\pm\sin{\theta_{\tilde q}} \; g_p^{R,L} \;.
\end{eqnarray}
For convenience we introduce the abbreviations
\begin{equation}
\label{eq:lambda} 
\lambda_j^{\pm}=(g_s^j)^2\pm (g_p^j)^2 \; \; \mbox{and} \; \; \lambda_j^A=2 g_s^j g_p^j \;.
\end{equation}
The SM QCD Feynman rules which we need to calculate the $t \bar t$ cross
sections are given in Fig.~\ref{fig:sm_gqq}.

To obtain the Feynman rules with reversed ``fermion flow'' for
diagrams containing Dirac fermions (Fig.~\ref{fig:feyntwo},
$g\bar{q}q$-vertex and quark propagator in Fig.~\ref{fig:sm_gqq}) one
has to replace in the above rules the strings of Dirac matrices
$\Gamma_i\ = \ 1,\, i\gamma_5,\, \gamma_\mu\gamma_5,\, \gamma_\mu,\,
\sigma_{\mu\nu}$ by
\begin{eqnarray}\label{Gamma-Eig}
\Gamma'_i\  = \ \eta_i\Gamma_i\ = \ C\Gamma_i^TC^{-1}\quad \mbox{with}\quad \eta_i\ = \ \left\{
  \begin{array}{rl} 1 & \quad \mbox{for}\quad \Gamma_i\ =\ 1,\, i\gamma_5,\,
    \gamma_\mu\gamma_5   \\ -1 & \quad \mbox{for}\quad \Gamma_i\ = \ \gamma_\mu,\,
    \sigma_{\mu\nu}   \end{array} \right. ~,
\end{eqnarray}
where $C$ denotes the charge conjugation operator. Further rules for
dealing with external spinors can be found in
Ref.~\cite{Denner:1992vz}.


\color{black}
%
%
%
%
\section{Analytical NLO SQCD corrections to $q\overline{q} \to t\overline{t}$}\label{sec:qqtt_analytic}

\begin{figure}[h]
\begin{center}
\includegraphics[width=3.1cm,
  keepaspectratio]{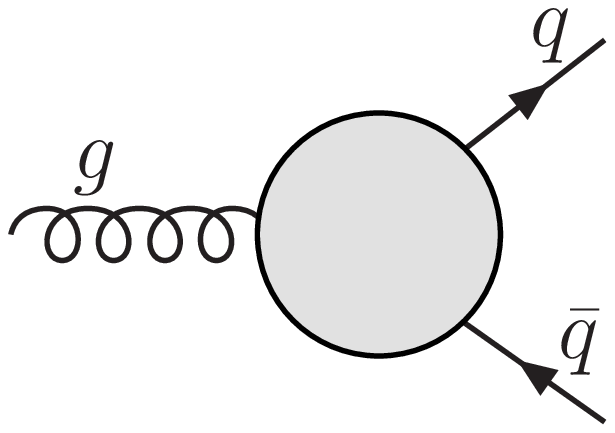}
\hspace*{-.2cm} $\begin{array}{c} \mbox{\bf  =} \\[50pt]\end{array}$
\hspace*{.02cm}
\includegraphics[width=3.1cm,
  keepaspectratio]{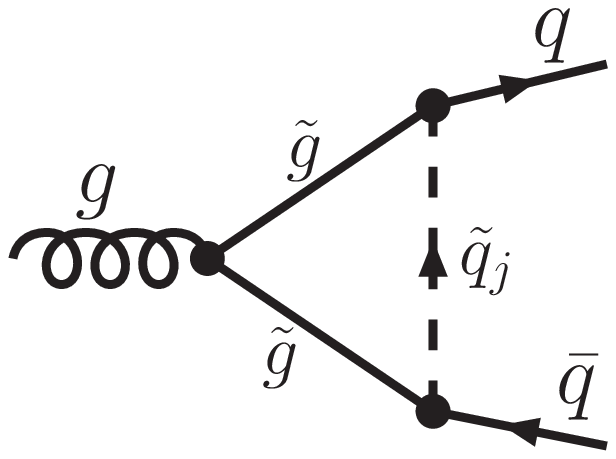}
\hspace*{-.3cm} $\begin{array}{c} \mbox{\bf  +} \\[50pt]\end{array}$
\hspace*{-.2cm}
\includegraphics[width=3.1cm,
  keepaspectratio]{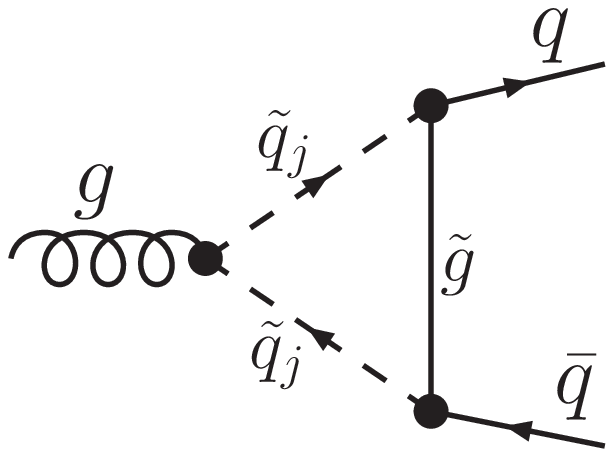}
\hspace*{-.3cm} $\begin{array}{c} \mbox{\bf  +} \\[50pt]\end{array}$
\hspace*{-.1cm}
\includegraphics[width=2.8cm,
  keepaspectratio]{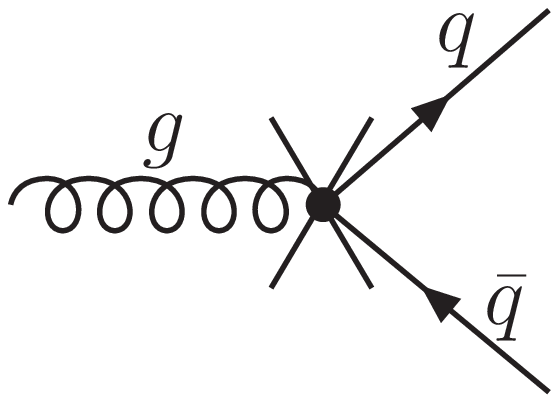}
\\[-20pt]
\end{center}
\mbox{}\\[-53pt]
\caption{\emph{Renormalized gluon-quark-quark vertex correction at NLO Susy-QCD. 
Graphs containing squarks are summed over the squark mass eigenstates j=L,R(1,2) (no(with) mixing).}}
\label{fig:gtt_renorm}
\end{figure}

With the Feynman rules presented in Appendix A 
and the counterterms of the renormalization
procedure as described in Section~\ref{sec:renormalization} 
the SQCD one-loop corrections to
the $\qqa$ subprocess can be given in the compact form of
Eq.~(\ref{eq:qqannihilation}).  The explicit expressions of the UV
finite (after renormalization) gluon self-energy contribution, $\hat
\Pi(\sd)$, is given in Eq.~(\ref{eq:subgluon}) of Section~\ref{sec:renormalization}.  
The renormalized $gq\bar q$ vertex
is paramatrized in terms of UV finite (after renormalization)
formfactors, $F_V,F_M,G_A$ (introduced in
Ref.~\cite{Beenakker:1993yr}), describing the vector, magnetic and
axial vector parts, respectively.  The SQCD one-loop vertex
corrections of Fig.~\ref{fig:gtt_renorm} modify these formfactors as
follows:
\begin{eqnarray}
\frac{\alpha_s}{4\pi} F_V(\sd,m_q)&=& \frac{\alpha_s}{4\pi}
\sum_{j=1,2}\left\{ 
- \frac{1}{6} \lambda^+_j C_{00}(\sd,m_{\tilde g},m_{\tilde q_j},m_{\tilde q_j}) \right.
\nonumber \\ 
&+& \left.
\frac{3}{4} \left[ \lambda_j^+ (B_0(\sd,m_{\tilde g},m_{\tilde g})-(2 C_{00}+(m_q^2+m_{\tilde g}^2-m_{\tilde q_j}^2)
C_0)(\sd,m_{\tilde q_j},m_{\tilde g},m_{\tilde g})) 
\right. \right. \nonumber\\
&-& \left. \left. \lambda_j^- 2 m_{\tilde g}\, m_q C_0(\sd,m_{\tilde q_j},m_{\tilde g},m_{\tilde g})\right] \right\}
+ \delta Z_V 
\\ 
\frac{\alpha_s}{4\pi} F_M(\sd,m_q)&=& \frac{\alpha_s}{4\pi}
\sum_{j=1,2} \left\{  \lambda^+_j  \frac{m_q^2}{6} 
(C_1+C_2+C_{11}+C_{22}+2\,C_{12})(\sd,m_{\tilde g},m_{\tilde q_j},m_{\tilde q_j})
 \right. \nonumber\\
&-& \left.
\lambda_j^- \frac{1}{6}\, m_{\tilde g}\,m_q (C_0+C_1+C_2)(\sd,m_{\tilde g},m_{\tilde q_j},m_{\tilde q_j}) \right. 
\\
&+& \left. \frac{3}{2} m_q^2 \!\left[ \lambda_j^+ 
(C_1+C_2+C_{11}+C_{22}+\!2C_{12})
+\lambda_j^- 
\frac{m_{\tilde g}}{m_q}
(C_1+C_2)
\right]\!\!(\sd,m_{\tilde q_j},m_{\tilde g},m_{\tilde g}) \!\right\}\nonumber
\\
\frac{\alpha_s}{4\pi} G_A(\sd,m_q)&=& \frac{\alpha_s}{4\pi}
\sum_{j=1,2} \left\{  \frac{1}{6} \lambda_j^A C_{00}(\sd,m_{\tilde g},m_{\tilde q_j},m_{\tilde q_j}) 
+  \frac{3}{4} \lambda_j^A \left[
-B_0(\sd,m_{\tilde g},m_{\tilde g}) \right. \right.
\nonumber\\
&+& \left. \left.\!(2 C_{00}-\!(m_q^2-m_{\tilde g}^2+m_{\tilde q_j}^2)
C_0-2 m_q^2 (C_1+C_2))(\sd,m_{\tilde q_j},m_{\tilde g},m_{\tilde g}) \right]\!
\right\} +\delta Z_A 
\end{eqnarray}
with $\delta Z_{V,A}$ of Eq.~(\ref{eq:renorm}) and  $\lambda^{\pm}_j$, $\lambda_j^A$ defined in Eq.~(\ref{eq:lambda}). 
The two point functions, $B_{0}(\sd,m_1,m_2)=B_{0}(\sd,m^2_1,m^2_2)$, and the 
three point functions, $[C_{l},C_{lm}](\sd,m_1,m_2,m_3)=[C_{l},C_{lm}](m_q^2,\sd,m_q^2,m^2_1,m^2_2,m^2_3)$, follow the notation of
Ref.~\cite{Hahn:1998yk}.
%
%

\begin{figure}[h]
\begin{center}
  \includegraphics[width=4.4cm,
  keepaspectratio]{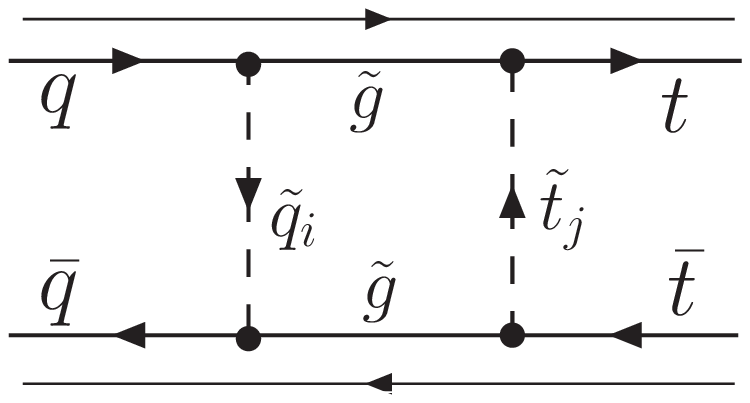}
\hspace*{1.5cm}
  \includegraphics[width=4.4cm,
  keepaspectratio]{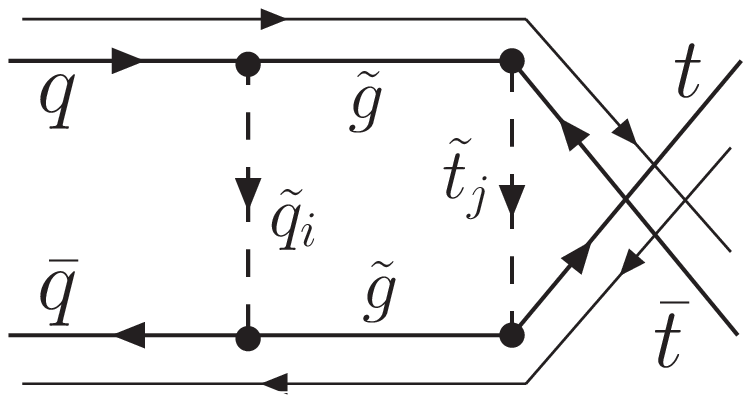}\\
\vspace*{-2pt}
\hspace*{-.2cm}(a)\hspace*{5.5cm}(b)\\
\end{center}
\vspace*{-20pt}
\caption{\emph{(a) Direct box diagram  
and (b) crossed box diagram contributing to
$q\overline{q}\to t\overline{t}$ at NLO SQCD, where we explicitly
indicate our choice for the 
fermion flow (see Ref.~\cite{Denner:1992vz}).}}\label{fig:qqttboxes_flow}
\end{figure}

Finally, we provide the explicit expressions for the direct ($B_t$)
and crossed box diagrams ($B_u$) in $\qqa$.  To determine the relative
sign between the direct and crossed box diagrams (as well as between
box and Born contributions) we apply the rules of
Ref.~\cite{Denner:1992vz}.  Fixing the reference order as $t \bar t
\bar q q$ and choosing the fermion flow as indicated in
Fig.~\ref{fig:qqttboxes_flow} we only need to assign an additional
minus sign to the direct box contribution.  Because of differences in
previous calculations of the direct and crossed box diagrams in
$q\bar{q}$ annihilation, we also provide the explicit expressions for
the corresponding matrix elements, $\delta{\cal M}_{box}^t$ and $\delta{\cal M}_{box}^u$, and
of the Born matrix element ${\cal M}_B^{q\bar q}$:
\begin{eqnarray}
\delta{\cal M}^t_{box} \!\! &=& \label{M_t-channel}
\!-\!\! \int\! \!\frac{d^4k}{\!X(p_2)\!}\,
\overline{v}_k(p_3) \,\Gamma_{1,kl}^{a,i}\; 
i(-\kslash\!-\!\pslash_3\!+\!m_{\tilde{g}})
\,\Gamma^{a,j}_{2,on}\, v_n(p_1) 
\overline{u}_r(p_2) \,\Gamma_{1,ro}^{b,j}\; i(\pslash_4\!-\!\kslash\!+\!m_{\tilde{g}})
\,\Gamma_{2,lm}^{b,i} \,u_m(p_4) \nonumber\\
\delta{\cal M}_{box}^u \!\! &=& \label{M_u-channel}
\!\!\!\int\! \!\frac{d^4k}{\!X\!(p_1)\!}\,
\overline{v}_k(p_3) \,\Gamma_{1,kl}^{a,i}\, \, i(-\kslash\!-\!\pslash_3
\!+\!m_{\tilde{g}})
\!\left(\Gamma_{1,ro}^{a,j}\right)'\! v_r(p_2)
\overline{u}_n(p_1)\Big(\!\Gamma_{2,on}^{b,j}\!\Big)' i
(\pslash_4\!-\!\kslash\!+\!m_{\tilde{g}})
\Gamma_{2,lm}^{b,i} \,u_m(p_4)\nonumber\\
{\cal M}_B^{q\bar q} \!\! &=& \label{M_born}
\overline{u}_r(p_2)(-ig_sT^{a}_{rn}\gamma^{\mu})
v_n(p_1)\Bigl(-i\frac{g_{\mu\nu}}{\hat{s}}\delta^{ab}\Bigr)
\overline{v}_k(p_3)
(-ig_sT^{b}_{km}\gamma^{\nu})u_m(p_4) \!\!\!\!\!
\end{eqnarray}
with $X(p_i) = (2\pi)^4 [k^2-m^2_{\tilde{q}_i}][(p_4-k)^2-m_{\tilde{g}}^2][(p_3+k)^2-m_{\tilde{g}}^2]
[(p_i+k-p_4)^2-m^2_{\tilde{t}_j}]$, $\hat{s} = (p_3+p_4)^2$
and $i,j=1,2(L,R)$ denoting squark indices, and $a,b,k\!$~-~$\!\!r$ color indices.
The couplings $\Gamma_{(1,2),kl}^{a,j}$ are specified in Fig.~\ref{fig:feyntwo}
and the momenta of the external particles
are defined in Fig.~\ref{fig:qqbarlo}.
The matrix element for the direct box diagram, $\delta{\cal M}_{box}^t$, of Eq.~(\ref{M_t-channel})
receives an additional minus sign because
of the permutation parity of the spinors with respect to the 
Born diagram of Fig.~\ref{fig:qqbarlo} that has been chosen as the reference diagram. 
\\
\underline{Direct box}\\ The Feynman diagram of the direct box is
shown in Fig.~\ref{fig:qqttboxes_flow}(a).  The corresponding spin-like contribution
to the NLO SQCD matrix element squared of Eq.~(\ref{eq:qqannihilation}) reads
(with $z=\cos\hat\theta$)
\begin{eqnarray}
\label{eq:btlike} 
\lefteqn{B_t(\sd,\td,\lt=1/2,\ltb=1/2)=B_t(\sd,\td,\lt=-1/2,\ltb=-1/2)=
\sum_{i,j} \frac{\mt \sd}{16} (1-z^2)  } 
\nonumber\\
&& \times \left\{
\,A_3^{\tilde{q}_i\tilde{t}_j} \mt 
 \left[ \,2 \mt^2 ( D_0+2 D_2 +  D_{22} + 2  D_{13}  )
-2 \sd \beta_t^2 (D_1 + D_{12} )-\sd (1+\beta_t^2) D_{11}
-4 D_{00}\right]\right.
\nonumber\\
& & \hspace*{.5cm} + \left. A_2^{\tilde{t}_j} m_{\tilde g} 
\left[\, 4 \mt^2(D_0+D_2)-2 \sd \beta_t^2 D_1\right]
+A_1^{\tilde{q}_i\tilde{t}_j} 2 m_{\tilde g}^2 \mt  D_0 
\right\}
\end{eqnarray}
with $[D_l,D_{lm}]=[D_l,D_{lm}](\mt^2,0,0,\mt^2,\hat{t},\hat{s},m^2_{\tilde{t}_j},m^2_{\tilde{g}},m^2_{\tilde{q}_i},m^2_{\tilde{g}})$ defined in Ref.~\cite{Hahn:1998yk} and depending on the squark flavors $i$ and $j$.
The spin-unlike parts of the direct box can be written as follows
\begin{eqnarray} 
\label{eq:btunlike} 
B_t(\sd,\td,\lt=1/2,\ltb=-1/2)&=& (B_t^a + B^b_t )(\sd,\td)
\nonumber\\ 
B_t(\sd,\td,\lt=-1/2,\ltb=1/2)&=& (B_t^a - B^b_t)(\sd,\td)
\end{eqnarray}
with
\begin{eqnarray} 
B_t^a(\sd,\td)&=&\sum_{i,j}  \frac{\sd^2}{16}\left\{
A_3^{\tilde{q}_i\tilde{t}_j} \left[-\mt^2 \beta_t 4 z D_1 
+ \frac{1}{2}\mt^2 (1-2 \beta_t z+z^2) (2 D_2 + D_0 + D_{22})
\right. \right. \nonumber\\ 
&-& \!\left. \left. 
 (1+2 \beta_t z+z^2) (\mt^2 D_{11}+D_{00}  )
-4\mt^2 \beta_t  z D_{12}
+\frac{1}{4} \sd (1+\beta_t^2) (1+2 \beta_t z+z^2) D_{13}
 \right] \right.
\nonumber\\
&+&\! \left. A_2^{\tilde{t}_j} m_{\tilde g}\mt \! \left[ 
-\beta_t 4 z D_1
+(1-2 \beta_t z+z^2) (D_2+D_0)\right]
+ \frac{1}{2} A_1^{\tilde{q}_i\tilde{t}_j} m_{\tilde g}^2  (1-2 \beta_t z+z^2)  D_0 
\right\}
\nonumber \\
B_t^b(\sd,\td)&=&\sum_{i,j} 
\frac{\sd^2}{16}\left\{ 
A_4^{\tilde{q}_i\tilde{t}_j} \left[-2 \mt^2 \beta_t (1+z^2)( D_1+D_{12}) 
-(\beta_t+2 z+\beta_t z^2) (\mt^2 D_{11}+D_{00}) \right.\right.
\nonumber\\
&+&\left. \left. 
\frac{1}{4} \sd (1+\beta_t^2) (\beta_t+2 z+\beta_tz^2) D_{13}
-\frac{1}{2} \mt^2 (\beta_t-2z+\beta_t z^2) (D_0 +2 D_2+D_{22}) \right] \right.
\nonumber\\
&+& \left.  A_2^{\tilde{t}_j} A_2^{\tilde{q}_i} m_{\tilde g}\mt \left[ 
-2 \beta_t (1+z^2) D_1
-(\beta_t-2 z+\beta_t z^2) (D_2+D_0)\right]
\right. \nonumber \\
&+& \left .\frac{1}{2} A_5^{\tilde{q}_i\tilde{t}_j} m_{\tilde g}^2 (\beta_t-2 z+\beta_t z^2)  D_0 
\right\}
\end{eqnarray}
%
%
The coefficients $A_k$ are defined by the mixing matrices
of the squarks occuring inside the loop.
In case of a real mixing matrix these coefficients
can be expressed in terms of the squark mixing angles 
$\theta_{\tilde{q}}$ and $\theta_{\tilde{t}}$\,:
\begin{eqnarray}\label{qqttboxcoeff}
A_1^{\tilde{q}_i\tilde{t}_j} &=& 2\cdot\left\{ \begin{array}{rl}
    \cos^2{\theta_{\tilde{q}}} \sin^2{\theta_{\tilde{t}}}+
    \sin^2{\theta_{\tilde{q}}}\cos^2{\theta_{\tilde{t}}} & \quad \mbox{if}
    \quad i=j \\
    \cos^2{\theta_{\tilde{q}}}\cos^2{\theta_{\tilde{t}}}
    +\sin^2{\theta_{\tilde{q}}} \sin^2{\theta_{\tilde{t}}} & \quad\mbox{if}
    \quad i\not= j \end{array} \right.,  \nonumber \\
A_2^{\tilde{t}_j}\ \, &=& \hspace{.5cm} 
    (-1)^j \sin{2\theta_{\tilde{t}}}\ , \nonumber\\
A_2^{\tilde{q}_i}\ \, &=& \hspace{.5cm} 
    {}- (-1)^i \cos{2\theta_{\tilde{q}}}\ , \nonumber\\
A_3^{\tilde{q}_i\tilde{t}_j} &=& 2\cdot\left\{ \begin{array}{rl}
    \cos^2{\theta_{\tilde{q}}}\cos^2{\theta_{\tilde{t}}}
    +\sin^2{\theta_{\tilde{q}}} \sin^2{\theta_{\tilde{t}}} & \quad \mbox{if}
    \quad i=j \\
\cos^2{\theta_{\tilde{q}}} \sin^2{\theta_{\tilde{t}}}+
    \sin^2{\theta_{\tilde{q}}}\cos^2{\theta_{\tilde{t}}} & \quad \mbox{if}
    \quad i\not=j\end{array} \right. .
\nonumber \\
A_4^{\tilde{q}_i\tilde{t}_j} &=& -\left[
    (-1)^j \cos{2\theta_{\tilde{t}}}+
    (-1)^i \cos{2\theta_{\tilde{q}}} \right]
\nonumber\\
A_5^{\tilde{q}_i\tilde{t}_j} &=& -\left[
    (-1)^j \cos{2\theta_{\tilde{t}}}
    -(-1)^i\cos{2\theta_{\tilde{q}}} \right]
\end{eqnarray}
%
%
\underline{Crossed box}\\
The Feynman diagram of the crossed box is shown in Fig.~\ref{fig:qqttboxes_flow}(b).
The corresponding spin-like contributions, 
$B_u(\sd,\td,\lt=1/2,\ltb=1/2)=B_u(\sd,\td,\lt=-1/2,\ltb=-1/2)$, and 
spin-unlike contributions, $B_u(\sd,\td,\lt=1/2,\ltb=-1/2)$ and
$B_u(\sd,\td,\lt=-1/2,\ltb=1/2)$, 
to the NLO SQCD matrixelement squared of Eq.~(\ref{eq:qqannihilation}) can
be obtained from the expressions for the direct box of 
Eqs.~(\ref{eq:btlike}),~(\ref{eq:btunlike})
by performing the following replacements: 
\begin{eqnarray} 
z &\rightarrow& {}-z \nonumber \\
A_2^{\tilde{q}_i}  &\rightarrow&   {}-A_2^{\tilde{q}_i}\nonumber \\
A_1^{\tilde{q}_i\tilde{t}_j} &\leftrightarrow& A_3^{\tilde{q}_i\tilde{t}_j} \nonumber \\
A_4^{\tilde{q}_i\tilde{t}_j} &\leftrightarrow& A_5^{\tilde{q}_i\tilde{t}_j}  
\end{eqnarray} 
and the four point functions $D_l$ and $D_{lm}$ are evaluated at
$[D_l,D_{lm}]=[D_l,D_{lm}](\mt^2,0,0,\mt^2,\hat{u},\hat{s},m^2_{\tilde{t}_j},m^2_{\tilde{g}},m^2_{\tilde{q}_i},m^2_{\tilde{g}})$. 

%
%
%
%
\section{Analytic expressions for the NLO SQCD corrections to $gg \to t\overline{t}$}
\label{sec:ggtt_analytic}

\begin{figure}[h]
\begin{center}
\hspace*{-.3cm}
\includegraphics[width=3.1cm,
  keepaspectratio]{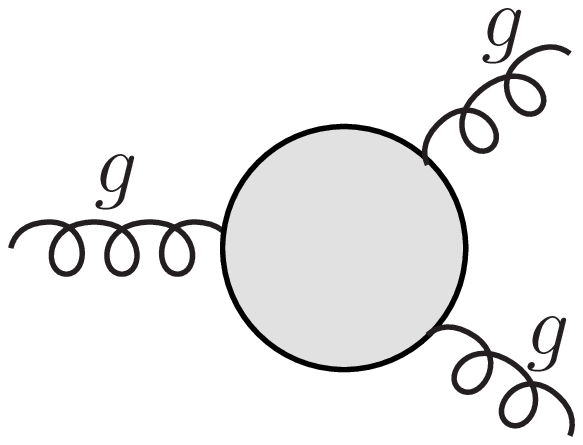}
\hspace*{-.42cm} $\begin{array}{c} \mbox{\bf  =} \\[50pt]\end{array}$
\hspace*{-.22cm}
\includegraphics[width=3.1cm,
  keepaspectratio]{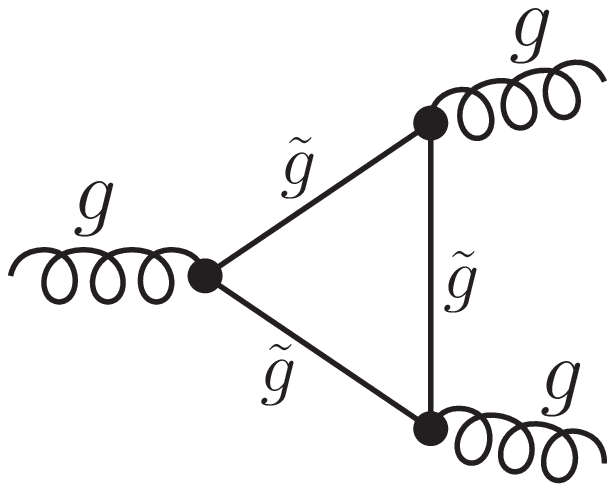}
\hspace*{-.37cm} $\begin{array}{c} \mbox{\bf  +} \\[50pt]\end{array}$
\hspace*{-.27cm}
\includegraphics[width=3.1cm,
  keepaspectratio]{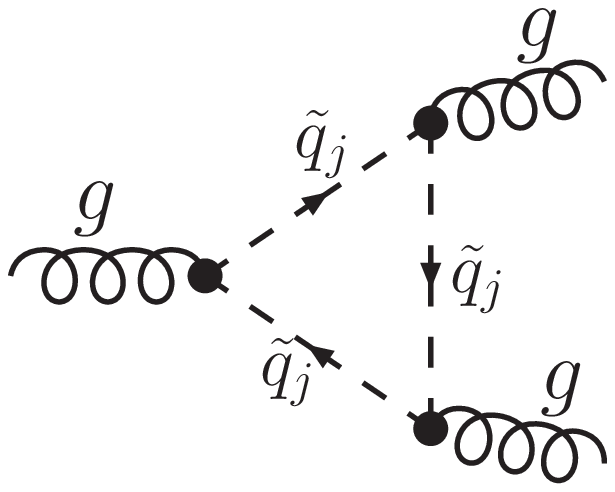}
\hspace*{-.37cm} $\begin{array}{c} \mbox{\bf  +} \\[50pt]\end{array}$
\hspace*{-.27cm}
\includegraphics[width=3.1cm,
  keepaspectratio]{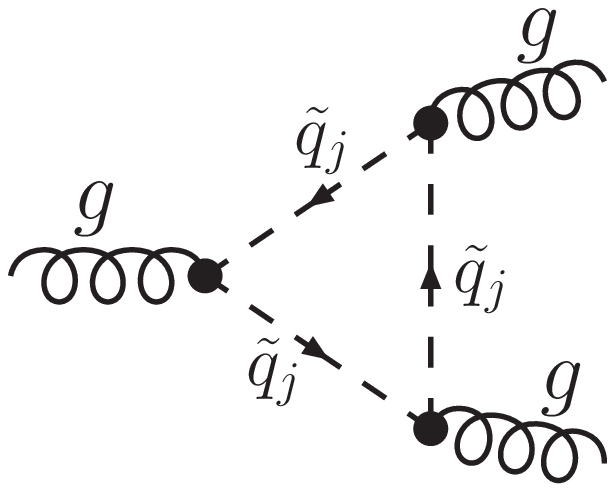}
\hspace*{-.37cm} $\begin{array}{c} \mbox{\bf  +} \\[50pt]\end{array}$
\hspace*{-.27cm}
\includegraphics[width=2.7cm,
  keepaspectratio]{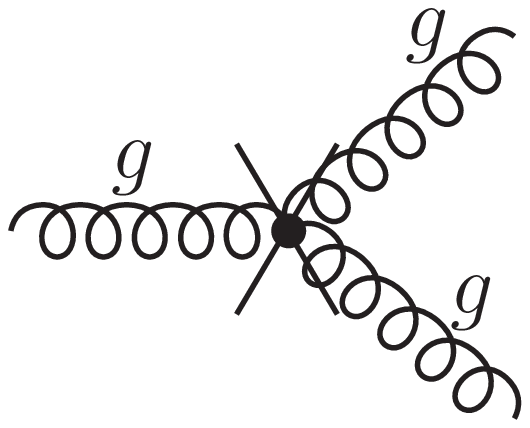}
\hspace*{-.52cm}
\\[-20pt]
\end{center}
\mbox{}\\[-53pt]
\caption{\emph{Renormalized gluon-gluon-gluon vertex correction at NLO SQCD. 
Graphs containing squarks are summed 
over the squark-mass eigenstates j=L,R(1,2) (no(with) mixing) and the 
quark flavors q=\{u,d,s,c,b,t\}.}}
\label{fig:ggg_renorm}
\end{figure}

The Feynman-diagrams of Figs.~\ref{fig:ggtt_gen_vertself}
and~\ref{fig:ggtt_boxes} represent the SQCD ${\cal O}(\alpha_s)$
contributions to the gluon fusion subprocess consisting of the
following contributions:
\begin{itemize}
\item
the modification of the $g\ttbar$-vertex in the $s$- and
$t(u)$-production channels described by the UV finite (after renormalization)
form factors $F_V,F_M,G_A$
and $\rho_i^{V,(t,u)},\sigma_i^{V,(t,u)}$, respectively, 
\item
the modification of the $ggg$ vertex ($\rho_2^{V,s}$) 
and the gluon self-energy
($\hat\Pi$) in the $s$ production channel,
\item
the self-energy insertion to
the off-shell top quark propagator in the $t(u)$-production
channel ($\rho_i^{\Sigma,(t,u)}, \sigma_i^{\Sigma,(t,u)}$), 
\item
and the
box diagrams of the $t$ and $u$ production channels, 
($\rho_{i,(a,b,c)}^{\Box,t},\rho_{12}^{\Box,gg\tilde t\tilde t},
\sigma_{i,(a,b,c)}^{\Box,t}$) and  
($\rho_{i,(a,b)}^{\Box,u},\sigma_{i,(a,b)}^{\Box,u}$), respectively.   
\end{itemize}
The form factors for the $gt\bar t$ vertex, $F_V,F_M,G_A$, and the
subtracted gluon self-energy, $\hat\Pi$, are provided in Appendix B and
Section~\ref{sec:renormalization}, respectively.
The contribution of the renormalized $ggg$ vertex of
Fig.~\ref{fig:ggg_renorm} to the NLO SQCD matrix element of
Eq.~(\ref{eq:ggfusion}) reads as follows:
\begin{eqnarray}
\frac{\alpha_s}{4\pi}\rho_2^{V,s}(\sd) &=& \frac{\alpha_s}{4\pi}\left\{  \sum_{q=u,c,t \atop d,s,b} \sum_j 
\left[\frac{1}{6\sd} (\sd+8 m_{\tilde{q}_j}^2) B_0(\sd,m^2_{\tilde q_j},m^2_{\tilde q_j})
-\frac{4m_{\tilde q_j}^2}{3\sd} B_0(0,m^2_{\tilde q_j},m^2_{\tilde q_j}) \right.
\right.
\nonumber \\
&+& \left. \left. m_{\tilde q_j}^2 C_0(0,\sd,0,m^2_{\tilde q_j},m^2_{\tilde q_j},m^2_{\tilde q_j})+\frac{5}{18} \right] 
-6 \left[\frac{1}{3\sd}(4m_{\tilde g}^2-\sd) B_0(\sd,m^2_{\tilde g},m^2_{\tilde g})
\right. \right. \nonumber\\
&-& \left. \left. \frac{4m_{\tilde g}^2}{3\sd}
B_0(0,m^2_{\tilde g},m^2_{\tilde g})+m_{\tilde g}^2 C_0(0,\sd,0,m^2_{\tilde g},m^2_{\tilde g},m^2_{\tilde g})+\frac{5}{18}\right]\right\}+\delta Z_1
\end{eqnarray}
with $\delta Z_1$ of Eq.~(\ref{eq:deltaz1}) and the $B_0$ and $C_0$ functions 
in the convention of Ref.~\cite{Hahn:1998yk}.
%
%

\subsection{Parity conserving coefficients to the SMEs $M^{V,t}_i$}
\label{sec:parconv}
 
The $t$-channel parity conserving coefficients to the SMEs $M_i^{V,t}$
read as follows:\\ \underline{Vertex corrections:}\\ The vertex
corrections consist of the stop-stop-gluino, $\rho_{i,ttg}^{V,t}$, the
gluino-gluino-stop, $\rho_{i,ggt}$, contributions and the counterterm,
$\rho_{i,CT}$, as shown in Fig.~\ref{fig:gtt_renorm}:
\begin{equation}
\rho_i^{V,t}= \rho_{i,ttg}^{V,t}+\rho_{i,ggt}^{V,t}+\rho_{i,CT}^{V,t}
\end{equation}
with
\begin{eqnarray}
\rho_{1,ttg}^{V,t}& =& -\sum_{j=1,2} \frac{1}{3} \lambda_j^+ C_{00} \nonumber \\
\rho_{4,ttg}^{V,t}& =& \sum_{j=1,2} 
\frac{1}{6} \lambda^+_j  (\mt^2-\td) \left(C_2+C_{22}+C_{12} \right)
\nonumber\\
\rho_{11,ttg}^{V,t}& =& -\rho_{1,ttg}^{V,t}
\nonumber\\
\rho_{14,ttg}^{V,t}& =& \sum_{j=1,2} 
\frac{1}{6}\left\{ \lambda^+_j (C_{11}+C_{22}+2C_{12}+C_1+C_2)
- \lambda^-_j \frac{m_{\tilde g}}{\mt}(C_0+C_1+C_2)\right\}
\nonumber\\
\rho_{16,ttg}^{V,t}& =& -4\rho_{14,ttg}^{V,t} 
\end{eqnarray}
with $[C_l,C_{lm}]=[C_l,C_{lm}](m_t^2,0,\td,m^2_{\tilde g},m^2_{\tilde t_j},m^2_{\tilde t_j})$ 
of Ref.~\cite{Hahn:1998yk}, and
\begin{eqnarray}
\rho_{1,ggt}^{V,t}& =& \sum_{j=1,2} 
\frac{3}{2}\left\{ \lambda^+_j 
\left[ (m_{\tilde t_j}^2-\mt^2-m_{\tilde g}^2) C_0+
(\td-\mt^2) C_2-2 C_{00}+B_0(0,m^2_{\tilde g},m^2_{\tilde g})\right] \right. \nonumber \\
&-& \left. 2\lambda^-_j m_{\tilde g} \mt C_0\right\}
\nonumber \\
\rho_{4,ggt}^{V,t}& =& \sum_{j=1,2} 
\frac{3}{2} \lambda^+_j (\mt^2-\td) \left[C_2+C_{22}+C_{12} \right]
\nonumber\\
\rho_{11,ggt}^{V,t}& =& \sum_{j=1,2} 
\frac{3}{2}\left\{ \lambda^+_j 
\left[ (\td-m_{\tilde t_j}^2+m_{\tilde g}^2) C_0+
(\td-\mt^2) C_1+2 C_{00}-B_0(0,m^2_{\tilde g},m^2_{\tilde g})\right]
\right. \nonumber \\
&+& \left. \lambda^-_j \frac{m_{\tilde g}}{\mt} (\td+\mt^2) C_0\right\}
\nonumber\\
\rho_{14,ggt}^{V,t}& =&  \sum_{j=1,2}
\frac{3}{2}\left\{ \lambda^+_j (C_{11}+C_{22}+2C_{12}+C_1+C_2)
+ \lambda^-_j \frac{m_{\tilde g}}{\mt}(C_1+C_2)\right\}
\nonumber\\
\rho_{16,ggt}^{V,t}& =& -4\rho_{14,ggt}^{V,t}
\end{eqnarray}
with $[C_l,C_{lm}]=[C_l,C_{lm}](m_t^2,0,\td,m^2_{\tilde t_j},m^2_{\tilde g},m^2_{\tilde g})$ 
and the counterterm
\begin{eqnarray}
\frac{\alpha_s}{4\pi}\rho_{1,CT}^{V,t}& =&
-\frac{\alpha_s}{4\pi}\rho_{11,CT}^{V,t} = 2 \delta Z_V
\end{eqnarray}
with $\delta Z_{V}$ of Eq.~(\ref{eq:renorm}). The parameters 
$\lambda_j^{\pm}$ are given in  Eq.~(\ref{eq:lambda}).\\
\underline{Top quark self-energy insertion:}\\
The self-energy insertion in the off-shell top-quark propagator
as shown in Fig.~\ref{fig:ggtt_gen_vertself} 
with the SQCD one-loop corrections shown in Fig.~\ref{fig:qself_renorm} 
is described by:
\begin{eqnarray}
\frac{\alpha_s}{4\pi}\rho_1^{\Sigma,t}& = & -(\hat t+\mt^2) (\Sigma_V(\hat t)
+\delta Z_V)
-  2 \mt^2 (\Sigma_S(\hat t)-\delta Z_V-
\Sigma_S(\mt^2)-\Sigma_V(\mt^2))
\nonumber\\
\frac{\alpha_s}{4\pi}\rho_{11}^{\Sigma,t} & = &  2 \hat t (\Sigma_V(\hat t)
+\delta Z_V)
\nonumber\\
&+ &  (\hat t+\mt^2) (\Sigma_S(\hat t)-\delta Z_V-
\Sigma_S(\mt^2)-\Sigma_V(\mt^2)) \; ,
\end{eqnarray}
where the SQCD one-loop contribution to the vector and scalar parts of
the top quark self-energy, $\Sigma_{V,S}$, and $\delta Z_V$ are
provided in Section~\ref{sec:renormalization}.\\ 
\underline{Box contribution:} \\ 
The box diagrams of Fig.~\ref{fig:ggtt_boxes} can also be parametrized in
terms of coefficients to the SMEs $M_i^{V,t}$ as follows:\\
{\bf Box a:} $\tilde g-\tilde g-\tilde g-\tilde t_j$
\begin{eqnarray}
\rho_{1,a}^{\Box,t} &=& +\sum_{j=1,2} \left\{ \lambda_j^+ \, [C_0 +
(m_{\tilde t_j}^2-\mt^2-m_{\tilde g}^2) \, D_0 +(m_{\tilde g}^2-\mt^2) \, D_2 -6 D_{00} \right.
\nonumber \\
&-  &  \left. \td \, (2 D_{22}+D_{222}) -2\, (\td+\mt^2) \,
(D_{12}+D_{122})-6 D_{002}-2 \mt^2 D_{112} +(\sd-2\mt^2)\,
D_{123}] \right.
\nonumber \\
&+  & \left. \lambda_j^-\, [-2 m_{\tilde g} \mt \, D_0] \right\}
\nonumber \\
\rho_{2,a}^{\Box,t} & = & - \sum_{j=1,2}  2 \lambda_j^+\, [ D_{00}+D_{002}]
\nonumber \\
\rho_{4,a}^{\Box,t} & = & +\sum_{j=1,2} 2\lambda_j^+ \, [(\mt^2-m_{\tilde g}^2)\, (D_1+D_2)
+2 D_{00} + \mt^2 \, ( 2 D_{11} +2 D_{13}+D_{111})
\nonumber \\
& + & (\td +\mt^2)\, D_{22} +(\td +5\mt^2)\,
D_{12}+2 (D_{001}+ 2 D_{002})
\nonumber\\
& + & \td \, D_{222} +(\td+3 \mt^2) \,D_{112} + (3\td+2 \mt^2) \,
D_{122}
 + (3\mt^2-\sd) \, D_{113} +  (\td-\sd+3\mt^2)
\, D_{123} ]
\nonumber \\
\rho_{6,a}^{\Box,t} & = & +\sum_{j=1,2}  2\lambda_j^+ \, [D_{22}+2D_{12}+D_{222}+
2D_{112}+4D_{122}+2D_{123}]
\nonumber \\
\rho_{11,a}^{\Box,t} & = & -\sum_{j=1,2} \left\{2 \lambda_j^+ \, [C_0
+(m_{\tilde t_j}^2-m_{\tilde g}^2) \, D_0 +(\td+2\mt^2-m_{\tilde g}^2) \, D_1
\right. \nonumber \\
& + & \left. (\td+\mt^2)\,(D_{11}+D_{13}+D_{112}+D_{123})
\right. \nonumber\\
& + & \left. \td \, (D_2+2D_{12}+D_{122})-2 D_{00}
+6 D_{001} +\mt^2 \, D_{111}-(\sd-3\mt^2) \, D_{113} ]
\right. \nonumber \\
&-  &  \left. \lambda_j^-\, \frac{m_{\tilde g}}{\mt}\, [(m_{\tilde g}^2+\mt^2-m_{\tilde t_j}^2)\, D_0
-2(\td+\mt^2)\, D_1-2\td \, D_2-C_0] \right\}
\nonumber \\
\rho_{12,a}^{\Box,t} & = & +\sum_{j=1,2} 4 \left\{ \lambda_j^+ \,[D_{00}+2D_{001}+D_{002}]
+ \lambda_j^-\, \frac{m_{\tilde g}}{\mt}\,D_{00} \right\}
\nonumber \\
\rho_{14,a}^{\Box,t}& = & +\sum_{j=1,2} 2 \left\{ \lambda_j^+ \, [2D_1+D_2+2D_{11}+D_{22}+4D_{12} +2D_{13}] \right.
\nonumber \\
& + & \left.  \lambda_j^-\, \frac{m_{\tilde g}}{\mt}\, [2D_1+D_2] \right\}
\nonumber\\
\rho_{16,a}^{\Box,t} & = & -\sum_{j=1,2}  4 \left\{ \lambda_j^+\, 
[4 D_1+2 D_2+3D_{22}+12 D_{12}+2D_{111}+D_{222} \right.
\nonumber \\ 
& + & \left. 6\,(D_{11}+D_{13}+D_{112}+D_{113}+D_{122}+D_{123})] \right.
\nonumber\\
&+  & \left. \lambda_j^- \, \frac{m_{\tilde g}}{\mt}\, [4D_1+2D_2+2D_{11}
+D_{22}+4D_{12}+2D_{13}] \right\}
\end{eqnarray}
where the three- and four-point functions are denoted by
$C_0 \equiv C_0(0,0,\sd,m^2_{\tilde g},m^2_{\tilde g},m^2_{\tilde g})$ and
$[D_l,D_{lm},D_{lmn}] = [D_l,D_{lm},D_{lmn}](m_t^2,0,0,m_t^2,\td,\sd,m^2_{\tilde t_j},m^2_{\tilde g},m^2_{\tilde g},m^2_{\tilde g})$, respectively.\\
{\bf Box b:} $\tilde t_j-\tilde t_j-\tilde t_j-\tilde g$
\begin{eqnarray}
\rho_{2,b}^{\Box,t} & = & - \sum_{j=1,2}  2\lambda_j^+ D_{002}
\nonumber\\
\rho_{4,b}^{\Box,t} & = & - \sum_{j=1,2}  4 \lambda_j^+
[D_{00}+2 D_{001}+D_{002}]
\nonumber\\
\rho_{6,b}^{\Box,t} & = & + \sum_{j=1,2}  2 \lambda_j^+
[D_2+D_{222}+2 (2 D_{12}+D_{22}+D_{112}+D_{123}+2 D_{122})]
\nonumber\\
\rho_{12,b}^{\Box,t} & = & - \sum_{j=1,2} 4 \left\{  - \lambda_j^+
[2 D_{001}+D_{002}]
+ \lambda_j^- \frac{m_{\tilde g}}{\mt}D_{00} \right\}
\nonumber\\
\rho_{16,b}^{\Box,t} & = & - \sum_{j=1,2} 4 \left\{ \lambda_j^+
[D_2+D_{222}+2 (D_1+2D_{11}+4 D_{12}+2 D_{13}+D_{22}+D_{111} \right.
\nonumber\\
&+ &  \left. 3 D_{112}+3 D_{123}+3 D_{113}+3 D_{122})] \right.
\nonumber\\
&- & \left. \lambda_j^- \frac{m_{\tilde g}}{\mt} [D_0+D_{22}+2 (2 D_1+D_2
+D_{11}+2 D_{12}+D_{13})] \right\}\; ,
\end{eqnarray}
with the 4-point functions 
$[D_l,\!D_{lm},\!D_{lmn}]\!=\![D_l,\!D_{lm},\!D_{lmn}](m_t^2,0,0,m_t^2,\td,\sd,m^2_{\tilde g},m^2_{\tilde t_j},m^2_{\tilde t_j},m^2_{\tilde t_j})$.\\
{\bf Box c:} $\tilde t_j-\tilde t_j-\tilde g-\tilde g \quad{\rm and}\quad \tilde g-\tilde g-\tilde t_j-\tilde t_j$
\begin{eqnarray}
\rho_{1,c}^{\Box,t}&=& - \sum_{j=1,2}  4  \lambda_j^+ D_{00}
\nonumber\\
\rho_{2,c}^{\Box,t}&=& - \sum_{j=1,2} 2  \lambda_j^+ \left[2  D_{00}+2  D_{002}+2  D_{003} \right]
\nonumber\\
\rho_{4,c}^{\Box,t}&=& - \sum_{j=1,2} 2  \lambda_j^+ 
\left[-2  D_{00}+(m_{\tilde g}^2-\mt^2) (D_1+D_2)-
 \mt^2 (2  D_{11}+3  D_{12}+D_{22}) \right.
\nonumber\\
&-& \left. \td (D_{12}+D_{22})-
2  D_{001}-2  D_{002}-\mt^2 (D_{111}+2  D_{112}-
2  D_{123}-D_{113}+D_{122}-D_{223}) \right. \nonumber \\
&-& \left. \sd (D_{123}+D_{113})-
\td (D_{112}+2  D_{123}+D_{113}+D_{222}+2  D_{122}+D_{223})\right]
\nonumber\\
\rho_{6,c}^{\Box,t}&=& + \sum_{j=1,2}  4 \lambda_j^+ \left[D_{12}+D_{22}+D_{112}+2  D_{123}+
          D_{113}+D_{222}+2  D_{122}+D_{223} \right]
\nonumber\\
\rho_{11,c}^{\Box,t}&=& + \sum_{j=1,2}  4  \lambda_j^+ D_{00}
\nonumber\\
\rho_{12,c}^{\Box,t}&=& + \sum_{j=1,2} 4  \left\{\lambda_j^+ \left[2  D_{00}+2  D_{001}+2  D_{002}\right]+
          \lambda_j^- \frac{m_{\tilde g}}{\mt} 2  D_{00}\right\}
\nonumber\\
\rho_{14,c}^{\Box,t}&=& + \sum_{j=1,2}  2 \left\{\lambda_j^+ \left[D_1+D_2+D_{11}+2  
D_{12}+D_{22}\right]+
          \lambda_j^-  \frac{m_{\tilde g}}{\mt} (D_1+D_2) \right\}
\nonumber\\
\rho_{16,c}^{\Box,t}&=& - \sum_{j=1,2}  8  \left\{\lambda_j^+ \left[D_1+D_2+2  D_{11}+
4  D_{12}+2  D_{22}+
          D_{111}+3  D_{112}+D_{222}+3  D_{122}\right]\right.
\nonumber\\
&+&\left. \lambda_j^- \frac{m_{\tilde g}}{\mt} \left[D_1+D_2+D_{11}+2  D_{12}+D_{22}\right]\right\} \; .
\end{eqnarray}
with $[D_l,\!D_{lm},\!D_{lmn}]\!=\![D_l,\!D_{lm},\!D_{lmn}](m_t^2,0,m_t^2,0,\td,\hat{u},m^2_{\tilde t_j},m^2_{\tilde g},m^2_{\tilde g},m^2_{\tilde t_j})$. \\[10pt]
{\bf Triangle-box:} $\tilde t_j-\tilde t_j-\tilde g$ 
\begin{eqnarray}
\rho_{12}^{\Box,gg\tilde t\tilde t}&=& \sum_{j=1,2} \left\{  \lambda_j^+ (C_1+C_2) - \lambda_j^- \frac{m_{\tilde g}}{\mt} C_0\right\}(m_t^2,\sd,m_t^2,m^2_{\tilde g},m^2_{\tilde t_j},m^2_{\tilde t_j}) \; .
\end{eqnarray}

\subsection{Parity violating coefficients to the SMEs $M^{A,t}_i$}
\label{sec:parviol}

The $t$-channel parity violating coefficients to the SMEs $M^{A,t}_i$ read as
follows:\\ 
\underline{Vertex corrections:}
\begin{equation}
\sigma_i^{V,t}= \sigma_{i,ttg}^{V,t}+\sigma_{i,ggt}^{V,t}+\sigma_{i,CT}^{V,t}
\end{equation}
with 
\begin{eqnarray}
\sigma_{1,ttg}^{V,t}& =& - \sum_{j=1,2} 
\frac{1}{3} \lambda_j^A C_{00} \nonumber \\
\sigma_{4,ttg}^{V,t}& =& +\sum_{j=1,2} 
\frac{1}{6} \lambda_j^A  (\mt^2-\td) \left[C_2+C_{22}+C_{12} \right]
\nonumber\\
\sigma_{14,ttg}^{V,t}& =& -\sum_{j=1,2} 
\frac{1}{6} \lambda_j^A \left[C_{11}-C_{22}+C_1-C_2)\right]
\end{eqnarray}
with $[C_l,C_{lm}]=[C_l,C_{lm}](m_t^2,0,\td,m^2_{\tilde g},m^2_{\tilde t_j},m^2_{\tilde t_j})$ and
\begin{eqnarray}
\sigma_{1,ggt}^{V,t}& =& +\sum_{j=1,2} \frac{3}{2} \lambda_j^A 
\left[ (m_{\tilde t_j}^2+\mt^2-m_{\tilde g}^2) C_0+2 \mt^2 C_1
+(\td+\mt^2) C_2-2 C_{00}+B_0(0,m^2_{\tilde g},m^2_{\tilde g})\right]
\nonumber \\
\sigma_{4,ggt}^{V,t}& =& +\sum_{j=1,2} \frac{3}{2} 
\lambda_j^A (\mt^2-\td) \left[C_2+C_{22}+C_{12} \right]
\nonumber\\
\sigma_{14,ggt}^{V,t}& =& -\sum_{j=1,2} \frac{3}{2}
 \lambda_j^A \left[C_{11}-C_{22}+C_1-C_2\right]
\end{eqnarray}
with $[C_l,C_{lm}]=[C_l,C_{lm}](m_t^2,0,\td,m^2_{\tilde t_j},m^2_{\tilde g},m^2_{\tilde g})$ 
and the counterterm
\begin{eqnarray}
\frac{\alpha_s}{4\pi} \sigma_{1,CT}^{V,t}& =& {}- 2 \delta Z_A \; ,
\end{eqnarray}
where the renormalization constant, $\delta Z_A$, is given in Eq.~(\ref{eq:renorm}). The parameters 
$\lambda_j^A$ are given in  Eq.~(\ref{eq:lambda}).\\
\underline{Top quark self-energy insertion:}
\begin{eqnarray}
\frac{\alpha_s}{4\pi} \sigma_1^{\Sigma,t} & = & (\hat t-m_t^2)(\Sigma_A(\hat t)+\delta Z_A)
\end{eqnarray}
with the axial vector part of the top quark self-energy, $\Sigma_A$, of 
Eq.~(\ref{eq:topself}). \\
\underline{Box contribution:}\\ 
{\bf Box a:} $\tilde g-\tilde g-\tilde g-\tilde t_j$
\begin{eqnarray}
\sigma_{1,a}^{\Box,t} &=& \sum_{j=1,2} \lambda_j^A 
\, \left[C_0+(m_{\tilde t_j}^2+m_t^2-m_{\tilde g}^2)\, D_0
+(m_{\tilde g}^2+m_t^2)\, D_2-6\, D_{00} \right.
\nonumber\\
&-& \left. \hat t \, (2 D_{22}+D_{222})+4 m_t^2\, D_1
-2 (\hat t+m_t^2)\, (D_{12}+D_{122})
-6\, D_{002}-2 \, m_t^2 D_{112}
\right.
\nonumber\\
&+& \left. (\hat s-2 m_t^2) \, D_{123} \right]
\nonumber\\
\sigma_{2,a}^{\Box,t} &=& -2 \sum_{j=1,2} \lambda_j^A \, [D_{00}+D_{002}]
\nonumber\\
\sigma_{4,a}^{\Box,t} &=& \sum_{j=1,2} 2 \lambda_j^A 
\left[2 D_{00}+m_t^2\, (2 (D_{11}+D_{13})+D_{111}) \right.
\nonumber\\
&+&\left. (m_t^2-m_{\tilde g}^2)\, (D_1+D_2)+(\hat t+5 m_t^2)\, D_{12}
+2 D_{001}+4 D_{002}+\hat t \, D_{222}+(\hat t+3 m_t^2)\, D_{112}
\right.
\nonumber\\
&+& \left. (3 \hat t+2 m_t^2)\, D_{122}+(3 m_t^2-\hat s)\, D_{113}
+(\hat t-\hat s+3 m_t^2) \, D_{123}+(\hat t+m_t^2) \, D_{22} \right]
\nonumber\\
\sigma_{6,a}^{\Box,t} &=& \sum_{j=1,2} 2 \lambda_j^A 
\, [D_{22}+2 D_{12}+D_{222}+2 D_{112}
+4 D_{122}+2 D_{123}]
\nonumber\\
\sigma_{14,a}^{\Box,t} &=& \sum_{j=1,2} 2 \lambda_j^A 
\, [D_2+2 D_{12}+D_{22}]
\end{eqnarray}
with the three- and four-point functions are denoted by
$C_0 \equiv C_0(0,0,\sd,m^2_{\tilde g},m^2_{\tilde g},m^2_{\tilde g})$ and
$[D_l,D_{lm},D_{lmn}] = [D_l,D_{lm},D_{lmn}](m_t^2,0,0,m_t^2,\td,\sd,m^2_{\tilde t_j},m^2_{\tilde g},m^2_{\tilde g},m^2_{\tilde g})$, respectively.\\
{\bf Box b:} $\tilde t_j-\tilde t_j-\tilde t_j-\tilde g$\\[-22pt]
\begin{eqnarray}
\sigma_{2,b}^{\Box,t} & = & -\sum_{j=1,2} 2 \lambda_j^A
D_{002}
\nonumber\\
\sigma_{4,b}^{\Box,t} & = & -\sum_{j=1,2} 4 \lambda_j^A
[D_{00}+2 D_{001}+D_{002}]
\nonumber\\
\sigma_{6,b}^{\Box,t} & = & \sum_{j=1,2} 2 \lambda_j^A
[D_2+D_{222}+2 (2 D_{12}+D_{22}+D_{112}+D_{123}+2 D_{122})]
\end{eqnarray}
with the 4-point functions 
$[D_l,\!D_{lm},\!D_{lmn}]\!=\![D_l,\!D_{lm},\!D_{lmn}](m_t^2,0,0,m_t^2,\td,\sd,m^2_{\tilde g},m^2_{\tilde t_j},m^2_{\tilde t_j},m^2_{\tilde t_j})$.\\[2pt]
{\bf Box c:} $\tilde g-\tilde g-\tilde t_j-\tilde t_j \quad {\rm and}\quad \tilde t_j-\tilde t_j-\tilde g-\tilde g$\\[-23pt]
\begin{eqnarray}
\sigma_{1,c}^{\Box,t} & = & -\sum_{j=1,2}  4 \lambda_j^A  D_{00}
\nonumber \\
\sigma_{2,c}^{\Box,t} & = & -\sum_{j=1,2}  4 \lambda_j^A
[D_{00}+D_{002}+D_{003}]
\nonumber \\
\sigma_{4,c}^{\Box,t} & = & \sum_{j=1,2} 2 \lambda_j^A
[2 D_{00}+2 D_{001}+2D_{002}-m_{\tilde g}^2 (D_1+D_2)
\nonumber\\
&+&\mt^2 (D_1+D_2+2 D_{11}+3D_{12}+D_{22}+D_{111}+2D_{112}-2D_{123}-D_{113}
+D_{122}-D_{223})
\nonumber\\
&+& \td (D_{12}+D_{22}+D_{112}+2D_{123}+D_{113}+D_{222}+2 D_{122}+D_{223})
+\sd (D_{123}+D_{113})]
\nonumber \\
\sigma_{6,c}^{\Box,t} & = & \sum_{j=1,2}  4\lambda_j^A
[D_{12}+D_{22}+D_{112}+2D_{123}+D_{113}+D_{222}+2D_{122}+D_{223}]
\nonumber \\
\sigma_{14,c}^{\Box,t} & = & \sum_{j=1,2}  2\lambda_j^A
[D_{1}+D_2+D_{11}+2D_{12}+2D_{13}+D_{22}+2D_{23}]
\end{eqnarray}\\[-16pt]
with $[D_l,\!D_{lm},\!D_{lmn}]\!=\![D_l,\!D_{lm},\!D_{lmn}](m_t^2,0,m_t^2,0,\td,\hat{u},m^2_{\tilde t_j},m^2_{\tilde g},m^2_{\tilde g},m^2_{\tilde t_j})$.

Finally, the $u$-channel contributions
$\rho_i^{(V,\Sigma),u},\rho_{i,(a,b)}^{(\Box),u}$ and
$\sigma_i^{(V,\Sigma),u}, \sigma_{i,(a,b)}^{(\Box),u}$ can be obtained
from the $t$-channel form factors by replacing $\hat t$ with $\hat u$.

\subsection{The color factors}
\label{sec:color}

The summation (average) over the color degrees of freedom
in the course of the derivation of the parton cross sections
leads to the following color factors:
\begin{eqnarray}
\label{eq:colorstu}
c^s(1) & = &\sum_{a,b,c;\atop j,l} (f_{abc} T_{jl}^c) \; (f_{abd} T_{jl}^d)^*
 =  3 \sum_{c,d}\delta_{cd} \; \tr\{T^c T^d\} = 12 \nonumber\\
c^t(1) & = & \sum_{a,b,c;\atop j,l}(iT_{jm}^a T_{ml}^b)
\; (f_{abc} T_{jl}^c )^* = c^s(2) = -6  \nonumber\\
c^u(1) & = &\sum_{a,b,c;\atop j,l}
(iT_{jm}^b T_{ml}^a)\; (f_{abc} T_{jl}^c)^*
= c^s(3) = 6 \nonumber\\
c^t(2) & = & -\sum_{a,b,c;\atop j,l}
(iT_{jm}^a T_{ml}^b)\;(-iT_{jm}^a T_{ml}^b)^*
= c^u(3) = \frac{16}{3} \nonumber\\
c^u(2) & = & -\sum_{a,b,c;\atop j,l}
(iT_{jm}^b T_{ml}^a)\; (-iT_{jm}^a T_{ml}^b)^*
= c^t(3) = -\frac{2}{3} 
\end{eqnarray}
and\\[-15pt]
\begin{eqnarray}
\label{eq:colorbox}
c^t_a(1)&=&\sum_{a,b,c;\atop j,l} (\frac{3}{2} (-iT_{jm}^a T_{ml}^b)-\frac{1}{4}(i\delta_{ab} \delta_{ij}))
\; (f_{abc} T^c_{jl})^* =\frac{3}{2} c^t(1) \nonumber\\
c^t_a(2)&=&\frac{3}{2} c^t(2)-\frac{1}{4}\sum_{a,b,c;\atop j,l}(i\delta_{ab} \delta_{ij})
\; (-iT_{jm}^a T_{ml}^b)^* =\frac{3}{2} c^t(2)+1 \nonumber\\
c^t_a(3)&=&\frac{3}{2} c^t(3)-\frac{1}{4}\sum_{a,b,c;\atop j,l}(i\delta_{ab} \delta_{ij})
\; (-iT_{jm}^b T_{ml}^a)^*=\frac{3}{2} c^t(3)+1 \nonumber\\
c^t_b(1)&=&\frac{1}{6} c^t(1) \nonumber\\
c^t_b(2)&=&\frac{1}{6} c^t(2)+\frac{1}{4}\sum_{a,b,c;\atop j,l}(i\delta_{ab} \delta_{ij})
\; (-iT_{jm}^a T_{ml}^b)^* =\frac{1}{6} c^t(2)-1 \nonumber\\
c^t_b(3)&=&\frac{1}{6} c^t(3)+\frac{1}{4}\sum_{a,b,c;\atop j,l}(i\delta_{ab} \delta_{ij})
\; (-iT_{jm}^b T_{ml}^a)^*=\frac{1}{6} c^t(3)-1 \nonumber\\
c^t_c(1)&=&0 \nonumber\\
c^t_c(2)&=&c^t_c(3)=-\frac{1}{4}\sum_{a,b,c;\atop j,l}(i\delta_{ab} \delta_{ij})
\; (-iT_{jm}^a T_{ml}^b)^* =1 \nonumber\\
c^u_a(1)&=&\frac{3}{2} c^u(1) \nonumber\\
c^u_a(2)&=&\frac{3}{2} c^u(2)-\frac{1}{4}\sum_{a,b,c;\atop j,l}(i\delta_{ab} \delta_{ij})
\; (-iT_{jm}^a T_{ml}^b)^*=\frac{3}{2} c^u(2)+1 \nonumber\\
c^u_a(3)&=&\frac{3}{2} c^u(3)-\frac{1}{4}\sum_{a,b,c;\atop j,l}(i\delta_{ab} \delta_{ij})
\; (-iT_{jm}^b T_{ml}^a)^*=\frac{3}{2} c^u(3)+1 \nonumber\\
c^u_b(1)&=&\frac{1}{6} c^u(1)  \nonumber \\
c^u_b(2)&=&\frac{1}{6} c^u(2)+\frac{1}{4}\sum_{a,b,c;\atop j,l}(i\delta_{ab} \delta_{ij})
\; (-iT_{jm}^a T_{ml}^b)^*=\frac{1}{6} c^u(2)-1 \nonumber\\
c^u_b(3)&=&\frac{1}{6} c^u(3)+\frac{1}{4}\sum_{a,b,c;\atop j,l}(i\delta_{ab} \delta_{ij})
\; (-iT_{jm}^b T_{ml}^a)^*=\frac{1}{6} c^u(3)-1 \; .
\end{eqnarray}


%
%
\bibliographystyle{apsrev}
\bibliography{ppttsqcd}

%
%
\end{document}